\documentclass[a4paper,12pt]{report}

\usepackage{amsfonts}
\usepackage{amsmath}
\usepackage{epsf}
\usepackage{texdraw}
\usepackage{amscd}
\usepackage{makeidx}

\newcounter{stuff}

\newtheorem{prop}[stuff]{Proposition}

\newtheorem{thm}[stuff]{Theorem}
\newtheorem{defi}[stuff]{Definition}

\newtheorem{lem}[stuff]{Lemma}

\newcommand{\bra}[1]{\ensuremath{\,\langle#1\vert}}
\newcommand{\ket}[1]{\ensuremath{\,\vert#1\/\rangle}}

\newcommand{\comm}[2]{\ensuremath{[\, #1\, ,\, #2\,]}}
\newcommand{\ip}[2]{\ensuremath{\,\langle#1\vert#2\/\rangle}}
\newcommand{\su}[2][]{\ensuremath{\mathfrak{su}^{#1}_{#2}}}
\newcommand{\End}[1]{\ensuremath{\mathrm{End}\left(#1\right)}}
\newcommand{\tr}{\ensuremath{\mathrm{Tr}}}
\newcommand{\Arg}{\ensuremath{\mathrm{Arg}}}
\newcommand{\Hyper}[4]{{ }_2F_1\left(\begin{array}{c} #1\;;\;#2\\ 
#3\end{array};#4\right)}

\newcommand{\C}{\mathbb{C}}
\newcommand{\R}{\mathbb{R}}
\newcommand{\N}{\mathbb{N}}
\newcommand{\Z}{\mathbb{Z}}
\newcommand{\dint}{\ensuremath{\int}}

\renewcommand{\Im}{\mathrm{Im}\,}
\renewcommand{\Re}{\mathrm{Re}\,}
\renewcommand{\hom}[3][]{\ensuremath{\mathrm{Hom}_{#1}\left(#2,\, #3\right)}}
\renewcommand{\arraystretch}{1.5}

\numberwithin{stuff}{chapter}
\title{A State Sum Model for\\(2+1) Lorentzian Quantum Gravity} 
\author{by Stefan Davids, BSc.~MSc.}
\date{Thesis submitted to the University of Nottingham\\ 
for the degree of Doctor of Philosophy,\\October, 2000}

\begin{document}
\renewcommand{\arraystretch}{0.7}

\pagenumbering{roman}
\maketitle
\tableofcontents

\begin{abstract}

A state sum model based on the group $SU(1,1)$ is defined.
Investigations of its geometry and asymptotics suggest it is a good
candidate for modelling $2+1$ Lorentzian quantum gravity.

\end{abstract}

\listoffigures\listoftables\newpage
\pagenumbering{arabic}

\chapter{Introduction}

The purpose of this work is to propose a state sum model, derived from
the group $SU(1,1)$, as a candidate for $2+1$ Lorentzian quantum
gravity.

Topological quantum field theories\cite{A89} have been thought for
some time to provide a reasonable mathematical framework for
investigating models which might provide good candidates for quantum
gravity\cite{Ba95}.  In essence their formulation naturally allows
relations between algebraic and geometric manifestations of objects;
the idea is that this blending of algebra and geometry should be a
natural framework in which to interpolate between the discrete
algebraic world of quantum mechanics and the smooth geometric world of
general relativity.  The general notion of a topological quantum field
theory and the state sum model is discussed in section \ref{tqft}.

Since this framework is very general the real interest in topological
quantum field theories comes from the formulation of examples, their
properties and geometry.  In terms of examples perhaps the most
important is the Ponzano-Regge model\cite{PR}, which predates the
formal notion of a topological quantum field theory by around twenty
years.  This model is discussed in section \ref{pr-sect}.

As the state sum to be proposed is based on some crucial facts about
$SU(1,1)$ representation theory, the group $SU(1,1)$ is the subject of
chapter \ref{SU11-CHAP}.  Its representation theory, mostly due to
Bargmann\cite{Ba47}, is discussed in section \ref{rep}.  The geometry
of any state sum model is essentially based on the tensor structure of
the category of its representations.  For $SU(1,1)$ the tensor
structure of its representation series, as defined by its
Clebsch-Gordon coefficients, is discussed in section \ref{CGC}.

The essence of any three dimensional state sum with a geometric
meaning is a correspondence, given by the underlying topological
quantum field theory, between tetrahedra in the (simplicial) manifold
of interest and the Racah coefficients, or 6j symbols, defined by the
representation theory.  The Racah coefficients for $SU(1,1)$ are thus
the subject of chapter \ref{RACAH}.

In section \ref{Racah} a definition of the Racah coefficient for the
$SU(1,1)$ representation series of interest is provided and some
elementary relations presented.  The definition given in section
\ref{Racah} is then shown to give a well defined function in section
\ref{converg}.  In section \ref{rac-CGC} some symmetries of the Racah
coefficient are investigated which allow it to be exhibited in a
closed form for some of the cases of interest in section \ref{rac}.

The definition of the proposed state sum model is given in chapter
\ref{Quantum-Geometry}.  In section \ref{Geom} the relation between
irreducible representations of $SU(1,1)$ and three dimensional
Lorentzian vectors is given.  Sections \ref{state} and
\ref{inv} are devoted to the statement of the proposed state sum and
a discussion of its topological invariance.

The most important part of any proposed model for quantum gravity is
being able to make contact with existing theories in some limit.
Indeed the importance of the Ponzano-Regge model stems from its
ability to recover a formula that looks like a path integral from the
state sum for a given manifold in a suitable asymptotic limit.
Chapter \ref{ASYMPT} is devoted to investigating the asymptotic limit
of the state sum defined in section \ref{state} in certain special
cases.

Finally our conclusions, and areas where further work is needed, are
presented in chapter \ref{CON}.

\chapter{TQFT's and Ponzano-Regge}

In this chapter topological quantum field theories (TQFT's) are
defined in terms of the Atiyah axioms.  In section \ref{tqft} we
define the notion of a state sum model and illustrate it with a simply
finite group state sum in two dimensions.  In section \ref{pr-sect}
the important formal\footnote{it is formal since it is not a true
TQFT; it fails to be completely topologically invariant} TQFT
known as the Ponzano-Regge model is introduced.

\section{TQFTs and State Sums}\label{tqft}

The formal notion of a TQFT is due to Atiyah\cite{A89} and defines a
natural way of associating Hilbert space and manifolds.  Axiomatically
one has the following

\begin{defi}[Atiyah]
A TQFT of dimension $d$ is a map, $Z$, that associates to each
$d$-dimensional closed, oriented manifold $M$, a Hilbert space
$Z\left(M\right)$, and to each $d+1$-dimensional oriented manifold,
$N$, a vector $Z\left(N\right)\in Z\left(\partial N\right)$ where
$\partial N$ denotes the boundary of $N$, such that the following
axioms are satisfied
\begin{itemize}
\item $Z$ is a functor from the category of $d$-dimenional manifolds,
with morphisms given by cobordism, to the category of Hilbert spaces,
with the morphisms being normal maps of Hilbert spaces. 
\item $Z$ is involutory where the involution in the category of
manifolds is orientation reversal and in the category of the Hilbert
spaces it is duality
\item $Z$ is multiplicative in the sense that a  disjoint union of
manifolds is mapped to the tensor product of the associated Hilbert
spaces.  That is $Z\left(M_1\sqcup M_2\right) =
Z\left(M_1\right)\otimes Z\left(M_2\right)$
\end{itemize}
\end{defi}
One may consider either the category of smooth or piecewise linear
manifolds.  The Hilbert space may also be weakened to a finitely
generated module over some ground ring.

There are three non-triviality axioms imposed, namely that
$Z\left(\phi\right) = \mathbb{F}$ (for $\phi$ the empty $d$-manifold
and $\mathbb{F}$ the underlying field) and $Z\left(\phi\right) = 1$
(for $\phi$ the empty $d+1$-manifold).  

The third concerns the cobordism $M\times I$ with $M$ some
$d$-dimensional manifold and $I$ the unit interval.  The
multiplicativeness of $Z$ implies that $Z\left(M\times I\right)$ is a
vector in the Hilbert space $Z\left(M\right)\otimes
Z\left(M\right)^\star$, or equivalently a morphism in the space
$\End{Z\left(M\right)}$.  It is easy to see, by composing cobordisms,
that $Z\left(M\times I\right)$ is idempotent and it is normally taken
to be the identity.

The theory as defined by the above axioms is a very broad one, yet
already it is possible to see certain characteristics that one would
want in a theory of quantum gravity, such as topological invariance
and the correct rules for composing quantum amplitudes\cite{Ba95}.
However our primary concern is in a specific class of examples of
TQFT's defined by \emph{state sums}\footnote{although the ones of most
interest here are not true TQFT's since they are not finite on every
manifold or completely topologically invariant}.  These define a
functor from the category of piecewise linear manifolds to the
category of Hilbert spaces which is dependent simply on some initial
data such as a group, category or Hopf algebra.

Typically a state sum consists of assigning some algebraic data to
simplices and then gluing the simplices together in a consistent way
to form a PL-manifold.  The gluing of the simplices translates to
multiplying their partition functions, $Z$, together with data
labelling common lower dimensional simplices constrained to be equal.
Finally a sum is taken over all possible labellings that are
consistent with some constraint arising from higher dimensional
simplices.  Their main advantage over other classes of TQFT is the
reliance simply on \emph{local} data and as such it should have more
of a flavour of quantum gravity (since general relativity is governed
by local observers) than a TQFT that relies on non-local data.

The simplest non-trivial example concerns a closed triangulated
2-manifold and a finite group as the initial data.  The state sum
consists of assigning group elements to the edges in the triangulation
in such a way that they multiply around each triangle to give the
identity.  Since the manifold is closed the state sum should yield a
number (an element of the underlying field associated to the empty
boundary) and the number is simply the number of ways this labelling
may be done, divided by the order of the finite group to the power of
the number of vertices in the triangulation.

\begin{figure}
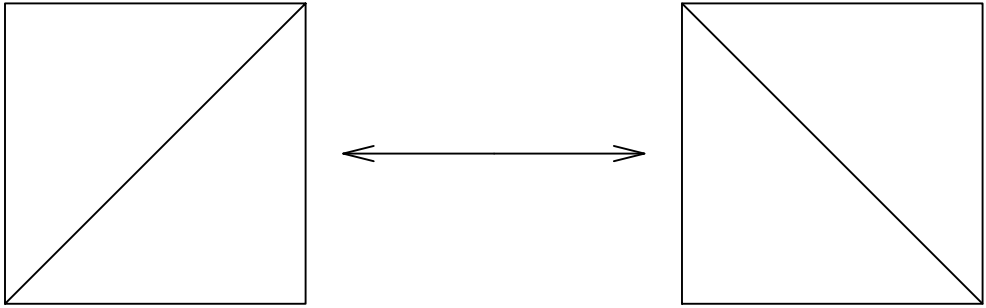

\begin{center}
\begin{texdraw}
\drawdim{mm}
\arrowheadtype t:V
\move(0 0)\lvec(40 0)\lvec(40 40)\lvec(0 40)\lvec(0 0)
\lvec(40 40)
\move(90 0)\lvec(130 0)\lvec(130 40)\lvec(90 40)\lvec(90 0)
\move(130 0) \lvec(90 40)
\move(65 20)\avec(45 20)
\move(65 20)\avec(85 20)
\end{texdraw}
\caption[The 2-2 Pachner move]
{The 2-2 Pachner move.  Two triangles sharing a common edge are
transformed into two different triangles sharing a common
edge.\label{2-2}}
\end{center}
\end{figure}

\begin{figure}
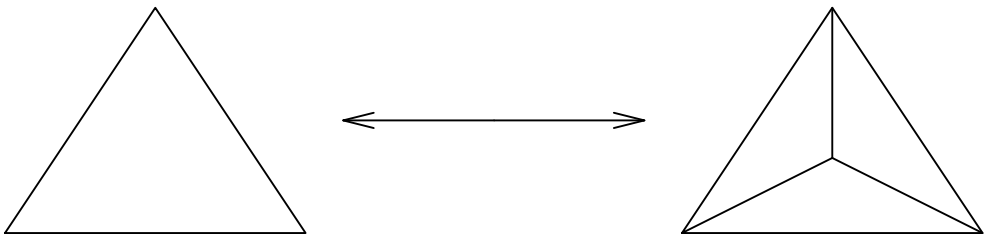

\begin{center}
\begin{texdraw}
\drawdim{mm}
\arrowheadtype t:V
\move(0 0)\lvec(40 0)\lvec(20 30)\lvec(0 0)

\move(90 0)\lvec(130 0)\lvec(110 30)\lvec(90 0)
\lvec(110 10)\move(130 0)\lvec(110 10)\move(110 30)\lvec(110 10)
\move(65 15)\avec(45 15)
\move(65 15)\avec(85 15)
\end{texdraw}
\caption[The 3-1 Pachner move]
{The 3-1 Pachner move.  Three triangles, each sharing a common edge,
are transformed into one triangle by removing the internal vertex and
associated edges.\label{3-1}}
\end{center}
\end{figure}

To show it is topologically invariant one needs the Pachner
moves\cite{Pa91} in two dimensions.  There are only two unique moves
here; the 2 to 2 move, shown in figure \ref{2-2}, and the 3 to 1 move
(or Barycentric subdivision) shown in figure \ref{3-1}.  Topological
invariance is simply the fact that the above state sum is invariant
under these two moves on the triangulation.

The 2 to 2 Pachner move leaves the state sum trivially invariant since
the internal edge has its label fixed by the boundary data and this is
common to both sides of the equation in figure \ref{2-2}.  The 3 to 1
move is less trivial, here on the right hand side of figure
\ref{3-1} it is easily verified that all bar one of the internal edges
are fixed by the boundary data and this one edge has a free choice of
labelling.  Thus one acquires a factor equal to the order of the
group, which precisely cancels the factor of one over the order of the
group arising from the extra internal vertex as required.

With some generalisations, in three dimensions this simple example
eventually ends up covering a large class of state sum models.  For
instance one may generalise the finite group to a finite\footnote{here
finite means that the co-integral on the Hopf algebra determines a
finite sum} involutory Hopf algebra and recover the state sum version
of Kuperberg's model\cite{K91} as discussed in \cite{BW95a}.  More
interesting is using the equivalent dual data, that is the category of
representations of the group (or Hopf algebra) and consider a Lie
group rather than a finite group.  This leads to the Ponzano-Regge
model which is the subject of the next section.

\section{The Ponzano-Regge Model}\label{pr-sect}

The Ponzano-Regge model formulated in \cite{PR} predated the whole
idea of TQFT's and state sums by nearly twenty years.  While not a
true TQFT itself it is nevertheless interesting because of its close
connection to physics.

The state sum is defined for the group $SU\left(2\right)$, the 3
simplices (tetrahedra) in the simplicial decomposition of the manifold
have their edges labeled by the finite dimensional unitary
representations of $SU\left(2\right)$ (essentially non-negative half
integers).  Each 3-simplex is then labeled with the relevant
\su{2} Racah coefficient (which is essentially a complex number
determined by the six representations labeling the 3-simplex's six
edges).

\begin{figure}[htb]
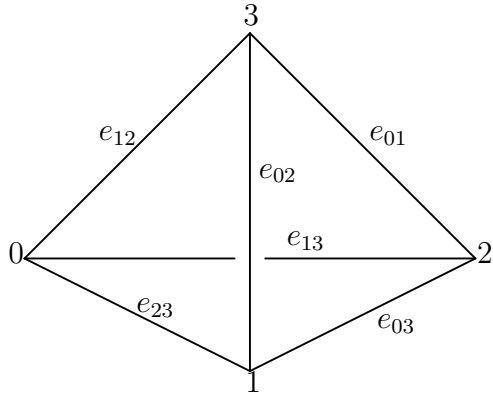

\begin{center}
\begin{texdraw}
\drawdim{mm}
\lvec(-30 15) \move(-15 7) \htext{$e_{23}$} 
\move(0 0) \lvec(30 15) \move(17 5) 
\htext{$e_{03}$} \move(30 15)
\lvec(2 15)\move(-2 15)\lvec(-30 15) 
\move(5 16) \htext{$e_{13}$} \move(-30 15)
\lvec(0 45) \move(-20 30) \htext{$e_{12}$} \move(0 45) 
\lvec(0 0) \move(1.3 25) \htext{$e_{02}$} \move(0 0)
\move(0 45) \lvec(30 15) \move(16 30)\htext{$e_{01}$}
\move(-0.5 -3) \htext{1}
\move(-32 14.2) \htext{0}
\move(30.3 14.2) \htext{2}
\move(-0.8 46) \htext{3}
\end{texdraw}
\end{center}
\caption[A labeled tetrahedron]
{A labeled tetrahedron.  The edge $e_{kl}$ is determined by deleting
the $k$-th and $l$-th vertices.\label{tetrahedron0}}
\end{figure}

To make this more precise, consider a labeled tetrahedron, as in
figure \ref{tetrahedron0} with an arbitrary order to the vertices.  To
each edge there is an direction assigned which is induced by the
ordering of the vertices.  Similarly each face has a direction induced
by the ordering of its vertices. An edge or face is said to be
positively oriented if the direction induced by the tetrahedron agrees
with the direction induced by the ordering of its vertices; otherwise
it is negatively oriented.

The four faces of the tetrahedron are then labeled by the tensor
product of the three representations labeling the edges.  Thus the
face formed by the vertices 0, 1 and 2 is labeled by the Hilbert
space $e_{23}\otimes e_{03}\otimes e_{13}^\star$ or equivalently the
space of morphisms
\hom{e_{23}\otimes e_{03}}{e_{13}}.  This second space is parameterised
by the Clebsch-Gordon coefficients\cite{AngMom} for \su{2} which
determine the decomposition of a tensor product representation in
terms of a direct sum of representations.

In the tetrahedron each edge will appear in one space of morphisms
with a positive orientation and once with a negative orientation,
since two faces of the tetrahedron are positively oriented and two are
negatively oriented.  Thus the positively oriented tetrahedron,
say\footnote{the choice of orientation is normally induced by whether
it agrees or disagrees with the orientation determined by the given
PL-manifold}, must correspond to the Hilbert space
\begin{multline}
\left(e_{23}\otimes e_{03}\otimes e_{13}^\star\right)\bigotimes
\left(e_{13}\otimes e_{01}\otimes e_{12}^\star\right)\bigotimes\\
\left(e_{03}\otimes e_{01}\otimes e_{02}^\star\right)^\star\bigotimes
\left(e_{23}\otimes e_{02}\otimes e_{12}^\star\right)^\star
\end{multline}
but it is clear that this is simply the space of maps from
$\hom{e_{23}\otimes e_{03}}{e_{13}}
\otimes\hom{e_{13}\otimes e_{01}}{e_{12}}$ to 
$\hom{e_{03}\otimes e_{01}}{e_{02}}\otimes
\hom{e_{23}\otimes e_{02}}{e_{12}}$ and as such it provides a change
of basis between the two possible ways of decomposing the tensor
product of three representations $e_{23}\otimes e_{03}\otimes e_{01}$
into the direct sum of representations $e_{12}$.  This change of basis
is precisely determined by the Racah coefficient for the given
representation series, namely $
\left\{\begin{array}{ccc}e_{23} & e_{03} & e_{13}\\
e_{01} & e_{12} & e_{02} \end{array}\right\}$ 

A precise definition of the Racah coefficient in terms of a
representation theoretic view point is given in section
\ref{Racah} in the context of \su{1,1} (although of course similar
ideas apply to \su{2}) and a specific formula is given in equation
\ref{6jdef}.  

It is a remarkable thing that the Racah coefficient is
non-zero\footnote{To be more precise it is zero when the
representation labels cannot be the edge lengths of a tetrahedron; the
converse is not quite true since the Racah coefficient is oscillatory
within the general domain of being interpretable as a tetrahedron.}
when the representation labels could be interpreted as edge lengths of
a tetrahedron.  If the tetrahedron has a real volume then it may be
identified with a tetrahedron embedded in Euclidean space, while an
imaginary volume allows it to be identified with a tetrahedron
embedded in Lorentzian space with all faces and edges space-like.
Triangle inequalities are satisfied for the edges determining the
faces of the tetrahedron in both cases.

To establish our conventions, the Racah coefficient is determined by
the equation
\begin{multline}
\left[\begin{array}{ccc}
j_1 & j_2 & j_{12}\\m_1 & m_2 & m_{12}
\end{array}\right]
\left[\begin{array}{ccc}
j_{12} & j_3 & j\\m_{12} & m_3 & m
\end{array}\right]\\ 
= \sum_{j_{23},\,m_{23}}\left(2j_{23}+1\right)
\left\{\begin{array}{ccc}
j_1 & j_2 & j_{12}\\j_3 & j & j_{23} 
\end{array}\right\}
\left[\begin{array}{ccc}
j_1 & j_{23} & j\\m_1 & m_{23} & m
\end{array}\right]
\left[\begin{array}{ccc}
j_2 & j_3 & j_{23}\\m_2 & m_3 & m_{23}
\end{array}\right]
\end{multline}
so that it differs by a phase $\left(-1\right)^{j_1+j_2+j+j_3}$ when
compared to the 6j symbol, for which Ponzano and Regge use the same
notation\footnote{The decision to use the phase conventions of the
Racah coefficient stems from a desire to make analogies between this
\su{2} case and the case for \su{1,1} where the Racah coefficient is a
more natural object.}.

The state sum is then formed as follows, for a given fixed simplicial
decomposition of a closed 3-manifold choose an arbitrary ordering of
the vertices.  To each tetrahedron assign a Racah coefficient whose
representation labels correspond to the edge lengths of the
tetrahedron in question and product the Racah coefficients for every
tetrahedron in the decomposition.  This product is then weighted by a
factor $2j+1$ (the dimension of the representation) for each edge,
labeled by a representation $j$ and finally a sum is take over all
representations labeling all the edges.

Thus for the closed 3-manifold $\mathfrak{M}$, the value of the state
sum (or partition function by analogy with statistical mechanics),
$Z\left(\mathfrak{M}\right)$, is given by
\begin{multline}\label{pr-ss}
Z\left(\mathfrak{M}\right) = \sum_{\mathrm{edges\ }j}
\left(2j+1\right)\left(-1\right)^{2j}
\prod_{\mathrm{faces}}\left(-1\right)^{a+b+c}\\
\prod_\mathrm{tetrahedra}
\left(-1\right)^{j_1+j_2+j_3+j}
\left\{\begin{array}{ccc} j_1 & j_2 & j_{12}\\
j_3 & j & j_{23} \end{array}\right\}
\end{multline}
where $a$, $b$, $c$ are the representations labeling the given face
and $j_1$, $j_2$, $j_3$, $j_{12}$, $j_{23}$ and $j$ are the
representations labeling the given tetrahedra.

Since we have used the phase convention corresponding to the 6j symbol
equation \ref{pr-ss} is invariant under reordering of vertices as
the 6j symbol is invariant under the 144 classical Regge symmetries.
If the Racah coefficient phase conventions had been used then any
reordering of vertices would have required a phase, depending on the
exact symmetry the reordering induces on the various labels, be
associated to the various Racah coefficients determined by the
simplicial decomposition of the manifold.

As with the toy state sum involving a finite group in the previous
section one must show that it is indeed a topological invariant by
checking invariance under the Pachner moves, and also that it is
invariant under relabeling of the vertices since the original choice
of labeling was arbitrary.  The latter follows from the known
symmetries of the Racah coefficient, the 144 Regge
symmetries\cite{Reg59}.

\begin{figure}[hbt]
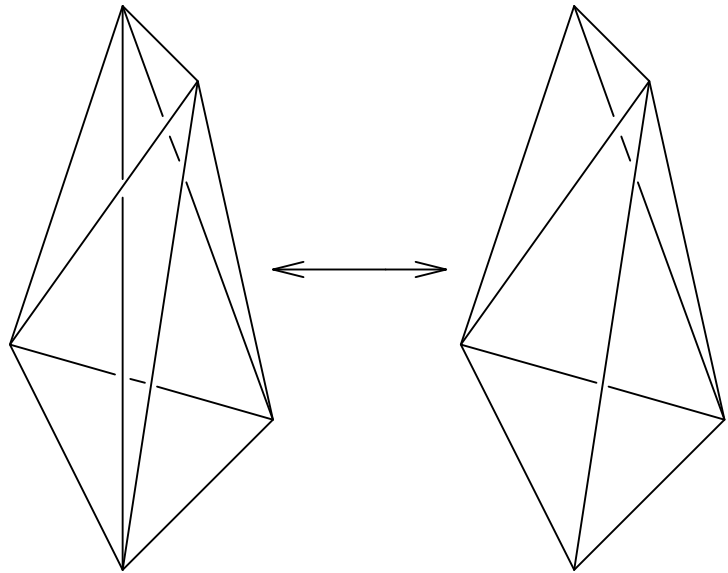

\begin{center}
\begin{texdraw}
\drawdim{mm}
\arrowheadtype t:V
\move(0 0)\lvec(0 49.5)\move(0 52)\lvec(0 75)
\move(0 0)\lvec(20 20)\lvec(8.5 51.6)\move(7.5 54.4)
\lvec(6.3 57.7)\move(5.3 60.4)\lvec(0 75)
\move(0 0)\lvec(-15 30)\lvec(0 75)
\move(0 0)\lvec(10 65)\lvec(0 75)
\move(20 20)\lvec(10 65)
\lvec(-15 30)
\lvec(-1 26)\move(1 25.4)\lvec(3 24.9)\move(4.6 24.4)\lvec(20 20)

\move(60 0)\lvec(80 20)\lvec(68.5 51.6)\move(67.5 54.4)
\lvec(66.3 57.7)\move(65.3 60.4)\lvec(60 75)
\move(60 0)\lvec(45 30)\lvec(60 75)
\move(60 0)\lvec(70 65)\lvec(60 75)
\move(80 20)\lvec(70 65)
\lvec(45 30)
\lvec(63 24.9)\move(64.6 24.4)\lvec(80 20)
\move(35 40)\avec(20 40)
\move(35 40)\avec(43 40)
\end{texdraw}
\caption[The 3-2 Pachner move]
{The 3-2 Pachner move.  Three tetrahedra sharing a common edge are
replaced by two tetrahedra sharing a common face.\label{3-2}}
\end{center}
\end{figure}

\begin{figure}[hbt]
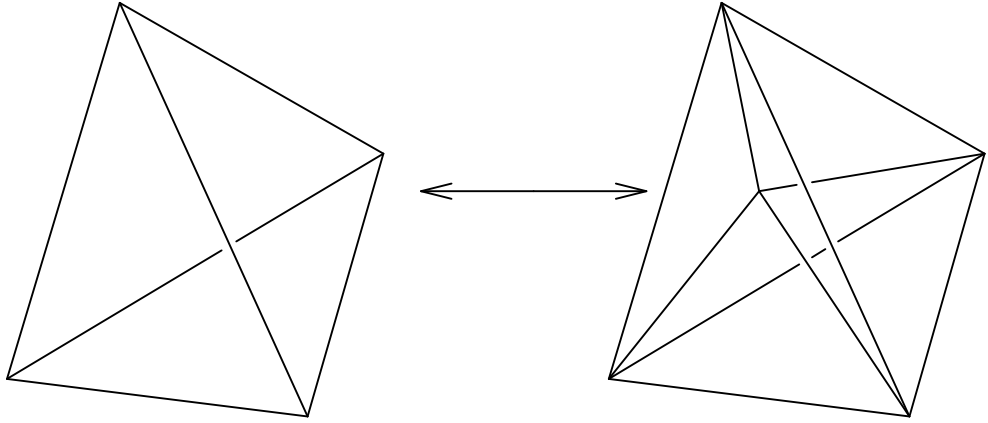

\begin{center}
\begin{texdraw}
\drawdim{mm}
\arrowheadtype t:V
\move(0 0)\lvec(15 50)
\move(0 0)\lvec(40 -5)
\move(0 0)\lvec(28.5 17.1)\move(30.5 18.3)\lvec(50 30)
\move(15 50)\lvec(40 -5)
\move(15 50)\lvec(50 30)
\move(50 30)\lvec(40 -5)
\move(80 0)\lvec(95 50)
\move(80 0)\lvec(120 -5)
\move(80 0)\lvec(105.3 15.2)\move(107 16.2)
\lvec(108.8 17.3)\move(110.5 18.3)\lvec(130 30)
\move(95 50)\lvec(120 -5)
\move(95 50)\lvec(130 30)
\move(130 30)\lvec(120 -5)
\move(100 25)\lvec(80 0)
\move(100 25)\lvec(95 50)
\move(100 25)\lvec(105 25.8)\move(107 26.2)\lvec(130 30)
\move(100 25)\lvec(120 -5)
\move(70 25)\avec(55 25)
\move(70 25)\avec(85 25)
\end{texdraw}
\caption[4-1 Pachner Move]{The 4-1 Pachner move caused by
adding or removing an internal vertex\label{4-1-pachner}}
\end{center}
\end{figure}

The Pachner moves in three dimensions consist of the 3 to 2
move, shown in figure \ref{3-2}, where three tetrahedra sharing a
common edge are replaced by two tetrahedra sharing a common face, and
the four to one move (barycentric subdivision), shown in figure
\ref{4-1-pachner} where adding an internal transforms one tetrahedron
into four tetrahedra each sharing a common face with two others.

The 3 to 2 Pachner move when written in terms of Racah coefficients is
precisely the Biedenharn-Elliot relation\cite{B53}
\begin{multline}
\sum_{j_{23}}\,\left(2j_{23}+1\right)\,
\left\{\begin{array}{ccc}
j_2 & j_3 & j_{23} \\
j_4 & j_{234} & j_{34}
\end{array}\right\}
\left\{\begin{array}{ccc}
j_1 & j_{23} & j_{123} \\
j_4 & j & j_{234}
\end{array}\right\}
\left\{\begin{array}{ccc}
j_1 & j_2 & j_{12} \\
j_3 & j_{123} & j_{23}
\end{array}\right\}\\ = 
\left\{\begin{array}{ccc}
j_1 & j_2 & j_{12} \\
j_{34} & j & j_{234}
\end{array}\right\}
\left\{\begin{array}{ccc}
j_{12} & j_3 & j_{123} \\
j_4 & j & j_{34}
\end{array}\right\}
\end{multline}

As shown in \cite{PR} it is also possible to derive an equation
relating four Racah coefficients to one Racah coefficient, using
orthogonality (see for instance equation 9.8.3 of \cite{AngMom}) of
the Racah coefficients, that is
\begin{equation}
\sum_{j_{12}}\left(2j_{12}+1\right)\,
\left\{\begin{array}{ccc}
j_1 & j_2 & j_{12}\\
j_3 & j & j_{23}
\end{array}\right\}
\left\{\begin{array}{ccc}
j_1 & j_2 & j_{12}\\
j_3 & j & j_{23}^\prime
\end{array}\right\} = \frac{\delta_{j_{23},j_{23}^\prime}}
{\left(2j_{23}+1\right)}
\end{equation}
and the Biedenharn-Elliot relation above.  This may be written
\begin{multline}\label{su2-4-1}
\sum_{j_{23}\,j_{234}}
\,\left(2j_{23}+1\right)\,\left(2j_{234}+1\right)
\left\{\begin{array}{ccc}
j_1 & j_2 & j_{12}\\ j_3 & j_{123} & j_{23}
\end{array}\right\}
\left\{\begin{array}{ccc}
j_1 & j_{23} & j_{123}\\j_4 & j & j_{234}
\end{array}\right\}\\
\left\{\begin{array}{ccc}
j_2 & j_3 & j_{23}\\ j_4 & j_{234} & j_{34}
\end{array}\right\}
\left\{\begin{array}{ccc}
j_1 & j_2 & j_{12}\\j_{34} & j & j_{234}
\end{array}\right\}
= \left\{\begin{array}{ccc}
j_{12} & j_3 & j_{123}\\ j_4 & j & j_{34}
\end{array}\right\}
\frac{1}{2j_{12}+1}
\end{multline}

For this algebraic relation to translate correctly into the geometric
4 to 1 Pachner move one must sum out the four internal edges that
connect the extra internal vertex to the four boundary vertices. Thus
it is clear a sum must be taken over $j_1$ and $j_2$ in addition to
the two already present (for consistency the sums are weighted with
the dimension of the representation as the other two are).  The right
hand side of equation \ref{su2-4-1} is thus written
\begin{equation}
\sum_{j_1\,j_2}\frac{\left(2j_1+1\right)\left(2j_2+1\right)}{2j_{12}+1}
\left\{\begin{array}{ccc}
j_{12} & j_3 & j_{123}\\ j_4 & j & j_{34}
\end{array}\right\}
\end{equation}
As discussed in \cite{PR}, the fact that $j_1$, $j_2$ and $j_{12}$
satisfy mutual triangle inequalities ensures that the sum becomes
\begin{equation}\label{divergent-pr}
\sum_{j_1\,j_2}\frac{\left(2j_1+1\right)\left(2j_2+1\right)}{2j_{12}+1}
= \sum_n\left(2n+1\right)^2
\end{equation}
Thus it is this quantity (clearly divergent as a sum) which plays the
role of the order of the finite group in the finite group state sum
presented previously and it is this quantity's divergence that means
that the Ponzano-Regge model is not a true TQFT. 

While the Ponzano-Regge model may not be a true TQFT there is an
intriguing connection with physics when one takes the semi-classical
limit of the model.  As discussed in \cite{Ba95} a good candidate
theory for quantum gravity must satisfy two constraints roughly
equating to being able to recover general relativity when $\hbar\to 0$
and quantum mechanics when $G\to 0$.  Planck's constant $\hbar$ is
used as fundamental unit of length in the theory, thus the \su{2}
representation labeled by $j$ labeling an edge corresponds to an
edge length of $\left(j+\frac{1}{2}\right)\hbar$ associated to the
edge in the geometric representation.  One takes the semi-classical
asymptotic limit by letting $\hbar\to 0$ while holding the actual edge
lengths of the tetrahedron fixed, which then corresponds to the
asymptotic limit of all the representation labels $j\to\infty$.

The asymptotic limit of a tetrahedron may be derived from a
remarkable asymptotic formula of Ponzano and Regge
\begin{equation}\label{pr-asymp+}
\lim_{e_{kl}\to\infty}
\left(-1\right)^{e_{23}+e_{03}+e_{01}+e_{12}}
\left\{\begin{array}{ccc}e_{23} & e_{03} & e_{13}\\
e_{01} & e_{12} & e_{02} \end{array}\right\}
 = \frac{1}{\sqrt{12\pi V}}
\cos\left(\sum_{\substack{l =0\\k < l}}^3 
m_{kl}\theta_{kl}+\frac{\pi}{4}\right)
\end{equation}
where each $m_{kl} = e_{kl}+\frac{1}{2}$.  The quantities
$\theta_{kl}$ are the (exterior) dihedral angles. These are defined as
the angle between the two outward normals to the faces which
share the edge joining vertex $k$ to vertex $l$.  Finally the quantity
$V$ is the volume of the tetrahedron under consideration, which may be
given in terms of the edge lengths via the Cayley determinant
\begin{equation}
V^2 = \frac{1}{2^3\left(3!\right)^2}
\left|\begin{array}{ccccc}
0 & j_{34}^2 & j_{24}^2 & j_{23}^2 & 1\\
j_{34}^2 & 0 & j_{14}^2 & j_{13}^2 & 1\\
j_{24}^2 & j_{14}^2 & 0 & j_{12}^2 & 1\\
j_{23}^2 & j_{13}^2 & j_{12}^2 & 0 & 1\\
1 & 1 & 1 & 1 & 0\\
\end{array}\right|
\end{equation}
$V^2>0$ for the region where equation \ref{pr-asymp+} provides a good
asymptotic approximation.  While Ponzano and Regge did not provide a
rigorous proof of this, a similar formula, which is asymptotic to the
Ponzano-Regge formula was proved in \cite{Rob98} while evidence for
the Ponzano-Regge formula itself was presented in \cite{Flude}.

The argument of the cosine in equation \ref{pr-asymp+} may be
recognised as precisely the Regge action for a tetrahedron. As shown
in \cite{PR}, the stationary phase approximation, which is expected to
dominate the classical theory, gives equations of motion that imply
the angles around each edge add up to integer multiples of $2\pi$
which is precisely what would be expected from a three dimensional
theory of quantum gravity; that is the simplicial manifold has a
partition function that is dominated by terms that enforce the
required condition for it to be mapped locally onto flat Euclidean
space.

The actual formula in equation \ref{pr-asymp+} may be written
\begin{equation}
\cos\mathfrak{L} = \frac{1}{2}\left(e^{i\mathfrak{L}} 
+ e^{-i\mathfrak{L}}\right)
\end{equation}
with $\mathfrak{L}$ the Regge action for a tetrahedron.  Thus it may
be regarded as the sum of the usual path integral exponential of an
action for a positively oriented and a negatively oriented
tetrahedron.  Note, however, that this is the path-integral that is
usually associated to an action in Lorentzian space since it is
$\exp\left\{i\times\mathrm{action}\right\}$ rather than the more usual
exponentially decaying $\exp\left\{- \mathrm{action}\right\}$ one
normally associates to Euclidean path integrals.

When $V^2 < 0$ the asymptotic formula in equation \ref{pr-asymp+} must
be modified to yield
\begin{multline}\label{pr-asymp-}
\lim_{e_{kl}\to\infty}
\left(-1\right)^{e_{23}+e_{03}+e_{01}+e_{12}}
\left\{\begin{array}{ccc}e_{23} & e_{03} & e_{13}\\
e_{01} & e_{12} & e_{02} \end{array}\right\}\\
 = \frac{1}{2\sqrt{12\pi\left|V\right|}}
\cos\phi\exp\left\{-\left|\sum_{\substack{l =0\\k < l}}^3 
m_{kl}\Im\theta_{kl}\right|\right\}
\end{multline}
where 
$$
\phi = \sum_{\substack{l =0\\k < l}}^3 \left(m_{kl}-\frac{1}{2}\right)
\Re\theta_{kl}
$$
and is always a multiple of $\pi$ so that the cosine factor is
essentially just a phase.  The `dihedral angles' here are still as
before, but here they turn out to be complex due to the negative
volume squared of the `tetrahedron'.  It may be shown\cite{BaFo93}
that this corresponds to a Lorentzian tetrahedron, with all edges and
faces space-like.  The imaginary parts of the `dihedral angles' may be
interpreted as Lorentz boosts in this context.  Again the asymptotic
formula is in the form of a path integral for which steepest descents,
in this case, leads to equations of motion implying that the
Lorentzian equivalent of the deficit angle around each edge vanishes,
as shown in \cite{BaFo93}.  

Again it should be noted that the path integral is the wrong `type'
for the geometry.  Since one is dealing with Lorentzian tetrahedra
when $V^2 < 0$ it would be expected that the path integral be of the
form $\exp\left\{i\times\mathrm{action}\right\}$ rather than the
Euclidean exponentially decaying type path integral that occurs.  Thus
while the Ponzano-Regge model provides an intriguing connection
between the representation theory of groups and quantum geometry, one
doesn't quite recover the expected asymptotic formulae to make contact
with what would be expected from physics.

Finally it should be remarked that one can make the Ponzano-Regge
model finite, and hence a proper TQFT.  Recall that the divergence of
equation \ref{divergent-pr} is precisely what governs its failure to
be a true TQFT since if that quantity were finite one could divide by
it, raised to the power of the number of vertices in the simplicial
decomposition, to regulate the four to one move.

Ponzano and Regge did propose a recipe for making the state sum finite
in \cite{PR}; roughly speaking one restricted the state sum to $j <
j_{\mathrm{max}}$ for some $j_{\mathrm{max}}$.  This clearly makes
equation \ref{divergent-pr} finite since it's now a finite sum.  The
problem with this is that topological invariance is lost unless one
uses the full series of $\su{2}$ representations and even if one is
willing to lose this in the hope of regaining it as
$j_{\mathrm{max}}\to\infty$ it is still extremely unobvious that the
state sum tends to something finite as a function of
$j_{\mathrm{max}}$ when $j_{\mathrm{max}}\to\infty$.

A more fruitful way of regulating the Ponzano-Regge model is to deform
the co-algebra structure of the underlying Lie algebra, thought of as
its universal enveloping algebra, as a Hopf algebra.  These objects are
commonly called quantum groups.  

When the deformation parameter is specialised to a root of unity the
category of representations becomes truncated, or more precisely there
will be some $j_{\mathrm{max}}$ such that any representation labeled
by $j > j_{\mathrm{max}}$ has (quantum) dimension zero.  This gives a
natural regularisation to the sum over representations in equation
\ref{pr-ss} which preserves all the algebraic relations that imply the
various Pachner moves, thus maintaining topological invariance.

The $\mathcal{U}_q\left(\mathfrak{sl}_2\right)$ version of the
Ponzano-Regge model is called the Turaev-Viro model\cite{TV92}.  This
was subsequently generalised to other finite\footnote{here finite
means precisely that the analogous equation to \ref{divergent-pr} is
finite.  In \cite{BW96} this is called the dimension of the category
of representations.}  Hopf algebras that satisfy some fairly mild
conditions in \cite{BW96}.

\chapter{$SU\left(1,1\right)$}\label{SU11-CHAP}

In this chapter the group $SU\left(1,1\right)$ is defined, along with
its unitary representations.  The approach to its representation
theory taken by Mukunda and Radhakrishnan in \cite{MR73} will be
discussed in section \ref{rep} as well as the exploitation of this
approach in
\cite{MR74i}, \cite{MR74ii}, \cite{MR74iii} and \cite{MR74iv} to
derive specific formulae for the Clebsch-Gordon coefficients for all
cases of coupling in the principal series in section \ref{CGC}.

\section{$SU(1,1)$ and its representation theory}\label{rep}

The group $SL\left(2,\C\right)$ is defined as complex $2\times 2$
matrices of the form
$$
\begin{pmatrix}
\alpha & \beta\\ \gamma & \delta
\end{pmatrix}
$$
with $\alpha\delta - \beta\gamma = 1$. $g\in SL\left(2,\C\right)$ acts
on hermitian matrices in the adjoint representation as
$$
\begin{pmatrix}
x_0 + x_1 & x_2 + ix_3\\ x_2 - ix_3 & x_0 - x_1
\end{pmatrix}\mapsto
g^\star\begin{pmatrix}
x_0 + x_1 & x_2 + ix_3\\x_2 - ix_3 & x_0 - x_1 &
\end{pmatrix} g
$$
where $\star$ denotes the hermitian adjoint.  The action preserves the
determinant $x_0^2 - x_1^2 - x_2^2 - x_3^2$, hence this defines an
action of $SL\left(2,\C\right)$ on $\R^4$ equipped with a Lorentzian
metric, which the action preserves.  The fact that $\pm\mathbb{I}\in
SL\left(2,\C\right)$ acts as the identity in this representation is
the well known fact that $SL\left(2,\C\right)$ is the double cover of
$SO\left(3,1\right)$, the (four dimensional) real Lorentz group.

The subgroup of $SL\left(2,\C\right)$ that preserves the coordinate
$x_4$ (or equivalently $x_2$) is defined by complex matrices of the
form
\begin{equation}
\left(\begin{array}{cc} \alpha & \beta\\
\bar{\beta} & \bar{\alpha}\end{array}\right)
\end{equation}
for $\alpha$, $\beta\in\C$ with $\left|\alpha\right|^2 -
\left|\beta\right|^2 = 1$. This is $SU\left(1,1\right)$ since it
acts on $\C^2$ by unitary transformations preserving the indefinite
metric $\left\|\mathbf{z}\right\|^2 = z_1^2 - z_2^2$.

It acts on matrices of the form
$$
\begin{pmatrix}
x_0 & x_1 + ix_2\\x_1 - ix_2 & x_0
\end{pmatrix}
$$
preserving the determinant $x_0^2 - x_1^2 - x_2^2$ and so defines an
action on $\R^3$ equipped with a Lorentzian metric, which it
preserves. Again $\pm\mathbb{I}$ acts as the identity and so
$SU\left(1,1\right)$ is the double cover of $SO\left(2,1\right)$, the
three dimensional Lorentz group.

The subgroup leaving the coordinate $x_2$ invariant are real $2\times 2$
matrices of the form
$$
\begin{pmatrix}
a & b\\ c & d
\end{pmatrix}
$$
with $ad-bc=1$ which is the group $SL\left(2,\R\right)$. Conjugation
by the element
$$
W = \frac{1}{\sqrt{2}}\begin{pmatrix}
1 & i\\i & 1
\end{pmatrix}\in SL\left(2,\C\right)
$$
transforms between these two subgroups, so if $g\in
SU\left(1,1\right)$ then $W g W^{-1}\in SL\left(2,\R\right)$. Since
the two groups are equivalent and one can work in either group as
convenient.

Finally, for completeness, the subgroup leaving the coordinate $x_0$
invariant is defined as
$$
\begin{pmatrix}
\alpha & \beta\\ -\bar{\alpha} & \bar{\beta}
\end{pmatrix}
$$
with $\left|\alpha\right|^2 +\left|\beta\right|^2=1$ and thus defines
the subgroup $SU\left(2\right)$.  This action preserves determinant
and hence the positive definite metric on $\R^3$ given by $x_1^2 +
x_2^2+x_3^2$.  It is thus the double cover of the Euclidean group
$SO\left(3\right)$ since, again, $\pm\mathbb{I}\in SU\left(2\right)$
acts as the identity on $\R^3$.

The Lie algebra \su{1,1} is generated by the following traceless
$2\times 2$ matrices\index{$\su{1,1}$}
$$
J_1 = -\frac{i}{2}\begin{pmatrix}
0 & -1\\1 & 0
\end{pmatrix}\;\;\;\;
J_2 = -\frac{i}{2}\begin{pmatrix}
0 & i\\i & 0
\end{pmatrix}\;\;\;\;
J_3 = -\frac{i}{2}\begin{pmatrix}
1 & 0\\0 & -1
\end{pmatrix}
$$
which have Lie bracket
\begin{equation}\label{comm-su11}
\begin{array}{lcr}
\comm{J_1}{J_2} & = & - J_3\\
\comm{J_3}{J_1} & = &  J_2\\
\comm{J_2}{J_3} & = &  J_1
\end{array}
\end{equation}
Setting $J_a = -iK_a$, the algebra may be given in terms of raising
and lowering operators as follows
\begin{equation}\label{su11bracket}
\comm{K_3}{K_-} = -K_-\;\;\; 
\comm{K_3}{K_+}=K_+\;\;\; \comm{K_+}{K_-}=-2K_3
\end{equation}
with
$$
K_+ = K_1 + iK_2\;\;\; K_- = K_1 - iK_2
$$
Let 
$$\mathfrak{H}_{\epsilon} = \left\{\ket{m}\;\left| \right. \;m = \epsilon+n,\;
n\in\mathbb{Z}\right\}\text{ for }\epsilon = 0,\frac{1}{2}
$$ 
be a Hilbert space with given Hilbert basis. Let \su{1,1} act on 
$\mathfrak{H}_\epsilon$ as
\begin{equation}\label{act_su11}
\begin{array}{ll}
K_\pm \ket{m} & = \pm\sqrt{(m\pm j)(m\mp j \pm 1)}\ket{m\pm 1}\\
K_3 \ket{m} & = m\ket{m}
\end{array}
\end{equation}

Then it is well known\cite{Ba47} that an inner product exists for
which the action given by equation \ref{act_su11} is unitary when $j$
and $\epsilon$ have the following values.

\begin{description}
\item[Continuous Series]
$j=\frac{1}{2}-is$ for $0< s <\infty$. Then the whole of
$\mathfrak{H}_\epsilon$ for $\epsilon = 0$ (even parity) and $\epsilon
= \frac{1}{2}$ (odd parity) is a carrier space for the unitary action
given by equation \ref{act_su11} and defines an irreducible
representation of \su{1,1}
\item[Discrete Series]
$j=\epsilon+\mathbb{Z}^+$. The action of \su{1,1} on the subspace of
$\mathfrak{H}_\epsilon$ with basis $\ket{\pm
j},\,\ket{\pm\left(j+1\right)},\dots$ is then unitary.  For each $j$
the subspace decomposes into two irreducible subspaces, the positive
discrete series, $\mathcal{D}_j^+$, is the subspace of
$\mathfrak{H}_\epsilon$ with basis labelled by
$\ket{j},\,\ket{j+1},\dots$ while the negative discrete series,
$\mathcal{D}_j^-$, is the subspace of $\mathfrak{H}_\epsilon$ with
basis labelled by $\ket{-j},\,\ket{-\left(j+1\right)},\dots$

One usually differentiates the two representations
$\mathcal{D}^\pm_\frac{1}{2}$ in the positive and negative discrete
series by calling them the
\emph{mock} or \emph{exceptional} discrete series since they do not appear
in the Plancherel decomposition and, thus, are not part of the
principal series.
\item[Exceptional Continuous Series]
For $0 < j < \frac{1}{2}$ the action on the whole of $\mathfrak{H}_0$
is unitary and irreducible.
\item[Trivial Representation]
The action on $\C$ with all three generators acting as $0$
\end{description}

The continuous and discrete series (excluding the mock discrete
series) are generally referred to as the principal series.

The decomposition of tensor products of the representations in the
principal series is well known\cite{Puk61}\cite{MR74i}. We have (all
direct sums are in integer steps)
\begin{subequations}
\begin{equation}\label{pp}
\mathcal{D}^\pm_{k^\prime}\otimes\mathcal{D}^\pm_{k^{\prime\prime}} = 
\bigoplus_{k\ge k^\prime + k^{\prime\prime}}\mathcal{D}^\pm_k
\end{equation}
\begin{equation}\label{pn}
\mathcal{D}^+_{k^\prime}\otimes\mathcal{D}^-_{k^{\prime\prime}} = 
\bigoplus_{k=k_{min}}^{k^\prime - k^{\prime\prime}}\mathcal{D}^+_k\oplus
\bigoplus_{k=k_{min}}^{-k^\prime + k^{\prime\prime}}\mathcal{D}^-_k\oplus
\dint\,ds\,\mathcal{C}^\epsilon_s
\end{equation}
where $k_{min}=1$, $\epsilon =0$ if $k+k^\prime$ is integer and
$k_{min}=\frac{3}{2}$, $\epsilon=\frac{1}{2}$ otherwise. 
\begin{equation}\label{pc}
\mathcal{D}^\pm_{k^\prime}\otimes\mathcal{C}
^{\epsilon^{\prime\prime}}_{s^{\prime\prime}} = 
\bigoplus_{k\ge k_{min}}\mathcal{D}^\pm_k\oplus
\dint\,ds\,\mathcal{C}^\epsilon_s
\end{equation}
where $k_{min}=1$, $\epsilon=0$ if
$k^\prime+\epsilon^{\prime\prime}$ is integer and
$k_{min}=\frac{3}{2}$, $\epsilon=\frac{1}{2}$ otherwise.  
\begin{equation}\label{cc}
\mathcal{C}^{\epsilon^\prime}_{s^\prime}
\otimes\mathcal{C}^{\epsilon^{\prime\prime}}_{s^{\prime\prime}} = 
\bigoplus_{k\ge k_{min}}\mathcal{D}^+_k\oplus
\bigoplus_{k\ge k_{min}}\mathcal{D}^-_k\oplus 2
\dint\,ds\,\mathcal{C}^\epsilon_s
\end{equation}
where $k_{min}=1$, $\epsilon=0$ if $\epsilon^\prime+
\epsilon^{\prime\prime}$ is integer and $k_{min}=\frac{3}{2}$,
$\epsilon=\frac{1}{2}$
The factor of $2$ in this last equation indicates
that the continuous series appears twice in this decomposition.
\end{subequations}

The representations $D^\pm_k$ are dual to the representations
$D^\mp_k$ while the continuous series is self dual.  

For our purposes it will be useful to pursue the approach to
$SU\left(1,1\right)$ representation theory developed by Mukunda and
Radhakrishnan in \cite{MR73}.  Here one chooses instead to diagonalise
one of the generators of the non compact $SO\left(1,1\right)$
subgroups, so that the action of either $K_1$ or $K_2$ is diagonal
(the actual choice made is $K_2$).

The representations of the principal series (the continuous,
$\mathcal{C}_s$, and the discrete series $\mathcal{D}_k^\pm$) may be
realised as follows.  To construct, say, the positive discrete series
one starts with a representation of \su{1,1} on two sets of harmonic
oscillators with generators $a_i$ and $a_i^\dagger$ for $i=1$,
$2$ and satisfying commutation relations
\begin{equation}\label{comm-osc}
\comm{a_i}{a_j^\dagger} = \delta_{i,j}\,,\;\;\;\;
\comm{a_i}{a_j}=\comm{a_i^\dagger}{a_j^\dagger} = 0
\end{equation}
The following representation then realises the commutation relations
in equation \ref{comm-su11}
\begin{align}
K_1 = & \frac{1}{4}\left(\left(a_1^\dagger\right)^2+\left(a_1\right)^2
+\left(a_2^\dagger\right)^2+\left(a_2\right)^2\right)\nonumber\\
K_2 = & -\frac{i}{4}\left(\left(a_1^\dagger\right)^2-\left(a_1\right)^2
+\left(a_2^\dagger\right)^2-\left(a_2\right)^2\right)\label{oscillate}\\
K_3 = & \frac{1}{2}\left(a^\dagger_1 a_1 + a^\dagger_2 a_2 +1 
\right)\nonumber
\end{align}

The oscillators may then by realised on the Hilbert space of square
integrable functions in two variables via the first order operators
\begin{equation} 
a_i = - \frac{i}{\sqrt{2}}\left(x_i + \frac{\partial}{\partial x_i}
\right)\,,\;\;\;\;
a_i^\dagger = \frac{i}{\sqrt{2}}\left(x_i - \frac{\partial}{\partial x_i}
\right)
\end{equation}

This then induces a representation of the generators of \su{1,1} as
second order differential operators on the Hilbert space of square
integrable functions in two variables via equation \ref{oscillate}.  A
transformation to polar coordinates means that the Casimir takes an
especially simple form, indeed one finds
$$
Q = K_1^2 + K_2^2 -K_3^2 = \frac{1}{4}\left(1 - \frac{\partial^2}
{\partial\theta^2}\right)
$$
where $\theta$ is the angular variable in the polar coordinates.
Thus the space breaks up into a direct sum of Hilbert spaces
since a general function of two variables may be expanded in a Fourier
decomposition of eigenfunctions of $Q$.
\begin{equation}
f\left(r,\theta\right) = \sum_{s=-\infty}^{\infty}
f_s\left(r\right)\frac{e^{-is\theta}}{\sqrt{2\pi}}
\end{equation}
with a similar decomposition of the norm of $f$ 
\begin{equation}
\left|f\left(r,\theta\right)\right|^2 = \sum_{s=-\infty}^{\infty}
\int_0^\infty\left|f_s\left(r\right)\right|^2r\,dr
\end{equation}
Thus the original Hilbert space of square integrable functions in
two variables decomposes as a direct sum of Hilbert spaces
$\mathfrak{H}\left(\mathcal{D}^+\right)$, being the Hilbert space of
functions on $\R^+$ with inner product given by
\begin{equation}
\ip{f}{g} = \int_0^\infty \bar{f}\left(r\right)g\left(r\right)
r\,dr
\end{equation}
The generators of \su{1,1} act by differential operators in $r$ alone
for each of these Hilbert spaces
\begin{align}
K_1 & = -\frac{1}{4}\left(r^2 + \frac{d^2}{dr^2} + \frac{1}{r}
\frac{d}{dr} - \frac{1}{r^2}\left(2k-1\right)^2\right)\nonumber\\
K_2 & = -\frac{i}{2}\left(r\frac{d}{dr}+1\right)\\
K_3 & = \frac{1}{4}\left(r^2 - \frac{d^2}{dr^2} - \frac{1}{r}
\frac{d}{dr} + \frac{1}{r^2}\left(2k-1\right)^2\right)\nonumber
\end{align}
where $k = \frac{1+\left|m\right|}{2}$ is the usual positive half
integer that enumerates the (positive) discrete series
representations.

The vectors giving the continuous `basis'\footnote{It is not a true
basis since the vectors are not actually in the Hilbert space because
they fail to be square integrable} of
$\mathfrak{H}\left(\mathcal{D}^+\right)$ in which $K_2$ acts
diagonally are then
$$
\ket{k,+,p} = \frac{1}{\sqrt{\pi}}r^{2ip-1}
$$
for $p\in\R$.  This is simply the components of a Fourier
decomposition of a general function in terms of eigenfunctions of
$K_2$, indeed for $f\in L^2\left(\R\right)$ one has (ignoring
normalisation factors)
$$
f\left(r\right) = \int_{-\infty}^\infty
\tilde{f}\left(p\right)r^{2ip-1}\, dp
$$
and a substitution of $r = e^\frac{k}{2}$ transforms this into the
traditional form for a Fourier transform.

The negative discrete series is represented similarly. The Hilbert
space, $\mathfrak{H}\left(\mathcal{D}^-\right)$, and inner product are
the same as $\mathfrak{H}\left(\mathcal{D}^+\right)$ but the
generators are represented slightly differently.  It turns out the
\su{1,1} generators in the negative discrete series look like
\begin{align}
K_1 & = \frac{1}{4}\left(r^2 + \frac{d^2}{dr^2} + \frac{1}{r}
\frac{d}{dr} - \frac{1}{r^2}\left(2k-1\right)^2\right)\nonumber\\
K_2 & = -\frac{i}{2}\left(r\frac{d}{dr}+1\right)\\
K_3 & = -\frac{1}{4}\left(r^2 - \frac{d^2}{dr^2} - \frac{1}{r}
\frac{d}{dr} + \frac{1}{r^2}\left(2k-1\right)^2\right)\nonumber
\end{align}

This follows from applying the following automorphism to \su{1,1}
\begin{equation}\label{auto-tau}
\tau : \begin{array}{cc}
K_1 & \mapsto -K_1\\
K_2 & \mapsto K_2\\
K_3 & \mapsto -K_3
\end{array}
\end{equation}
which clearly leaves the bracket in equation \ref{comm-su11} invariant
and has the effect of mapping the positive discrete series to the negative
discrete series.  This is simply duality.

Clearly the basis vectors are the same as for
$\mathfrak{H}\left(\mathcal{D}^+\right)$ as well, so one has
$$
\ket{k,-,p} = \frac{1}{\sqrt{\pi}}r^{2ip-1}
$$
again for $p\in\R$.

The continuous series is more subtle. It turns out to consist of pairs
of functions $f_1$, $f_2$, on $\R^+$ with the obvious inner
product. If $f\left(r\right) =
\left(f_1\left(r\right),f_2\left(r\right)\right)$ and $g\left(r\right) =
\left(g_1\left(r\right),g_2\left(r\right)\right)$ are two such pairs of
functions, then
\begin{equation}
\ip{f}{g} = \int_0^\infty\left(f_1\left(r\right)\bar{g}_1\left(r\right)
+f_2\left(r\right)\bar{g}_2\left(r\right)\right)r\,dr
\end{equation}
which defines the Hilbert space $\mathfrak{H}\left(\mathcal{C}\right)$.

The generators of \su{1,1} in this representation may be expressed
as the following operators on the given Hilbert space.
\begin{align}
K_1 & = \frac{1}{4}\left(r^2 + \frac{d^2}{dr^2} + \frac{1}{r}\frac{d}{dr}
+ {4s^2}{r^2}\right)\otimes\sigma_3\nonumber\\
K_2 & = -\frac{i}{2}\left(r\frac{d}{dr}+1\right)\otimes\mathbb{I}\\
K_3 & = \frac{1}{4}\left(-r^2 + \frac{d^2}{dr^2} + \frac{1}{r}\frac{d}{dr}
+ {4s^2}{r^2}\right)\otimes\sigma_3\nonumber
\end{align}
where $\sigma_3$ is the usual $2\times 2$ Pauli matrix.  This looks,
at first sight, like a reducible representation since the formal
expression for the generators is just two copies of the same thing
however when one exponentiates to the group $K_2$ and $K_3$ do not
preserve the two Hilbert spaces of square integrable functions.  The
actions of the exponentiated operators are calculated as explicit
integral kernels in \cite{MR73} and this doubling of the continuous
basis is discussed in \cite{MR67}.

The basis of $\mathfrak{H}\left(\mathcal{C}\right)$ may be written
$$
\ket{s,\epsilon, p,a} = \frac{1}{\sqrt{2\pi}}\left(\begin{array}{c}
1\\ a\end{array}\right) r^{2ip-1}
$$
for $a=\pm1$. The extra parameter $a$ arises because the basis
elements cannot be unambiguously labelled by the eigenvalue, $p$, of
the generator $K_2$.  

The Hilbert spaces and actions in the various representations were
chosen this way in \cite{MR73} so that the following orthonormality
relations would hold for the continuous `basis' vectors
\begin{align}
\ip{k,\pm,p}{k,\pm,p^\prime} & = \delta\left(p-p\prime\right)\\
\ip{s,\epsilon,p,a}{s,\epsilon,p^\prime,a^\prime} 
& = \delta\left(p-p\prime\right)\delta_{a,a^\prime} 
\end{align}

The dependency on the parity $\epsilon$ arises from considering the
boundary conditions as discussed in \cite{MR74i}.  Introduce
oscillators $b_i$ satisfying commutation relations
$$
\comm{b_i}{b^\dagger_j} = g_{ij}\;\;\;\;\;
\comm{b_i}{b_j} = 0\;\;\text{ for } i,j=1,2
$$
where the $b_i$ and $b_i^\dagger$ may be represented as differential
operators
$$
b_i = -\frac{i}{\sqrt{2}}\left(x_i + \frac{\partial}
{\partial x_i}\right)\;\;\;\;\;\;
b_i^\dagger = \frac{i}{\sqrt{2}}\left(x_i - \frac{\partial}
{\partial x_i}\right)
$$
and 
$$
\left(g\right)_{ij} = \begin{pmatrix}
1 & 0 \\ 0 & -1
\end{pmatrix}
$$
It may be verified that the representation  
\begin{align}
K_1 & = \frac{1}{2}\left(g^{ij}b_i^\dagger b_j+1\right)\nonumber\\
K_2 & = \frac{1}{4}g^{ij}\left(b^\dagger_i b^\dagger_j + b_i b_j\right)\\
K_3 & = -\frac{i}{4}g^{ij}\left(b^\dagger_i b^\dagger_j - b_i b_j\right)\nonumber
\end{align}
realises the commutation of \su{1,1}.  It is immediate that this is
invariant under the group $O(1,1)$ acting on the $(x_1,x_2)$-plane,
the generators of which commute with the $K_i$.  The full group
$O(1,1)$ is generated by the identity component
$$
\begin{pmatrix}
\cosh{\alpha} & \sinh{\alpha}\\ \sinh{\alpha} & \cosh{\alpha}
\end{pmatrix}
$$
and the discrete matrices
$$
P = \begin{pmatrix}
-1 & 0\\ 0 & -1
\end{pmatrix}\text{ and }
B = \begin{pmatrix}
1 & 0\\ 0 & -1
\end{pmatrix}
$$
One then introduces hyperbolic variables $r$ and $\eta$ to
parameterise the $(x_1,x_2)$-plane via
\begin{align}
x_1 & = \mathrm{sign}\left(x_2\right)r\sinh{\eta}\\
x_2 & = \mathrm{sign}\left(x_2\right)r\cosh{\eta}
\end{align}
if $\left|x_2\right| > \left|x_1\right|$, otherwise
\begin{align}
x_1 & = \mathrm{sign}\left(x_1\right)r\cosh{\eta}\\
x_2 & = \mathrm{sign}\left(x_1\right)r\sinh{\eta}
\end{align}
where in both cases $r\in\left[0,\infty\right)$ and
$\eta\in\left(-\infty,\infty\right)$.

This then leads to the break up of the $(x_1,x_2)$ plane into four
regions.  If the Hilbert Space of square integrable functions on this
space is denoted $\mathcal{C}$ then the operator $P$, which commutes
with the generators of \su{1,1} in this representations, breaks
$\mathcal{C}$ into even functions where $P=+1$, giving the even parity
part of the continuous series $\mathcal{C}^s_0$.  Odd functions, with
$P=-1$, then belong to the odd parity part of the continuous series
$\mathcal{C}^s_\frac{1}{2}$.

For each eigenspace of $P$ a given function is specified uniquely by
its values on the region $x_1 > \left|x_2\right|$ and $x_2
>\left|x_1\right|$ which then leads to a further decomposition so
that $f = f_1 + f_2$.

In the following the script letter $\mathcal{J}$ will be used to
denote a representation label $j$ when issues of parity, and
positivity for discrete representations, are important. Thus one
defines\index{$\mathcal{J}$}
$$
\mathcal{J}\equiv
\begin{cases}
\left(j,\pm\right) & \text{for } \mathcal{D}_j^\pm\text{ with }
2j-1\in\N\\
\left(\frac{1}{2}-is,\epsilon\right) & \text{for } 
\mathcal{C}_s^\epsilon \text{ with } 0<s<\infty
\end{cases}
$$
A delta function for the script letter, $\mathcal{J}$, is defined as
\begin{equation}\label{delta}
\index{$\delta\left(\mathcal{J},\mathcal{J}^\prime\right)$}
\delta\left(\mathcal{J},\mathcal{J}^\prime\right)\equiv
\begin{cases}
\frac{\delta_{k,\,k^\prime}}{\mu\left(k\right)} 
& \text{for } \mathcal{J},\mathcal{J}^\prime
\in\mathcal{D}_k^+\text{ or }\mathcal{J},\mathcal{J}^\prime
\in\mathcal{D}_k^-\\
\frac{\delta\left(s-s^\prime\right)}
{\mu\left(s\right)}\delta_{\epsilon,\,\epsilon^\prime}
& \text{for }\mathcal{J},\mathcal{J}^\prime\in\mathcal{C}_s^\epsilon\\
0 & \text{otherwise}
\end{cases}
\end{equation}
where $\mu\left(\mathcal{J}\right)$ is the Plancherel measure for the
representation $\mathcal{J}$ so that
$$
\dint\,\mathfrak{D}\mathcal{J}
\delta\left(\mathcal{J},\mathcal{J}^\prime\right) = 1
$$
where the notation $\dint\mathfrak{D}\mathcal{J}$ is shorthand for the
Plancherel decomposition\cite{Ba47}
\begin{multline}\label{Plancherel}
\dint\,\mathfrak{D}\mathcal{J}\,\equiv\sum_{k\in\mathcal{D}^+}\left(2k-1\right)
+\sum_{k\in\mathcal{D}^-}\left(2k-1\right)\\
+\int_{s\in\mathcal{C}^0}\frac{\mathrm{coth}\left(\pi s\right)}{4\pi s}\,ds\,
+\int_{s\in\mathcal{C}^\frac{1}{2}}\frac{\tanh\left(\pi s\right)}{4\pi s}\,ds
\end{multline}

 \section{Definition of the Clebsch-Gordon Coefficients}\label{CGC}

In \cite{MR74i}, \cite{MR74ii}, \cite{MR74iii} and \cite{MR74iv} the
Clebsch-Gordon coefficients $\left[\begin{array}{ccc} 
\mathcal{J}_1 & \mathcal{J}_2 & \mathcal{J}_{12}\\
p_1 & p_2 & p_{12} \end{array}\right] $ were constructed for all cases
of coupling in the principal series in a continuous basis.  The
Clebsch-Gordon coefficient is defined simply as the inner product of a
coupled continuous basis vector and an uncoupled one.  It thus provides
an integral kernel that transforms between the coupled and uncoupled
continuous bases.

Let $SU\left(1,1\right)$ act in the tensor product representation as
two copies of the group, that is $SU\left(1,1\right)^1\otimes
SU\left(1,1\right)^2$ where the superscripts will distinguish the two
copies.  The uncoupled continuous basis vectors, $\Phi$, span the
space $\mathcal{J}_1\otimes\mathcal{J}_2$ and are eigenfunctions with
respect to the action of the operators $K_2^1$ and $K_2^2$, with the
two respective Casimirs $Q^1$ and $Q^2$ acting as scalars.  The
uncoupled continuous basis is simply the product of the spanning
functions that define a continuous basis for $\mathcal{J}_1$ and
$\mathcal{J}_2$.  It satisfies orthonormality relations
\begin{equation}\label{uncouple-norm}
\ip{\Phi^{\mathcal{J},\,\mathcal{J}^\prime}_{p,\,p^\prime}}
{\Phi^{\mathcal{J}^{\prime\prime},\,
\mathcal{J}^{\prime\prime\prime}}
_{p^{\prime\prime},\,p^{\prime\prime\prime}}}
= \delta\left(\mathcal{J},\,\mathcal{J}^{\prime\prime}\right)
\delta\left(\mathcal{J}^\prime,\,\mathcal{J}^{\prime\prime\prime}\right)
\delta\left(p-p^{\prime\prime}\right)
\delta\left(p^\prime-p^{\prime\prime\prime}\right)
\end{equation}
For this, and subsequent equations where it is not directly relevant,
we have suppressed the extra discrete index $a=\pm1$ labelling the
double occurrence of the continuous basis in the continuous series and
the $\delta_{a,a^\prime}$ it gives rise to.

The coupled continuous basis vector, $\Psi$, is a spanning set for
$\mathcal{J}_1\otimes\mathcal{J}_2$ on which the total $K_2$
generator, given by $K_2^1+K_2^2$ acts diagonally, as well as the
total Casimirs $Q = Q^1 + Q^2$ acting as a scalar.  It satisfies
orthogonality relations
\begin{equation}\label{couple-norm}
\ip{\Psi^{\mathcal{J},\,\mathcal{J}^\prime\,\,\mathcal{J}_1}_{p_1}}
{\Psi^{\mathcal{J}^{\prime\prime},\,\mathcal{J}^{\prime\prime\prime}
\,\,\mathcal{J}_2}_{p_2}}
= \delta\left(\mathcal{J},\,\mathcal{J}^{\prime\prime}\right)
\delta\left(\mathcal{J}^\prime,\,\mathcal{J}^{\prime\prime\prime}\right)
\delta\left(\mathcal{J}_1,\,\mathcal{J}_2\right)
\delta\left(p_1-p_2\right)
\end{equation}

The Clebsch-Gordon coefficient is now defined as
\index{$\left[\begin{array}{ccc} 
\mathcal{J}_1 & \mathcal{J}_2 & \mathcal{J}_{12}\\
p_1 & p_2 & p_{12}\end{array}\right]$}
\begin{equation}
\ip{\Phi^{\mathcal{J}_1,\,\mathcal{J}_2}_{p_1,\,p_2}}
{\Psi^{\mathcal{J},\,\mathcal{J}^\prime\,\,\mathcal{J}_{12}}
_{p_{12}}}
= \delta\left(p_{12}-p_1-p_2\right)
\delta\left(\mathcal{J}_1,\mathcal{J}\right)
\delta\left(\mathcal{J}_2,\mathcal{J}^\prime\right)
\left[\begin{array}{ccc} 
\mathcal{J}_1 & \mathcal{J}_2 & \mathcal{J}_{12}\\
p_1 & p_2 & p_{12} \end{array}\right]
\end{equation}
while the adjoint Clebsch-Gordon coefficient satisfies
\begin{multline}
\ip{\Psi^{\mathcal{J}_1,\,\mathcal{J}_2\,\,\mathcal{J}_{12}}
_{p_{12}}}{\Phi^{\mathcal{J}_1,\,\mathcal{J}_2}_{p_1,\,p_2}} = 
\ip{\Phi^{\mathcal{J}_1,\,\mathcal{J}_2}_{p_1,\,p_2}}
{\Psi^{\mathcal{J},\,\mathcal{J}^\prime\,\,\mathcal{J}_{12}}
_{p_{12}}}^\star\\  = \delta\left(\mathcal{J}_1,\mathcal{J}\right)
\delta\left(\mathcal{J}_2,\mathcal{J}^\prime\right)
\delta\left(p_{12}-p_1-p_2\right)
\left[\begin{array}{ccc} 
\mathcal{J}_1 & \mathcal{J}_2 & \mathcal{J}_{12}\\
p_1 & p_2 & p_{12} \end{array}\right]^\star
\end{multline}

It turns out that every Clebsch-Gordon coefficient for the various
cases of coupling within the principal series may be written as linear
combinations of normalisations of integrals of the form
\begin{equation}
\int_0^1 t^\alpha\left(1-t\right)^\beta\Hyper{a}{b}{c}{t}\,dt
\end{equation}
where $\alpha$, $\beta$, $a$, $b$ and $c$ are constants that depend on
the particular case of coupling and $\Hyper{a}{b}{c}{t}$ is the
hypergeometric function\cite{HypGeo}.  Since these integrals will be
useful later, for completeness they will be reproduced in this
section.

\subsection{Orthogonality Relations} 

It is desirably first of all to derive some orthonormality relations
for the generic Clebsch-Gordon coefficient $\left[\begin{array}{ccc}
\mathcal{J}_1 & \mathcal{J}_2 & \mathcal{J}_{12}\\
p_1 & p_2 & p_{12} \end{array}\right]$.  

The operator $\mathcal{P}$, given by the kernel
\begin{equation}
\int\,\mathfrak{D}\mathcal{J}_1\,\mathfrak{D}\mathcal{J}_2\,dp_1\,dp_2
\ket{\Phi^{\mathcal{J}_1,\,\mathcal{J}_2}_{p_1,\,p_2}}
\bra{\Phi^{\mathcal{J}_1,\,\mathcal{J}_2}_{p_1,\,p_2}}
\end{equation}
forms the identity operator for the tensor product representation.
One then finds
\begin{multline}
\delta\left(\mathcal{J},\,\mathcal{J}^{\prime\prime}\right)
\delta\left(\mathcal{J}^\prime,\,\mathcal{J}^{\prime\prime\prime}\right)
\delta\left(\mathcal{J}_{12},\,\mathcal{J}_{12}^\prime\right)
\delta\left(p_{12}-p_{12}^\prime\right)
=\ip{\Psi^{\mathcal{J},\,\mathcal{J}^\prime\,\,\mathcal{J}_{12}}_{p_{12}}}
{\Psi^{\mathcal{J}^{\prime\prime},\,\mathcal{J}^{\prime\prime\prime}
\,\,\mathcal{J}^\prime_{12}}_{p^\prime_{12}}}\\=
\int\,\mathfrak{D}\mathcal{J}_1\,\mathfrak{D}\mathcal{J}_2\,dp_1\,dp_2
\ip{\Psi^{\mathcal{J},\,\mathcal{J}^\prime\,\,\mathcal{J}_{12}}_{p_{12}}}
{\Phi^{\mathcal{J}_1,\,\mathcal{J}_2}_{p_1,\,p_2}}
\ip{\Phi^{\mathcal{J}_1,\,\mathcal{J}_2}_{p_1,\,p_2}}
{\Psi^{\mathcal{J}^{\prime\prime},\,\mathcal{J}^{\prime\prime\prime}
\,\,\mathcal{J}^\prime_{12}}_{p^\prime_{12}}}\\=
\int\,\mathfrak{D}\mathcal{J}_1\,\mathfrak{D}\mathcal{J}_2\,dp_1\,dp_2\;
\delta\left(p_{12}-p_1-p_2\right)
\delta\left(\mathcal{J}_1,\mathcal{J}\right)
\delta\left(\mathcal{J}_2,\mathcal{J}^\prime\right)
\left[\begin{array}{ccc} 
\mathcal{J}_1 & \mathcal{J}_2 & \mathcal{J}_{12}\\
p_1 & p_2 & p_{12} \end{array}\right]^\star\\ \times
\delta\left(p_{12}^\prime-p_1-p_2\right)
\delta\left(\mathcal{J}_1,\mathcal{J}^{\prime\prime}\right)
\delta\left(\mathcal{J}_2,\mathcal{J}^{\prime\prime\prime}\right)
\left[\begin{array}{ccc} 
\mathcal{J}_1 & \mathcal{J}_2 & \mathcal{J}_{12}^\prime\\
p_1 & p_2 & p_{12}^\prime \end{array}\right]
\end{multline}

This give the following orthogonality relation
\begin{prop}
\begin{multline}\label{CGC-orth}
\int\,dp_1\,dp_2\; 
\delta\left(p_{12}-p_1-p_2\right)
\delta\left(p_{12}^\prime-p_1-p_2\right)
\left[\begin{array}{ccc} 
\mathcal{J}_1 & \mathcal{J}_2 & \mathcal{J}_{12}\\
p_1 & p_2 & p_{12} \end{array}\right]^\star
\left[\begin{array}{ccc} 
\mathcal{J}_1 & \mathcal{J}_2 & \mathcal{J}_{12}^\prime\\
p_1 & p_2 & p_{12}^\prime\end{array}\right]\\ = 
\delta\left(\mathcal{J}_{12},\,\mathcal{J}_{12}^\prime\right)
\delta\left(p_{12}-p_{12}^\prime\right)
\end{multline}
\end{prop}

One may proceed similarly by using the decomposition of the identity
given by the kernel
\begin{equation}
\int\;\mathfrak{D}\mathcal{J}\,\mathfrak{D}\mathcal{J}^\prime\,
\mathfrak{D}\mathcal{J}_{12}\,dp_{12}\,
\ket{\Psi^{\mathcal{J},\,\mathcal{J}^\prime\,\,\mathcal{J}_{12}}_{p_{12}}}
\bra{\Psi^{\mathcal{J},\,\mathcal{J}^\prime\,\,\mathcal{J}_{12}}_{p_{12}}}
\end{equation}

We have
\begin{multline}
\delta\left(\mathcal{J}_1,\,\mathcal{J}^{\prime}_1\right)
\delta\left(\mathcal{J}_2,\,\mathcal{J}^{\prime}_2\right)
\delta\left(p_1-p^{\prime}_1\right)
\delta\left(p_2-p^{\prime}_2\right) =
\ip{\Phi^{\mathcal{J}_1,\,\mathcal{J}_2}_{p_1,\,p_2}}
{\Phi^{\mathcal{J}^{\prime}_1,\,\mathcal{J}^{\prime}_2}
_{p^{\prime}_1,\,p^{\prime}_2}}\\=
\int\;\mathfrak{D}\mathcal{J}\,\mathfrak{D}\mathcal{J}^\prime\,
\mathfrak{D}\mathcal{J}_{12}\,dp_{12}\,
\ip{\Phi^{\mathcal{J}_1,\,\mathcal{J}_2}_{p_1,\,p_2}}
{\Psi^{\mathcal{J},\,\mathcal{J}^\prime\,\,\mathcal{J}_{12}}_{p_{12}}}
\ip{\Psi^{\mathcal{J},\,\mathcal{J}^\prime\,\,\mathcal{J}_{12}}_{p_{12}}}
{\Phi^{\mathcal{J}^{\prime}_1,\,\mathcal{J}^{\prime}_2}
_{p_1^\prime,\,p_2^\prime}}\\=
\int\;\mathfrak{D}\mathcal{J}\,\mathfrak{D}\mathcal{J}^\prime\,
\mathfrak{D}\mathcal{J}_{12}\,dp_{12}\,
\delta\left(p_{12}-p_1^\prime -p_2^\prime \right)
\delta\left(\mathcal{J}_1^\prime,\mathcal{J}\right)
\delta\left(\mathcal{J}_2^\prime,\mathcal{J}^\prime\right)
\left[\begin{array}{ccc} 
\mathcal{J}_1^\prime & \mathcal{J}_2^\prime & \mathcal{J}_{12}\\
p_1^\prime & p_2^\prime & p_{12} \end{array}\right]^\star\\ \times
\delta\left(p_{12}-p_1-p_2\right)
\delta\left(\mathcal{J}_1,\mathcal{J}\right)
\delta\left(\mathcal{J}_2,\mathcal{J}^{\prime}\right)
\left[\begin{array}{ccc} 
\mathcal{J}_1 & \mathcal{J}_2 & \mathcal{J}_{12}\\
p_1 & p_2 & p_{12} \end{array}\right]
\end{multline}
giving the orthogonality relation
\begin{prop}
\begin{multline}\label{CGC-orth2}
\int\,dp_{12}\,\dint\,\mathfrak{D}\mathcal{J}_{12}\,  
\delta\left(p_{12}-p_1-p_2\right)
\delta\left(p_{12}-p_1^\prime-p_2^\prime\right)\\
\left[\begin{array}{ccc} 
\mathcal{J}_1 & \mathcal{J}_2 & \mathcal{J}_{12}\\
p_1 & p_2 & p_{12} \end{array}\right]
\left[\begin{array}{ccc} 
\mathcal{J}_1^\prime & \mathcal{J}_2^\prime & \mathcal{J}_{12}\\
p_1^\prime & p_2^\prime & p_{12} \end{array}\right]^\star
= \delta\left(p_{1}-p_{1}^\prime\right)
\delta\left(p_{2}-p_{2}^\prime\right) 
\end{multline}
\end{prop}

\subsection{Coupling of $\mathcal{D}^+_{k_1}$ with $\mathcal{D}^+_{k_2}$} 

In this subsection the method and results of \cite{MR74i} will be
discussed to illustrate this way of constructing Clebsch-Gordon
coefficients.  This is the case where an uncoupled vector in the
tensor product Hilbert space of two positive discrete series (labelled
by $k_1$ and $k_2$) is inner producted with a coupled continuous basis vector
arising from the positive discrete series (labelled by $k_{12}$).

One constructs the tensor product Hilbert space $\mathcal{D}_{k_1}^+
\otimes\mathcal{D}_{k_2}^+$ as the Hilbert space of square integrable
functions in four variables.  As when constructing the discrete series
representation in section \ref{rep}, one represents the tensor product
action of \su{1,1} on a number of harmonic oscillators, four in this
case. The oscillators are represented on the Hilbert space the same
way as before. Thus

\begin{align}
K_1 = & \frac{1}{4}\left(a_\mu^\dagger a_\mu^\dagger
+a_\mu a_\mu\right)\nonumber\\
K_2 = & -\frac{i}{4}\left(a_\mu^\dagger a_\mu^\dagger
-a_\mu a_\mu\right)\\
K_3 = & \frac{1}{2}\left(a^\dagger_\mu a_\mu+2\right)\nonumber
\end{align}
where $\mu = 0,\,1,\,2,\,3$ and repeated indices are summed over.  The
indices 0 and 1 refer to the first copy of \su{1,1} while the second
carries indices 2 and 3.

One then needs to construct the coupled and uncoupled continuous basis
of the tensor product space since it is the overlap between these that
determines the Clebsch-Gordon coefficient.  In the uncoupled
continuous basis one requires only that the two Casimirs for the two
copies of \su{1,1} are diagonal, along with the operators $K_2^i$ for
$i=1$, 2, where the superscript will be used to keep track of the two
copies of \su{1,1}. The coupled continuous basis requires the two
Casimirs, $Q^i$, to be diagonal, so that only a single product is
involved, as well as the total pseudo-angular momentum operator $K_2$
and its Casimir, $Q$.

The uncoupled continuous basis $\Phi$ may be derived in roughly the
same way as the original continuous basis for each Hilbert space of
the discrete series in section \ref{rep}.  One changes variables from
the $x_\mu$ to radial and angular variables via

\begin{align*}
x_0 = & r\cos\frac{\beta}{2}\cos\phi & x_1 = & r\cos\frac{\beta}{2}\sin\phi\\
x_2 = & r\sin\frac{\beta}{2}\cos\psi & x_3 = & r\sin\frac{\beta}{2}\sin\psi
\end{align*}
for $0\le r <\infty$, $0\le\beta\le\pi$ and $0\le\psi,\,\phi\le 2\pi$.
This choice of variables to parameterise $\R^4$ allows the Casimir
$Q^1$ to be written as a function solely of
$\frac{\partial}{\partial\phi}$, while $Q^2$ becomes a function solely
of $\frac{\partial}{\partial\psi}$. The uncoupled continuous basis in
the representation $\mathcal{D}^+_{k_1}\otimes\mathcal{D}^+_{k_2}$ is
then
\begin{equation}\label{uncoup}\index{$\Phi^{k_1,\,k_2}_{p_1,\,p_2}$}
\Phi^{k_1,\,k_2}_{p_1,\,p_2} = \frac{1}{2\pi^2}
\left(r\cos\frac{\beta}{2}\right)^{2ip_1-1}
\left(r\sin\frac{\beta}{2}\right)^{2ip_2-1}
e^{-i\left(2k_1-1\right)\phi}
e^{-i\left(2k_2-1\right)\psi}
\end{equation}
Note that in section \ref{rep} the dependence on $k$ was moved from
the continuous basis vectors into the operators, but this has not been done here.
It may be verified that these continuous basis vectors are normalised to respect
the orthonormality relations of equation \ref{uncouple-norm}.

To find the coupled continuous basis the $\psi$ and $\phi$ dependence will be as
for the uncoupled continuous basis as one still requires the Casimirs $Q^1$ and
$Q^2$ to be diagonal.  The total Casimir has the following expression
as a differential operator 
\begin{multline}
Q = \frac{\partial^2}{\partial\beta^2} +
\cot\beta\frac{\partial}{\partial\beta} + \frac{1}{\sin^2\beta}
\left(\frac{\partial^2}{\partial\left(\phi+\psi\right)^2}
+\frac{\partial^2}{\partial\left(\phi-\psi\right)^2}\right.\\
\left. -2\cos\beta\frac{\partial^2}{\partial\left(\phi+\psi\right)
\partial\left(\phi-\psi\right)}\right)
\end{multline}
The solution to the eigenvalue problem in $\beta$ turns out to be
the familiar $D$ functions of angular momentum and so the coupled
continuous basis vector has the following form when normalised to obey the
orthogonality relations of equation \ref{couple-norm}
\begin{equation}\label{coup}\index{$\Psi^{k_1,\,k_2,\,\,k_{12}}_{p_{12}}$}
\Psi^{k_1,\,k_2,\,\,k_{12}}_{p_{12}}
= \sqrt{\frac{2k_{12}-1}{2}}\frac{1}{\pi^{\frac{3}{2}}}
e^{-i\left(2k_1-1\right)\phi}
e^{-i\left(2k_2-1\right)\psi}
d^{k_{12}-1}_{k_1+k_2-1,\,k_1-k_2}\left(\beta\right) r^{2ip_{12}-1}
\end{equation}
with 
\begin{multline}
d^{j}_{m,n}\left(\theta\right) =
 \sqrt{\frac{\left(j+m\right)!\left(j-n\right)!}
 {\left(j-m\right)!\left(j+n\right)!}}
\frac{\left(\cos\frac{\theta}{2}\right)^{-m-n}
\left(-\sin\frac{\theta}{2}\right)^{m-n}}
{\left(m-n\right)!}\\
\Hyper{j-n+1}{-j-n}{m-n+1}{\sin^2\frac{\theta}{2}}
\end{multline}
for $m\ge n$. In equation \ref{coup} $k_1$ has been assumed greater
than $k_2$ without loss of generality.

Thus the Clebsch-Gordon coefficient is finally defined via
\begin{multline}
\delta\left(p_{12}-p_1-p_2\right)
\left[\begin{array}{ccc} 
k_1 & k_2 & k_{12}\\
p_1 & p_2 & p_{12} \end{array}\right]
=\ip{\Phi^{k_1,\,k_2}_{p_1,\,p_2}}
{\Psi^{k_1,\,k_2\,\,k_{12}}_{p_{12}}}\\
= -\frac{1}{2\pi^3}\sqrt{\frac{2k_{12}-1}{2\pi}}
\int_0^\infty\int_0^\pi\int_0^{2\pi}\int_0^{2\pi}
r^3\sin\beta
\left(r\cos\frac{\beta}{2}\right)^{2ip_1-1}
\left(r\sin\frac{\beta}{2}\right)^{2ip_2-1}\\
d^{k_{12}-1}_{k_1+k_2-1,\,k_1-k_2}\left(\beta\right) r^{-2ip_{12}-1}
\;dr\,d\beta\,d\psi\,d\phi\,
\end{multline}

The integrals over $r$, $\phi$ and $\psi$ may be done immediately to
recover the following definition for the Clebsch-Gordon coefficient.

\begin{multline}
\left[\begin{array}{ccc} k_1 & k_2 & k_{12}\\p_1 & p_2 & p_{12}
\end{array}\right] = 
\sqrt{\frac{2k_{12}-1}{2\pi}}\\
\int_0^\pi \left(\cos^2\frac{\beta}{2}\right)^{-ip_1}
\left(\sin^2\frac{\beta}{2}\right)^{-ip_2}
d^{k_{12}-1}_{k_1+k_2-1,k_1-k_2}\left(\beta\right)\,d\beta
\end{multline}

Now making the substitution $t=\sin^2\frac{\theta}{2}$ one may rewrite the
essential integral in the form
\begin{multline}\label{PPP}
I\left(+,+,+\right) = 
\int_0^1 \left(1-t\right)^{-ip_1 -k_1}
t^{-ip_2+k_2-1}\\
\Hyper{k_{12}-k_1+k_2}{-k_{12}-k_1+k_2+1}
{2k_2}{t}
\end{multline}

For the integral in equation \ref{PPP} to be well defined one needs it
to be bounded at the end points.  Firstly one needs the behaviour of
the hypergeometric function as $t\to 0$ or $1$.  The hypergeometric
function is defined as following\cite{HypGeo}:

\begin{equation}\label{Hypergeom}
\Hyper{a}{b}{c}{z} = \sum_{n=0}^\infty
\frac{\left(a\right)_n\left(b\right)_n z^n}{\left(c\right)_n n!}
\end{equation}
where the notation $\left(a\right)_n$ is defined for each non negative
integer $n$ as
$$
\left(a\right)_n = \begin{cases}
\frac{\Gamma\left(a+n\right)}{\Gamma\left(a\right)} & 
\text{ for } a\ge 0\\
\frac{\Gamma\left(-a+n\right)}{\Gamma\left(-a\right)} & 
\text{ for } 0\le n\le -a\\
0 & \text{otherwise}
\end{cases}
$$

Asymptotically it is known to behave in the following way
\begin{equation}\label{hyper}
\Hyper{a}{b}{c}{\xi}\sim
B_1 + B_2 \left(1-\xi\right)^{c-a-b}  \text{ for } \xi\approx 1
\end{equation}
For $B_i$ constants that depend on $a$, $b$, and $c$.  If $\Re c-a-b >
0$ one may use Gauss's theorem to provide a value for the constant
$B_1$.
\begin{thm}[Gauss]\label{gauss}
Let $\Re c-a-b > 0$ then
$$
\Hyper{a}{b}{c}{1} = \frac{\Gamma\left(c\right)\Gamma\left(c-a-b\right)}
{\Gamma\left(c-a\right)\Gamma\left(c-b\right)}
$$
providing $c$ is not a negative integer or zero.
\end{thm}

Thus one finds
$$
B_1 = \frac{\Gamma\left(c\right)\Gamma\left(c-a-b\right)}
{\Gamma\left(c-a\right)\Gamma\left(c-b\right)}
$$
Note in particular that if either $c-a$ or $c-b$ is a non positive
integer then the constant $B_1$ must vanish by continuity.

Thus the hypergeometric function in equation \ref{PPP} displays the
following behaviour as $t\to 1$
\begin{equation}
\Hyper{k_{12}-k_1+k_2}{-k_{12}-k_1+k_2+1}
{2k_2}{t}\sim\left(1-t\right)^{2k_1-1} 
\end{equation}
and this is sufficient to ensure the integrand in equation \ref{PPP}
is bounded at $t=1$.

For the lower bound $t=0$ one simply has, from equation
\ref{Hypergeom}, that
\begin{equation}
\Hyper{k_{12}-k_1+k_2}{-k_{12}-k_1+k_2+1}
{2k_2}{0} = 1
\end{equation}
and so the integrand behaves like $t^{k_2-1}$ as $t\to 0$ which is
also clearly bounded.

Mukunda and Radhakrishnan go on to write the Clebsch-Gordon
coefficient as an ${ }_3F_2$ function in a closed form.  However for
our purposes the integral in equation \ref{PPP} is more useful since
it defines the dependence of the Clebsch-Gordon coefficient on the
continuous basis parameters, $p_1$ and $p_2$, in a way that is
suitable for asymptotic analysis.

\subsection{Coupling of $\mathcal{D}^+_{k_1}$ with $\mathcal{D}^-_{k_2}$} 

The next case of coupling to illustrate, where a positive and a
negative discrete series couple, is discussed more fully in
\cite{MR74ii}. The results will simply be sketched here in order
to derive the defining integrals which are of interest.  Again this
case is dependent on solving a second order differential equation
which determines when the Casimir for the total pseudo-angular
momentum is diagonal.  The solution, unsurprisingly, involves a
hypergeometric function in much the same way as before.

Thus, for example, when coupling to produce a representation in the
positive discrete series the coupled vector has the form
\begin{equation}\label{pncouple}
\Psi = \frac{1}{\pi}\sqrt{\frac{2k_{12}-1}{2\pi}} r^{2ip_{12}-2}
e^{i\left(k_1+k_2-1\right)\mu}e^{i\left(k_1-k_2\right)\nu}
d^{k_{12}}_{k_1+k_2-1,k_1-k_2}\left(\xi\right)
\end{equation}
with 
\begin{multline}
d^{j}_{m,n}\left(\xi\right) =
 \sqrt{\frac{\left(m-j\right)!\left(m+j-1\right)!}
 {\left(n-j\right)!\left(n+j-1\right)!}}
\frac{\left(\cosh\frac{\xi}{2}\right)^{-m-n}
\left(\sinh\frac{\theta}{2}\right)^{m-n}}
{\left(m-n\right)!}\\
\Hyper{j-n}{1-j-n}{m-n+1}{-\sinh^2\frac{\xi}{2}}
\end{multline}

Note that this particular expression for the $d$ matrix is only valid
for $\left|\sinh\frac{\xi}{2}\right|\le 1$ because that is the region
of convergence for this hypergeometric series. In order to actually
evaluate the integral the solution needs to be analytically continued.

Thus let the Gauss equation (see for instance \cite{HypGeo}) be written
\begin{equation}
z\left(1-z\right)\frac{d^2y}{dz^2}+\left[c-\left(1+a+b\right)z\right]
\frac{dy}{dz}-aby = 0
\end{equation}
with $-c\not\in\N$ and $\Re\left(c-a-b\right) > 1$.  A solution is
given by $u_1=\Hyper{a}{b}{c}{z}$ with the hypergeometric function
defined by equation \ref{Hypergeom} and the sum in this equation
converges everywhere for $\left|z\right|\le 1$\cite{HypGeo}. If
$2-c\not\in\N$ as well then there is a second solution $u_5 =
z^{1-c}\Hyper{1+a-c}{1+b-c}{2-c}{z}$ valid for the same
region\footnote{The case $c=1$ is not relevant to this discussion so
this degenerate case can be safely ignored}.

For $\left|z\right| > 1$ the sum diverges, however it is possible to
analytically continue and use either of the two following
solutions\cite{Kumar}
\begin{align}
u_3\left(z\right) = & \left(-z\right)^{-a}\Hyper{a}{a+1-c}{a+1-b}{z^{-1}}\\
u_4\left(z\right) = & \left(-z\right)^{-b}\Hyper{b+1-c}{b}{b+1-a}{z^{-1}}
\end{align}

For the case under consideration, where the coupled vector lies in the
positive discrete series, only the solution $u_1$ is valid when
$\left|z\right|\le 1$ since $u_5$ is undefined.  The negative discrete
series case is similar, however when the coupled vector lies in the
continuous series one can use both solutions (essentially the second
solution is the same as the first with $s_{12}\leftrightarrow -s_{12}$
where $s_{12}$ is the continuous parameter labelling the continuous
series representations).

The analytic continuation must be chosen to ensure the coupled vector
is square integrable with respect to $\xi$. For the continuation given
by $u_3$ the integrand behaves like
$\left(\sinh\frac{\xi}{2}\right)^{-k_{12}}$ while that for $u_4$
behaves as $\left(\sinh\frac{\xi}{2}\right)^{-1+k_{12}}$, hence one
needs the analytic continuation provided by $u_3$ with the actual analytic
continuation of $u_1$ given by a suitable normalisation to ensure
they agree when $\left|z\right|=1$.

One should note that these formulae are largely unchanged when the
coupled representation lies in the continuous series, the only change
is $k_{12}$ is replaced by $\frac{1} {2}+is_{12}$.  In this case one
may analytically continue using either $u_3$ or $u_4$ and the coupled
vector is \emph{not} square integrable, nor would one expect it to be
given the existence of the $\delta$ function
$\delta\left(\mathcal{J}_{12},
\mathcal{J}_{12}^\prime\right)$ in the
orthonormality relations of equation \ref{couple-norm}.

The hypergeometric function in the coupled vector in equation
\ref{pncouple} is given by
$$
\Hyper{k_{12}-k_1+k_2}{1-k_{12}-k_1+k_2}
{2k_2}{-\sinh^2\frac{\xi}{2}} 
$$ 
for  $\xi\le 2\sinh^{-1} 1$ and
$$
N\left(\sinh^2\frac{\xi}{2}\right)^{-k_{12}+k_1-k_2}
\Hyper{k_{12}-k_1+k_2}{1+k_{12}-k_1-k_2}
{2k_{12}}{\sinh^{-2}\frac{\xi}{2}}
$$
for $\xi\ge 2\sinh^{-1} 1$, where $N$ is some normalisation constant to
ensure the two solutions agree on $\left|z\right| = 1$.

So, as before, one has the determining integral for the Clebsch-Gordon
coefficient given by an inner product between coupled and uncoupled
vectors.  In this case though it is necessary to keep careful track of
what region the various analytic continuations of the hypergeometric
series are valid in.

\begin{multline}
\left[\begin{array}{ccc} k_1 & k_2 & k_{12}\\p_1 & p_2 & p_{12}
\end{array}\right] = \left(-1\right)^{2k_2-1}
\sqrt{\frac{2k_{12}-1}{2\pi}}\\ \int_0^\pi 
\left(\cosh^2\frac{\xi}{2}\right)^{-ip_1}
\left(\sinh^2\frac{\xi}{2}\right)^{-ip_2}
d^{k_{12}-1}_{k_1+k_2-1,k_1-k_2}\left(\xi\right)\,d\xi
\end{multline}
The determining integral is then of the form
\begin{multline}\label{PNpre1}
I\left(+,-,\mathcal{J}\right) = \int_0^{\sinh^{-1}\left(1\right)}
\left(\cosh^2\frac{\xi}{2}\right)^{-ip_1-k_1+\frac{1}{2}}
\left(\sinh^2\frac{\xi}{2}\right)^{-ip_2+k_2-\frac{1}{2}}\\
\Hyper{j_{12}-k_1+k_2}{1-j_{12}-k_1+k_2}
{2k_2}{-\sinh^2\frac{\xi}{2}}\,d\xi\\
+ \int_{\sinh^{-1}\left(1\right)}^\infty
\left(\cosh^2\frac{\xi}{2}\right)^{-ip_1-k_1+\frac{1}{2}}
\left(\sinh^2\frac{\xi}{2}\right)^{-ip_2+k_1-j_{12}-\frac{1}{2}}\\
\Hyper{j_{12}-k_1+k_2}{1+j_{12}-k_1-k_2}
{2j_{12}}{-\sinh^{-2}\frac{\xi}{2}}\,d\xi
\end{multline}
where in this, and subsequent equations of this form, the normalisation
factor that ensures the analytic continuation agrees on the unit
circle has been omitted, and $j_{12} = k_{12}$ for $\mathcal{J}$ in
the positive discrete series and $j_{12} = \frac{1}{2}+is$ for
$\mathcal{J}$ in the continuous series.  

The negative discrete series is the same as the positive discrete
series, except $k_1$ and $k_2$ must be swapped in equation
\ref{PN}. The indices satisfy $k_1+k_2-1\ge k_1-k_2\ge k_{12}\ge 1$
for $k_{12}$ in the positive discrete series, an analogous relation
for the negative discrete series and there are no restrictions for
$\mathcal{J}$ in the continuous series, save for $k_1,\,k_2\ge 1$.

The entire integral can now be transformed so that it resembles the
integral in equation \ref{PPP}. First note that the solution $u_1$ of
the hypergeometric equation satisfies the following
transformation\cite{Kumar}
\begin{equation}
\Hyper{a}{b}{c}{z} =
 \left(1-z\right)^{-b}\Hyper{c-a}{b}{c}{\frac{z}{z-1}}
\end{equation}
for $\Re z < \frac{1}{2}$, while that for $u_3$ satisfies
\begin{equation}
\left(-z\right)^{-a}\Hyper{a}{a+1-c}{a+1-b}{z^{-1}} =
\left(1-z\right)^{-a}\Hyper{a}{c-b}{a+1-b}{\frac{1}{z-1}}
\end{equation}
for $\left|1-z\right| > 1$.

Applying these to the integral in equation \ref{PNpre1} gives 
\begin{multline}\label{PNpre2}
I\left(+,-,\mathcal{J}\right) = \int_0^{\sinh^{-1}\left(1\right)}
\left(\cosh^2\frac{\xi}{2}\right)^{-ip_1-k_2+j_{12}-\frac{1}{2}}
\left(\sinh^2\frac{\xi}{2}\right)^{-ip_2+k_2-\frac{1}{2}}\\
\Hyper{k_2+k_1-j_{12}}{1-j_{12}-k_1+k_2}{2k_2}{\tanh^2\frac{\xi}{2}}
\,d\xi\\
+\int_{\sinh^{-1}\left(1\right)}^\infty
\left(\cosh^2\frac{\xi}{2}\right)^{-ip_1-k_2-j_{12}+\frac{1}{2}}
\left(\sinh^2\frac{\xi}{2}\right)^{-ip_2+k_{2}-\frac{1}{2}}\\
\Hyper{j_{12}-k_1+k_2}{j_{12}+k_1+k_2-1}{2j_{12}}
{\cosh^{-2}\frac{\xi}{2}}\,d\xi
\end{multline}
Finally a substitution of $t = \tanh^2\frac{\xi}{2}$ yields the desired
form for the integral from equation \ref{PNpre2}
\begin{multline}\label{PN}
I\left(+,-,\mathcal{J}\right) = \int_0^{\frac{1}{2}}
\left(1-t\right)^{ip_1+ip_2-j_{12}}
t^{-ip_2+k_2-1}\\
\Hyper{k_2+k_1-j_{12}}{1-j_{12}-k_1+k_2}{2k_2}{t}
\,dt\\
+\int_\frac{1}{2}^1
\left(1-t\right)^{+ip_1+ip_2+j_{12}-1}
t^{-ip_2+k_2-1}\\
\Hyper{j_{12}-k_1+k_2}{j_{12}+k_1+k_2-1}{2j_{12}}
{1-t}\,dt
\end{multline}

As before to be assured of convergence one must check the end points
of the integrand, at $t=0$ and $t=1$. At $t=0$, it is the first
integral in equation \ref{PN} that is relevant and regardless of the
status of $\mathcal{J}$, the integrand behaves like $t^{k_2-1}$ since
the hypergeometric function is unity at $t=0$ and so the lower limit
is clearly bounded as $k_2\ge 1$.

For the other end point at $t=1$, the behaviour can be derived from
the second integral in equation \ref{PN} and behaves like
$\left(1-t\right)^{j_{12}-1}$ since the hypergeometric function tends
to unity again.  If $\mathcal{J}$ is in the discrete series one has
$j_{12}=k_{12} \ge 1$ and the integrand is bounded at this limit;
for $\mathcal{J}$ in the continuous series it diverges as
$\left(1-t\right)^{-\frac{1}{2}}$.  This divergence is insufficient to
prevent the integral converging, although it will have consequences
when the asymptotic contribution of the Clebsch-Gordon coefficients is
estimated in section \ref{converg}.

\subsection{Coupling of $\mathcal{D}^+_{k_1}$ with 
$\mathcal{C}^\epsilon_{s_2}$} 

The next choice of coupling is the case of a positive discrete series
representation and a continuous series representation, and is
discussed in detail in \cite{MR74iii}.  Since two cases of coupling
have been discussed already and the details of the last two cases are
considerably more involved, the details will be omitted here in favour
of the determining integral, which will be needed later.

When these couple to give a representation in the positive discrete
series the resulting Clebsch-Gordon coefficient has its dependence on
the continuous continuous basis parameters governed by two integrals of the
following form.

\begin{multline}\label{PCPpre1}
I_1\left(+,C,+\right) = \int_0^\infty
\left(\sinh^2\frac{\xi}{2}\right)^{-is_2-ip_2}
\left(\cosh^2\frac{\xi}{2}\right)^{\frac{1}{2}-k_1-ip_1}\\
\Hyper{\frac{1}{2}-k_1+k_{12}-is_2}{-k_1-k_{12}+\frac{3}{2}-is_2}
{1-2is_2}{-\sinh^2\frac{\xi}{2}}\,d\xi
\end{multline}
and
\begin{multline}\label{PCPpre2}
I_2\left(+,C,+\right) = \int_0^\pi
\left(\sin^2\frac{\theta}{2}\right)^{is_2+ip_2}
\left(\cos^2\frac{\theta}{2}\right)^{-\frac{1}{2}+k_1-ip_1}\\
\Hyper{\frac{1}{2}+k_1-k_{12}-is_2}{k_1+k_{12}-\frac{1}{2}-is_2}
{1-2is_2}{\sin^2\frac{\theta}{2}}\,d\theta
\end{multline}

Equation \ref{PCPpre2} may simply be written in a form similar to
the other integrals as
\begin{multline}\label{PCP2}
I_2\left(+,C,+\right) = \int_0^1
t^{is_2-ip_2-\frac{1}{2}}
\left(1-t\right)^{-1+k_1-ip_1}\\
\Hyper{\frac{1}{2}+k_1-k_{12}-is_2}{k_1+k_{12}-\frac{1}{2}-is_2}
{1-2is_2}{t}\,dt
\end{multline}
while, after being careful with choosing the analytic continuation
as previously, equation \ref{PCPpre1} may be rewritten as
\begin{multline}\label{PCP1}
I_1\left(+,C,+\right) = \int_0^\frac{1}{2}
t^{-is_2-ip_2-\frac{1}{2}}
\left(1-t\right)^{-k_{12}+ip_1+ip_2}\\
\Hyper{\frac{1}{2}+k_1-k_{12}-is_2}{\frac{3}{2}-k_1-k_{12}-is_2}
{1-2is_2}{t}\,dt\\
+\int_\frac{1}{2}^1
t^{-is_2-ip_2-\frac{1}{2}}
\left(1-t\right)^{k_{12}-1+ip_1+ip_2}\\
\Hyper{\frac{1}{2}+k_1+k_{12}-is_2}{-\frac{1}{2}+k_1+k_{12}-is_2}
{k_{12}}{1-t}\,dt
\end{multline}

One should note that both integrands behave as (respectively)
$\left(1-t\right)^{k_1-1}\to 0$ and $\left(1-t\right)^{k_{12}-1}\to 0$
at the $t=1$ limit since $k_1\ge 1$. Similarly both exhibit a mild
divergence at the $t=0$ limit, diverging as $t^{-\frac{1}{2}}$ as
$t\to 0$.  Again the divergence is insufficient to cause problems with
convergence of the integrals.

When the two representations couple to give a element of the
continuous series there are three integrals, a linear combination of
which determines the Clebsch-Gordon coefficient. From \cite{MR74iii}
one has

\begin{multline}\label{PCCpre1}
I_1\left(+,C,C\right) = \int_0^\infty
\left(\sinh^2\frac{\xi}{2}\right)^{k_1-\frac{1}{2}-ip_1}
\left(\cosh^2\frac{\xi}{2}\right)^{is_2-ip_2}\\
\Hyper{k_1+is_2+is_{12}}{k_1+is_2-is_{12}}
{2k_1}{-\sinh^2\frac{\xi}{2}}\,d\xi
\end{multline}

\begin{multline}\label{PCCpre2}
I_2\left(+,C,C\right) = \int_0^\infty
\left(\sinh^2\frac{\xi}{2}\right)^{is_2-ip_2}
\left(\cosh^2\frac{\xi}{2}\right)^{k_1-\frac{1}{2}-ip_1}\\
\Hyper{k_1+is_2+is_{12}}{k_1+is_2-is_{12}}
{1+2is_2}{-\sinh^2\frac{\xi}{2}}\,d\xi
\end{multline}

\begin{multline}\label{PCCpre3}
I_3\left(+,C,C\right) = \int_0^\pi
\left(\sin^2\frac{\theta}{2}\right)^{is_2-ip_2}
\left(\cos^2\frac{\theta}{2}\right)^{k_1-\frac{1}{2}-ip_1}\\
\Hyper{k_1+is_2+is_{12}}{k_1+is_2-is_{12}}
{1+2is_2}{\sin^2\frac{\theta}{2}}\,d\theta
\end{multline}
where the hypergeometric functions in equations \ref{PCCpre1} and
\ref{PCCpre2} are to be understood in terms of their analytic 
continuations when $\left|\sinh^2\frac{\xi}{2}\right|>1$.

When written as the previous integrals have been, one finds equation
\ref{PCCpre1} transforms to
\begin{multline}\label{PCC1}
I_1\left(+,C,C\right) = \int_0^\frac{1}{2}
t^{k_1-1-ip_1}
\left(1-t\right)^{-is_{12}-\frac{1}{2}+ip_1+ip_2}\\
\Hyper{k_1-is_{12}-is_2}{k_1-is_{12}+is_2}
{2k_1}{t}\,dt\\
+\int_\frac{1}{2}^1
t^{k_1-1-ip_1}
\left(1-t\right)^{is_{12}-\frac{1}{2}+ip_1+ip_2}\\
\Hyper{k_1+is_{12}+is_2}{k_1+is_{12}-is_2}
{2is_{12}+1}{1-t}\,dt
\end{multline}
equation \ref{PCCpre2} becomes
\begin{multline}\label{PCC2}
I_2\left(+,C,C\right) = \int_0^\frac{1}{2}
t^{is_2-\frac{1}{2}-ip_2}
\left(1-t\right)^{-is_{12}-\frac{1}{2}+ip_1+ip_2}\\
\Hyper{1-k_1-is_{12}+is_2}{k_1-is_{12}+is_2}
{1+2is_2}{t}\,dt\\
+\int_\frac{1}{2}^1
t^{is_2-ip_2-\frac{1}{2}}
\left(1-t\right)^{is_{12}-\frac{1}{2}+ip_1+ip_2}\\
\Hyper{k_1+is_{12}+is_2}{1-k_1+is_{12}+is_2}
{2is_{12}+1}{1-t}\,dt
\end{multline}
while equation \ref{PCCpre3} transforms to
\begin{multline}\label{PCC3}
I_3\left(+,C,C\right) = \int_0^1
t^{is_2-ip_2-\frac{1}{2}}
\left(1-t\right)^{k_1-1-ip_1}\\
\Hyper{k_1+is_2+is_{12}}{k_1+is_2-is_{12}}
{1+2is_2}{t}\,dt
\end{multline}

Again it is easily verified that all three integrals converge,
although the integrand may diverge as $\frac{1}{\sqrt{t}}$ when $t\to
0$ at either the lower or upper limit.

\subsection{Coupling of $\mathcal{C}^\epsilon_{s_1}$ 
with $\mathcal{C}^{\epsilon^\prime}_{s_2}$} 

The final case of coupling involves two continuous series
representations coupling to give either an element of the discrete
series, or a third continuous series representation.  This is
discussed in detail in \cite{MR74iv} and again the details will be
omitted here.

When these couple to produce a representation in the positive discrete
series the continuous basis dependent part of the Clebsch-Gordon coefficient is
governed by a linear combination of two types of integral that arise
from the appropriate inner product.
\begin{multline}\label{CCPpre1}
I_1\left(C,C,+\right) = \int_0^\pi
\left(\cos^2\frac{\theta}{2}\right)^{-ip_1-is_1}
\left(\sin^2\frac{\theta}{2}\right)^{-ip_2+is_2}\\
\Hyper{k_{12}-is_1+is_2}{1-k_{12}-is_1+is_2}
{1+2is_2}{\sin^2\frac{\theta}{2}}\,d\theta
\end{multline}
and
\begin{multline}\label{CCPpre2}
I_2\left(C,C,+\right) = \int_0^\infty
\left(\cosh^2\frac{\xi}{2}\right)^{-ip_1-is_1}
\left(\sinh^2\frac{\xi}{2}\right)^{-ip_2+is_2}\\
\Hyper{k_{12}-is_1+is_2}{1-k_{12}-is_1+is_2}
{1+2is_2}{-\sinh^2\frac{\xi}{2}}\,d\xi
\end{multline}
Again the hypergeometric function in equation \ref{CCPpre2} is to be
understood in terms of its analytic continuation for
$\left|\sinh\frac{\xi}{2}\right| > 1$.

Equation \ref{CCPpre1} may be immediately rewritten as
\begin{multline}\label{CCP1}
I_1\left(C,C,+\right) = \int_0^1
\left(1-t\right)^{-ip_1-is_1-\frac{1}{2}}
t^{-ip_2+is_2-\frac{1}{2}}\\
\Hyper{k_{12}-is_1+is_2}{1-k_{12}-is_1+is_2}
{1+2is_2}{t}\,dt
\end{multline}
while by the same method as previously equation \ref{CCPpre2} may
eventually be given the form
\begin{multline}\label{CCP2}
I_2\left(C,C,+\right) = 
\int_0^\frac{1}{2}
\left(1-t\right)^{-k_{12}+ip_1+ip_2}
t^{-ip_2+is_2-\frac{1}{2}}\\
\Hyper{1-k_{12}+is_1+is_2}{1-k_{12}-is_1+is_2}
{1+2is_2}{t}\,dt\\+
\int_\frac{1}{2}^1
\left(1-t\right)^{k_{12}-1+ip_1+ip_2}
t^{-ip_2+is_2-\frac{1}{2}}\\
\Hyper{k_{12}-is_1+is_2}{k_{12}+is_1+is_2}
{2k_{12}}{1-t}\,dt
\end{multline}

When the two representations couple to produce a representation in the
continuous series the two defining integrals may be derived from
equations \ref{CCPpre1} and \ref{CCPpre2} via a simple replacement of
$k_{12}\to \frac{1}{2}+is_{12}$. Thus the defining integrals may be
written
\begin{multline}\label{CCC1}
I_1\left(C,C,C\right) = \int_0^1
\left(1-t\right)^{-ip_1-is_1-\frac{1}{2}}
t^{-ip_2+is_2-\frac{1}{2}}\\
\Hyper{\frac{1}{2}+is_{12}-is_1+is_2}{\frac{1}{2}-is_{12}-is_1+is_2}
{1+2is_2}{t}\,dt
\end{multline}
while by the same method as previously equation \ref{CCPpre2} may
eventually be given the form
\begin{multline}\label{CCC2}
I_2\left(C,C,C\right) = 
\int_0^\frac{1}{2}
\left(1-t\right)^{-\frac{1}{2}-is_{12}+ip_1+ip_2}
t^{-ip_2+is_2-\frac{1}{2}}\\
\Hyper{\frac{1}{2}-is_{12}+is_1+is_2}{\frac{1}{2}-is_{12}-is_1+is_2}
{1+2is_2}{t}\,dt\\+
\int_\frac{1}{2}^1
\left(1-t\right)^{is_{12}-\frac{1}{2}+ip_1+ip_2}
t^{-ip_2+is_2-\frac{1}{2}}\\
\Hyper{\frac{1}{2}+is_{12}-is_1+is_2}{\frac{1}{2}+is_{12}+is_1+is_2}
{1+2is_{12}}{1-t}\,dt
\end{multline}

The integrals given by equation \ref{CCP1}, \ref{CCP2},
\ref{CCC1} and \ref{CCC2} may be shown to converge using the same
methods as for the previous cases of coupling.

\chapter{The Racah coefficient}\label{RACAH}

In this chapter definitions and calculations of the Racah coefficient
are provided.  In section \ref{Racah} the $SU\left(1,1\right)$ Racah
coefficient is defined for the whole principal series in terms of a
double integral of Clebsch-Gordon coefficients in the
$SO\left(1,1\right)$ continuous basis of section \ref{CGC}. The
standard relations are easily derived from this definition.

This coefficient is shown to exist in section \ref{converg} by
explicitly determining the convergence of its defining integral.
Finally calculations are undertaken in sections \ref{rac-CGC} and
\ref{rac} to write the Racah coefficient, for the cases where all
representations are in the discrete series, in a closed form and
derive its symmetries.

\section{The Racah Coefficient}\label{Racah}

Consider the tensor product of three representations $\mathcal{J}_1$,
$\mathcal{J}_2$ and $\mathcal{J}_3$ and let the Casimir of the total
group define the representation $\mathcal{J}$.  Thus one is looking
for a map from $\mathcal{J}_1\otimes\mathcal{J}_2\otimes\mathcal{J}_3$
to a direct integral or sum of $\mathcal{J}$'s.

There are evidently two ways of computing this map.  The inner product
in which $\mathcal{J}_1$ and $\mathcal{J}_2$ are decomposed first may
be computed as follows; let the Clebsch-Gordon coefficient be
determined by 
\begin{equation}
\ip{\Phi^{\mathcal{J}_1,\,\mathcal{J}_2}_{p_1,\,p_2}}
{\Psi^{\mathcal{J},\,\mathcal{J}^\prime\,\,\mathcal{J}_{12}}
_{p_{12}}}
 = \delta\left(\mathcal{J}_1,\,\mathcal{J}\right)
\delta\left(\mathcal{J}_2,\,\mathcal{J}^\prime\right)
\delta\left(p_{12}-p_1-p_2\right)
\left[\begin{array}{ccc} 
\mathcal{J}_1 & \mathcal{J}_2 & \mathcal{J}_{12}\\
p_1 & p_2 & p_{12} \end{array}\right]
\end{equation}
So it defines a change of continuous basis from the uncoupled to the coupled
continuous basis (as in section \ref{CGC}) and hence an isomorphism
$$
\mathcal{J}_1\otimes\mathcal{J}_2\longrightarrow
\dint\mathfrak{D}\mathcal{J}_{12}\;\mathcal{J}_{12} 
$$
where a coupled continuous basis of each $\mathcal{J}_{12}$ is given
in terms of the uncoupled continuous basis as
\begin{equation}\label{1.2->12}
\ket{\Psi^{\mathcal{J}_1,\,\mathcal{J}_2\,\,\mathcal{J}_{12}}
_{p_{12}}}=\int\left[\begin{array}{ccc} 
\mathcal{J}_1 & \mathcal{J}_2 & \mathcal{J}_{12}\\
p_1 & p_2 & p_{12} \end{array}\right]
\ket{\Phi^{\mathcal{J}_1,\,\mathcal{J}_2}_{p_1,\,p_2}}
\,dp_1 dp_2
\end{equation}

For the space $\mathcal{J}_1\otimes\mathcal{J}_2\otimes\mathcal{J}_3$
an uncoupled continuous basis is given by the functions
$\Phi^{\mathcal{J}_1}_{p_1}
\Phi^{\mathcal{J}_2}_{p_2}\Phi^{\mathcal{J}_3}_{p_3}$, while 
the space $\mathcal{J}_{12}\otimes\mathcal{J}_3$ has a partially
coupled continuous basis given by the functions
$\Psi^{\mathcal{J}_1,\,\mathcal{J}_2\,\,\mathcal{J}_{12}}
_{p_{12}}\Phi^{\mathcal{J}_3}_{p_3}$ which are defined using equation
\ref{1.2->12}. 
There is a change of continuous basis between these two that implements the
isomorphism
$$
\mathcal{J}_1\otimes\mathcal{J}_2\otimes\mathcal{J}_3
\longrightarrow\mathcal{J}_{12}\otimes\mathcal{J}_3
$$

Finally there is a coupled continuous basis, given by the functions
$\Psi^{\left(\mathcal{J}_1\mathcal{J}_2\right)
\mathcal{J}_{3},\;\mathcal{J}}_p$, of the space $\mathcal{J}$. These arise
from an analogous direct integral decomposition of
$\mathcal{J}_{12}\otimes\mathcal{J}_3$ and may be explicitly computed
by an equation analogous to equation \ref{1.2->12}.

Hence the change of continuous basis from the uncoupled continuous basis
$\Phi^{\mathcal{J}_1,\,\mathcal{J}_2,\,\mathcal{J}_3}_{p_1,\,p_2,\,p_3}$
to the coupled continuous basis $\Psi^{\left(\mathcal{J}_1\mathcal{J}_2\right)
\mathcal{J}_{3},\;\mathcal{J}}_p$ which implements the
isomorphism
$$
\mathcal{J}_1\otimes\mathcal{J}_2\otimes\mathcal{J}_3
\longrightarrow\dint\mathfrak{D}\mathcal{J}\;\mathcal{J}
$$
may be explicitly realised as
\begin{multline}\label{rac-1}
\ket{\Psi^{\left(\mathcal{J}_1\mathcal{J}_2\right)
\mathcal{J}_{12}\mathcal{J}_{3},\;\mathcal{J}}_p}=\int
\left[\begin{array}{ccc} 
\mathcal{J}_1 & \mathcal{J}_2 & \mathcal{J}_{12}\\
p_1 & p_2 & p_{12} \end{array}\right]\\
\left[\begin{array}{ccc} 
\mathcal{J}_{12} & \mathcal{J}_3 & \mathcal{J}\\
p_{12} & p_3 & p \end{array}\right]
\ket{\Phi^{\mathcal{J}_1,\,\mathcal{J}_2,\,\mathcal{J}_3}_{p_1,\,p_2,\,p_3}}
\,dp_1\,dp_2\,dp_3\,dp_{12}
\end{multline}
where 
$$
\ket{\Psi^{\left(\mathcal{J}_1\mathcal{J}_2\right)
\mathcal{J}_{3},\;\mathcal{J}}_p}
= \int\mathfrak{D}\mathcal{J}_{12}\;
\ket{\Psi^{\left(\mathcal{J}_1\mathcal{J}_2\right)
\mathcal{J}_{12}\mathcal{J}_{3},\;\mathcal{J}}_p}
$$
and the
$\ket{\Psi^{\left(\mathcal{J}_1\mathcal{J}_2\right)
\mathcal{J}_{3},\;\mathcal{J}}_p}$ are a continuous basis for
each $\mathcal{J}$.

Similarly there is a change of continuous basis from the uncoupled
continuous basis
$\Phi^{\mathcal{J}_1,\,\mathcal{J}_2,\,\mathcal{J}_3}_{p_1,\,p_2,\,p_3}$
to the coupled continuous basis
$\Psi^{\mathcal{J}_1\left(\mathcal{J}_2
\mathcal{J}_{3}\right)\mathcal{J}_{23},\;\mathcal{J}}_p$ where 
the continuous basis of $\mathcal{J}_2\otimes\mathcal{J}_3$ is diagonalised
first.  Explicitly the change of continuous basis is given by
\begin{multline}\label{rac-2}
\ket{\Psi^{\mathcal{J}_1\left(\mathcal{J}_2
\mathcal{J}_{3}\right)\mathcal{J}_{23},\;\mathcal{J}}_p}=\int
\left[\begin{array}{ccc} 
\mathcal{J}_2 & \mathcal{J}_3 & \mathcal{J}_{23}\\
p_2 & p_3 & p_{23} \end{array}\right]\\
\left[\begin{array}{ccc} 
\mathcal{J}_1 & \mathcal{J}_{23} & \mathcal{J}\\
p_1 & p_{23} & p \end{array}\right]
\ket{\Phi^{\mathcal{J}_1,\,\mathcal{J}_2,\,\mathcal{J}_3}_{p_1,\,p_2,\,p_3}}
\,dp_1\,dp_2\,dp_3\,dp_{23}
\end{multline}
where again 
$$
\ket{\Psi^{\mathcal{J}_1\left(\mathcal{J}_2
\mathcal{J}_{3}\right),\;\mathcal{J}}_p} =
\dint\mathfrak{D}\mathcal{J}_{23}\;
\ket{\Psi^{\mathcal{J}_1\left(\mathcal{J}_2
\mathcal{J}_{3}\right)\mathcal{J}_{23},\;\mathcal{J}}_p}
$$
is a continuous basis for each $\mathcal{J}$.

The Racah coefficient $\left\{\begin{array}{ccc}
\mathcal{J}_1 & \mathcal{J}_2 & \mathcal{J}_{12}\\
\mathcal{J}_3 & \mathcal{J} & \mathcal{J}_{23}
\end{array}\right\}$ is defined as the unitary map between the two
different coupled continuous bases given by equations \ref{rac-1} and
\ref{rac-2}.  It is realised as
\index{$\left\{\begin{array}{ccc}
\mathcal{J}_1 & \mathcal{J}_2 & \mathcal{J}_{12}\\
\mathcal{J}_3 & \mathcal{J} & \mathcal{J}_{23}
\end{array}\right\}$}
\begin{equation}\label{RAC}
\left\{\begin{array}{ccc}
\mathcal{J}_1 & \mathcal{J}_2 & \mathcal{J}_{12}\\
\mathcal{J}_3 & \mathcal{J} & \mathcal{J}_{23}
\end{array}\right\}\delta\left(\mathcal{J},\mathcal{J}^\prime\right)
\delta\left(p-p^\prime\right)
=\ip{\Psi^{\mathcal{J}_1\left(\mathcal{J}_2
\mathcal{J}_{3}\right)\mathcal{J}_{23},\;\mathcal{J}^\prime}_{p^\prime}}
{\Psi^{\left(\mathcal{J}_1\mathcal{J}_2\right)
\mathcal{J}_{12}\mathcal{J}_{3},\;\mathcal{J}}_p}
\end{equation}
It is easy to see that the Racah coefficient is independent of the
specific realisation of the coupled and uncoupled continuous basis since it is
defined as an inner product.

Equation \ref{RAC} may be written, using equations \ref{rac-1} and
\ref{rac-2}, as a multiple integral of four Clebsch-Gordon coefficients
\begin{multline}\label{41}
\int\,dp_1\,dp_2\,dp_3\,dp_{12}\,dp_{23}\,
\delta\left(p_{12}-p_1-p_2\right)
\delta\left(p-p_{12}-p_3\right)
\delta\left(p_{23}-p_2-p_3\right)\\
\delta\left(p^\prime-p_{23}-p_1\right)
\left[\begin{array}{ccc} 
\mathcal{J}_1 & \mathcal{J}_2 & \mathcal{J}_{12}\\ 
p_1 & p_2 & p_{12} \end{array}\right]
\left[\begin{array}{ccc} 
\mathcal{J}_{12} & \mathcal{J}_3 & \mathcal{J}\\
p_{12} & p_3 & p\end{array}\right]
\left[\begin{array}{ccc} 
\mathcal{J}_2 & \mathcal{J}_3 & \mathcal{J}_{23}\\
p_2 & p_3 & p_{23}\end{array}\right]^\star\\
\left[\begin{array}{ccc} 
\mathcal{J}_1 & \mathcal{J}_{23} & \mathcal{J}^\prime\\
p_1 & p_{23} & p^\prime\end{array}\right]^\star
= \delta\left(\mathcal{J},\,\mathcal{J}^\prime\right)
\delta\left(p-p^\prime\right)
\left\{\begin{array}{ccc}
\mathcal{J}_1 & \mathcal{J}_2 & \mathcal{J}_{12}\\
\mathcal{J}_3 & \mathcal{J} & \mathcal{J}_{23}
\end{array}\right\}
\end{multline}
The convergence of this integral will be discussed in section
\ref{converg}.

\bigskip

Now using equation \ref{CGC-orth2} one may transform equation
\ref{41} into the so-called recoupling identity. 
Explicitly it is given in terms of Clebsch-Gordon coefficients by
\index{$\left\{\begin{array}{ccc}
\mathcal{J}_1 & \mathcal{J}_2 & \mathcal{J}_{12}\\
\mathcal{J}_3 & \mathcal{J} & \mathcal{J}_{23}
\end{array}\right\}$}
\begin{multline}\label{racah}
\int\,dp_{12}\,
\delta\left(p_{12}-p_1-p_2\right)
\delta\left(p-p_{12}-p_3\right)
\left[\begin{array}{ccc} 
\mathcal{J}_1 & \mathcal{J}_2 & \mathcal{J}_{12}\\
p_1 & p_2 & p_{12} \end{array}\right]
\left[\begin{array}{ccc} 
\mathcal{J}_{12} & \mathcal{J}_3 & \mathcal{J}\\
p_{12} & p_3 & p \end{array}\right]\\
= \dint\,\mathfrak{D}\mathcal{J}_{23}\;
\int\,dp_{23}\,
\delta\left(p_{23}-p_2-p_3\right)
\delta\left(p-p_{23}-p_1\right)\\
\left[\begin{array}{ccc} 
\mathcal{J}_2 & \mathcal{J}_3 & \mathcal{J}_{23}\\
p_2 & p_3 & p_{23} \end{array}\right]
\left[\begin{array}{ccc} 
\mathcal{J}_{23} & \mathcal{J}_1 & \mathcal{J}\\
p_{23} & p_1 & p \end{array}\right]
\left\{\begin{array}{ccc}
\mathcal{J}_1 & \mathcal{J}_2 & \mathcal{J}_{12}\\
\mathcal{J}_3 & \mathcal{J} & \mathcal{J}_{23}
\end{array}\right\}
\end{multline}

The Racah coefficient is then realised explicitly as a 
change of continuous basis since equation \ref{racah} determines a map
$$\alpha^{\mathcal{J}_1\, \mathcal{J}_2\, \mathcal{J}_3} : 
V_{\mathcal{J}_1}\otimes\left(V_{\mathcal{J}_2}\otimes V_{\mathcal{J}_3}\right)
\longrightarrow \left(V_{\mathcal{J}_1}\otimes 
V_{\mathcal{J}_2}\right)\otimes V_{\mathcal{J}_3}
$$

The inverse map is given by
$$
\alpha^{\mathcal{J}_1\, \mathcal{J}_2\, \mathcal{J}_3\;\star} : 
\left(V_{\mathcal{J}_1}\otimes V_{\mathcal{J}_2}\right)
\otimes V_{\mathcal{J}_3}
\longrightarrow V_{\mathcal{J}_1}\otimes
\left(V_{\mathcal{J}_2}\otimes V_{\mathcal{J}_3}\right)
$$
may be explicitly defined by
\begin{multline}\label{racah-adj}
\int\,dp_{23}\,
\delta\left(p_{23}-p_2-p_3\right)
\delta\left(p-p_{23}-p_1\right)
\left[\begin{array}{ccc} 
\mathcal{J}_2 & \mathcal{J}_3 & \mathcal{J}_{23}\\
p_2 & p_3 & p_{23} \end{array}\right]
\left[\begin{array}{ccc} 
\mathcal{J}_{23} & \mathcal{J}_1 & \mathcal{J}\\
p_{23} & p_1 & p \end{array}\right]\\
= \dint\,\mathfrak{D}\mathcal{J}_{12}\;
\int\,dp_{12}\,
\delta\left(p_{12}-p_1-p_2\right)
\delta\left(p-p_{12}-p_3\right)\\
\left[\begin{array}{ccc} 
\mathcal{J}_1 & \mathcal{J}_2 & \mathcal{J}_{12}\\
p_1 & p_2 & p_{12} \end{array}\right]
\left[\begin{array}{ccc} 
\mathcal{J}_{12} & \mathcal{J}_3 & \mathcal{J}\\
p_{12} & p_3 & p \end{array}\right]
\left\{\begin{array}{ccc}
\mathcal{J}_1 & \mathcal{J}_2 & \mathcal{J}_{12}\\
\mathcal{J}_3 & \mathcal{J} & \mathcal{J}_{23}
\end{array}\right\}^\star
\end{multline}
in the same way.

A variety of identities may be derived for the Racah coefficient in view
of equation \ref{racah}.
\begin{prop}[Orthogonality]\label{O}
$$
\dint\mathfrak{D}\mathcal{J}_{12}\,
\left\{\begin{array}{ccc}
\mathcal{J}_1 & \mathcal{J}_2 & \mathcal{J}_{12}\\
\mathcal{J}_3 & \mathcal{J} & \mathcal{J}_{23}
\end{array}\right\}
\left\{\begin{array}{ccc}
\mathcal{J}_1 & \mathcal{J}_2 & \mathcal{J}_{12}\\
\mathcal{J}_3 & \mathcal{J} & \mathcal{J}_{23}^\prime
\end{array}\right\}^\star = \frac{\delta(\mathcal{J}_{23},
\mathcal{J}_{23}^\prime)}
{\mu\left(\mathcal{J}_{23}\right)}
$$
\end{prop}

This follows from considering the identity map

$$\alpha^{\mathcal{J}_1\, \mathcal{J}_2\, \mathcal{J}_3}
\circ\alpha^{\mathcal{J}_1\, \mathcal{J}_2\, \mathcal{J}_3\,\star}:
V_{\mathcal{J}_1}\otimes\left(V_{\mathcal{J}_2}\otimes V_{\mathcal{J}_3}\right)
\rightarrow V_{\mathcal{J}_1}\otimes\left(V_{\mathcal{J}_2}\otimes 
V_{\mathcal{J}_3}\right)$$ 

Thus (the $\delta$ functions associated with each Clebsch-Gordon
coefficient have been suppressed for clarity)
\begin{multline*}
\int
\left[\begin{array}{ccc} 
\mathcal{J}_2 & \mathcal{J}_3 & \mathcal{J}_{23}\\
p_2 & p_3 & p_{23} \end{array}\right]
\left[\begin{array}{ccc} 
\mathcal{J}_{23} & \mathcal{J}_1 & \mathcal{J}\\
p_{23} & p_1 & p\end{array}\right]\,dp_{23}\\
 = \dint\,\mathfrak{D}\mathcal{J}_{12}\int\,
\left[\begin{array}{ccc} 
\mathcal{J}_1 & \mathcal{J}_2 & \mathcal{J}_{12}\\
p_1 & p_2 & p_{12} \end{array}\right]
\left[\begin{array}{ccc} 
\mathcal{J}_{12} & \mathcal{J}_3 & \mathcal{J}\\
p_{12} & p_3 & p \end{array}\right]
\left\{\begin{array}{ccc}
\mathcal{J}_1 & \mathcal{J}_2 & \mathcal{J}_{12}\\
\mathcal{J}_3 & \mathcal{J} & \mathcal{J}_{23}
\end{array}\right\}\,dp_{12}\\
 = \dint\,\mathfrak{D}\mathcal{J}_{12}\,
\mathfrak{D}\mathcal{J}_{23}^\prime\,\int\,
\left[\begin{array}{ccc} 
\mathcal{J}_2 & \mathcal{J}_3 & \mathcal{J}_{23}^\prime\\
p_2 & p_3 & p_{23}^\prime \end{array}\right]
\left[\begin{array}{ccc} 
\mathcal{J}_{23}^\prime & \mathcal{J}_1 & \mathcal{J}\\
p_{23}^\prime & p_1 & p\end{array}\right]\,dp_{23}^\prime\\
\left\{\begin{array}{ccc}
\mathcal{J}_1 & \mathcal{J}_2 & \mathcal{J}_{12}\\
\mathcal{J}_3 & \mathcal{J} & \mathcal{J}_{23}
\end{array}\right\}
\left\{\begin{array}{ccc}
\mathcal{J}_1 & \mathcal{J}_2 & \mathcal{J}_{12}\\
\mathcal{J}_3 & \mathcal{J} & \mathcal{J}_{23}^\prime
\end{array}\right\}^\star
\end{multline*}
which implies orthogonality.

\begin{figure}[htb!]
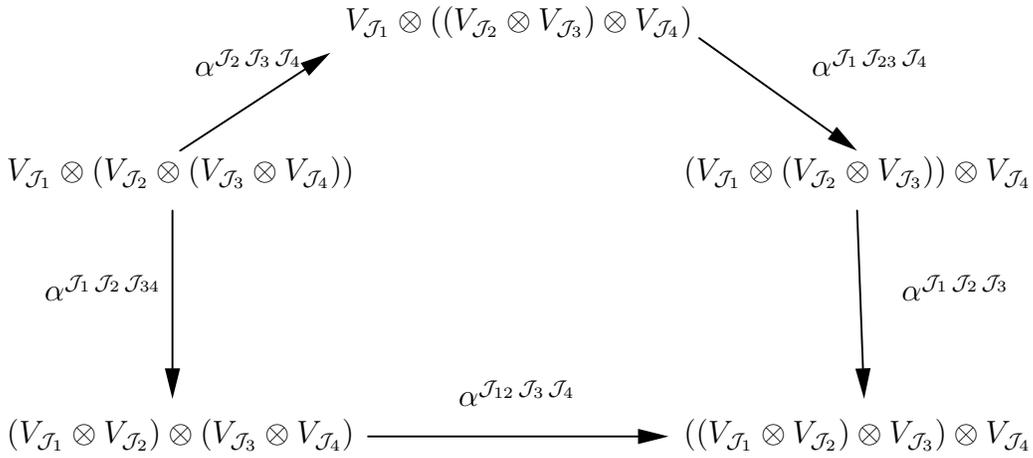

\vspace{1cm}
\begin{texdraw}
\drawdim{mm} 
\arrowheadtype t:F
\move(0 0)
\htext{$V_{\mathcal{J}_1}\otimes\left(V_{\mathcal{J}_2}
\otimes\left(V_{\mathcal{J}_3}\otimes V_{\mathcal{J}_4}\right)\right)$}
\move(23 5)
\avec(43 18)
\move(25 15)\htext{$\alpha^{\mathcal{J}_2\,\mathcal{J}_3\,\mathcal{J}_4}$}
\move(45 20)
\htext{$V_{\mathcal{J}_1}\otimes
\left(\left(V_{\mathcal{J}_2}\otimes V_{\mathcal{J}_3}\right)\otimes 
V_{\mathcal{J}_4}\right)$}
\move(92 20)
\avec(113 5)
\move(107 15)\htext{$\alpha^{\mathcal{J}_1\,\mathcal{J}_{23}\,\mathcal{J}_4}$}
\move(90 0)
\htext{$\left(V_{\mathcal{J}_1}\otimes\left(V_{\mathcal{J}_2}
\otimes V_{\mathcal{J}_3}\right)\right)\otimes V_{\mathcal{J}_4}$}
\move(22 -3)
\avec(22 -28)
\move(5 -15)\htext{$\alpha^{\mathcal{J}_1\,\mathcal{J}_2\,\mathcal{J}_{34}}$}
\move(113 -3)
\avec(114 -28)
\move(119 -15)\htext{$\alpha^{\mathcal{J}_1\,\mathcal{J}_2\,\mathcal{J}_3}$} 
\move(0 -35)
\htext{$\left(V_{\mathcal{J}_1}\otimes V_{\mathcal{J}_2}\right)
\otimes\left(V_{\mathcal{J}_3}\otimes V_{\mathcal{J}_4}\right)$}
\move(48 -33)
\avec(88 -33)
\move(60 -29)\htext{$\alpha^{\mathcal{J}_{12}\,\mathcal{J}_3\,\mathcal{J}_4}$}
\move(90 -35)
\htext{$\left(\left(V_{\mathcal{J}_1}\otimes V_{\mathcal{J}_2}\right)
\otimes V_{\mathcal{J}_3}\right)\otimes V_{\mathcal{J}_4}$}
\end{texdraw}
\caption{The Pentagon relation \label{pentagon}}
\end{figure}

\begin{prop}[Biedenharn-Elliot]\label{BE}
\begin{multline}\dint\,\mathfrak{D}\mathcal{J}_{23}\,
\left\{\begin{array}{ccc}
\mathcal{J}_2 & \mathcal{J}_3 & \mathcal{J}_{23} \\
\mathcal{J}_4 & \mathcal{J}_{234} & \mathcal{J}_{34}
\end{array}\right\}
\left\{\begin{array}{ccc}
\mathcal{J}_1 & \mathcal{J}_{23} & \mathcal{J}_{123} \\
\mathcal{J}_4 & \mathcal{J} & \mathcal{J}_{234}
\end{array}\right\}\\
\left\{\begin{array}{ccc}
\mathcal{J}_1 & \mathcal{J}_2 & \mathcal{J}_{12} \\
\mathcal{J}_3 & \mathcal{J}_{123} & \mathcal{J}_{23}
\end{array}\right\} = 
\left\{\begin{array}{ccc}
\mathcal{J}_1 & \mathcal{J}_2 & \mathcal{J}_{12} \\
\mathcal{J}_{34} & \mathcal{J} & \mathcal{J}_{234}
\end{array}\right\}
\left\{\begin{array}{ccc}
\mathcal{J}_{12} & \mathcal{J}_3 & \mathcal{J}_{123} \\
\mathcal{J}_4 & \mathcal{J} & \mathcal{J}_{34}
\end{array}\right\}
\end{multline}
\end{prop}

Biedenharn-Elliot follows from category theoretic considerations since the
Racah coefficient is the associator of the monoidal category of  unitary
representations of $SU(1,1)$. Thus it satisfies the Pentagon relation in figure
\ref{pentagon} from which the above can be read off. Similar three to two
relations may be derived using different starting and finishing points in the
pentagon relation.

One may also derive the following, again from the Pentagon
relation in figure \ref{pentagon}
\begin{prop}[Four to One]
\begin{multline}\label{4-1}
\dint
\,\mathfrak{D}\mathcal{J}_{12}\,\mathfrak{D}\mathcal{J}_{23}
\,\mathfrak{D}\mathcal{J}_{234}
\left\{\begin{array}{ccc}
\mathcal{J}_1 & \mathcal{J}_2 & \mathcal{J}_{12}\\ \mathcal{J}_3 & \mathcal{J}_{123} & \mathcal{J}_{23}
\end{array}\right\}
\left\{\begin{array}{ccc}
\mathcal{J}_1 & \mathcal{J}_{23} & \mathcal{J}_{123}\\\mathcal{J}_4 & \mathcal{J} & \mathcal{J}_{234}
\end{array}\right\}\\
\left\{\begin{array}{ccc}
\mathcal{J}_2 & \mathcal{J}_3 & \mathcal{J}_{23}\\ \mathcal{J}_4 & \mathcal{J}_{234} & \mathcal{J}_{34}
\end{array}\right\}
\left\{\begin{array}{ccc}
\mathcal{J}_1 & \mathcal{J}_2 & \mathcal{J}_{12}^\prime\\\mathcal{J}_{34} & \mathcal{J} & \mathcal{J}_{234}
\end{array}\right\}^\star
= \left\{\begin{array}{ccc}
\mathcal{J}_{12}^\prime & \mathcal{J}_3 & \mathcal{J}_{123}\\ \mathcal{J}_4 & \mathcal{J} & \mathcal{J}_{34}
\end{array}\right\}
\end{multline}
\end{prop}

Again, nine further four to one relations may be derived by considering
different starting and finishing points and different directions of  traversal
for the pentagon relation.

\section{On the existence of the Racah Coefficient}\label{converg}

In this section it will be shown that the Racah coefficient is
well defined by the integral in equation \ref{41}.

\begin{thm}\label{Racah-converg}
The Racah coefficient $\left\{\begin{array}{ccc}
\mathcal{J}_1 & \mathcal{J}_2 & \mathcal{J}_{12}\\
\mathcal{J}_3 & \mathcal{J} & \mathcal{J}_{23}
\end{array}\right\}$ as defined by
\begin{multline}\label{2p}
\int\,dp_{12}\,dp_{23}\,
\left[\begin{array}{ccc} 
\mathcal{J}_1 & \mathcal{J}_2 & \mathcal{J}_{12}\\ 
p - p_{23} & p_{12} + p_{23} - p & p_{12} \end{array}\right]
\left[\begin{array}{ccc} 
\mathcal{J}_{12} & \mathcal{J}_3 & \mathcal{J}\\
p_{12} & p - p_{12} & p\end{array}\right]\\
\left[\begin{array}{ccc} 
\mathcal{J}_2 & \mathcal{J}_3 & \mathcal{J}_{23}\\
p_{12} + p_{23} - p & p - p_{12} & p_{23}\end{array}\right]^\star
\left[\begin{array}{ccc} 
\mathcal{J}_1 & \mathcal{J}_{23} & \mathcal{J}^\prime\\
p^\prime - p_{23} & p_{23} & p^\prime\end{array}\right]^\star\\
= \delta\left(\mathcal{J},\,\mathcal{J}^\prime\right)
\left\{\begin{array}{ccc}
\mathcal{J}_1 & \mathcal{J}_2 & \mathcal{J}_{12}\\
\mathcal{J}_3 & \mathcal{J} & \mathcal{J}_{23}
\end{array}\right\}
\end{multline}
is a well defined function of its six indices
whose value is given by the integral in equation \ref{2p}.
\end{thm}

In order to prove this one needs to show the integral converges.  This
is governed by the asymptotics of the Clebsch-Gordon coefficients as
$p_{12}$ and $p_{23}$ become large. From the results of section
\ref{CGC} the Clebsch-Gordon coefficients for the various cases of
coupling can be expressed as integrals of the form
\begin{equation}\label{Int}
I = \int_0^1\,t^{\alpha^\prime}\left(1-t\right)^{\beta^\prime} 
{ }_2F_1
\left(\begin{array}{cc} a & b\\ & c\end{array} ; t\right)\,dt
\end{equation}
where the labels $\alpha^\prime$, $\beta^\prime$, a, b and c are
constants depending on the particular case of coupling;
$\alpha^\prime$ and $\beta^\prime$ may depend further on the continuous basis
elements under consideration.  The hypergeometric functions all
satisfy $c-a-b > 0$ so that they are bounded at $t=1$.

One type\footnote{These are integrals derived from the integrals
involving trigonometric functions} corresponds to the case where
$\alpha^\prime =
\alpha - ip_2$ and $\beta^\prime = \beta - ip_1$ while the 
other\footnote{These are derived from the integrals involving
hyperbolic functions} has $\alpha^\prime = \alpha - ip_2$ and
$\beta^\prime = \beta + ip_1 + ip_2$. The constants $\alpha$ and
$\beta$ depend only on the representation labels and our interest is
in the case where $p_1$ and $p_2$ are large with a certain relationship
between them.

The values of $\alpha$ and $\beta$ are given by the results of section
\ref{CGC} which, for convenience, are summarised in table
\ref{CGC-summ}.

\begin{table}[hbt!]
\begin{center}
{\renewcommand{\arraystretch}{1.1}
\begin{tabular}{|lc||lcl|}
\hline
Integral & from equation & 
\multicolumn{3}{c|}{Asymptotics of the Integrand}\\
& & at $t=0$ & & at $t=1$\\
\hline
$I\left(+,+,+\right)$ & (\ref{PPP}) 
& $t^{k_2-1-ip_2}$ && $\left(1-t\right)^{k_1-1-ip_1}$\\
$I\left(+,-,\mathcal{J}\right)$ & (\ref{PN}) &
$t^{k_2-1-ip_2}$ && $\left(1-t\right)^{j_{12}-1+ip_1+ip_2}$\\
$I_1\left(+,C,+\right)$ & (\ref{PCP1}) 
& $t^{-is_2-ip_2-\frac{1}{2}}$ &&  $\left(1-t\right)^{k_{12}-1+ip_1+ip_2}$\\
$I_2\left(+,C,+\right)$ & (\ref{PCP2})
& $t^{is_2-ip_2-\frac{1}{2}}$ && $\left(1-t\right)^{k_1-1-ip_1}$ \\
$I_1\left(+,C,C\right)$ & (\ref{PCC1})
& $t^{k_1-1-ip_1}$ && $\left(1-t\right)^{is_{12}-\frac{1}{2}+ip_1+ip_2}$ \\
$I_2\left(+,C,C\right)$ & (\ref{PCC2})
& $t^{is_2-\frac{1}{2}-ip_2}$ && 
$\left(1-t\right)^{is_{12}-\frac{1}{2}+ip_1+ip_2}$ \\
$I_3\left(+,C,C\right)$ & (\ref{PCC3})
& $t^{is_2-ip_2-\frac{1}{2}}$ && $\left(1-t\right)^{k_1-1-ip_1}$ \\
$I_1\left(C,C,+\right)$ & (\ref{CCP1}) 
& $t^{-ip_2+is_2-\frac{1}{2}}$ && $\left(1-t\right)^{-ip_1-is_1-\frac{1}{2}}$\\
$I_2\left(C,C,+\right)$ & (\ref{CCP2})
& $t^{-ip_2+is_2-\frac{1}{2}}$ && $\left(1-t\right)^{k_{12}-1+ip_1+ip_2}$ \\ 
$I_1\left(C,C,C\right)$ & (\ref{CCC1}) 
& $t^{-ip_2+is_2-\frac{1}{2}}$ && $\left(1-t\right)^{-ip_1-is_1-\frac{1}{2}}$\\
$I_2\left(C,C,C\right)$ & (\ref{CCC2})
& $t^{-ip_2+is_2-\frac{1}{2}}$ && $\left(1-t\right)^{is_{12}-\frac{1}{2}+ip_1+ip_2}$ \\ 
\hline
\end{tabular}}
\end{center}
\caption[Clebsch-Gordon coefficients]{\label{CGC-summ}The asymptotics
of the various integrands defining the Clebsch-Gordon coefficients at
the end points of the integral.  Here $\mathcal{J}$ may be any of $+$,
$-$ or $\mathcal{C}$.  One should note that the symmetries of the
Clebsch-Gordon coefficient under interchanging the first two
representations implies the existence of further integrals; for
instance the coupling
$\mathcal{D}^+\otimes\mathcal{D}^-\to\mathcal{J}$ has a further
integral that looks like $I\left(+,-,\mathcal{J}\right)$ except the
subscripts $1$ and $2$ are swapped.}
\end{table}

There are a number of theorems\cite{Asymp} which are of use in
determining the asymptotics of these integrals and will be reproduced
here for convenience.

\begin{thm}[Laplace Integrals\cite{Asymp}]\label{Laplace}
Let $f\left(t\right)$ be analytic in the region $a < \Arg\, t < b$,
$a<0$ and $b>0$, and at the origin, and satisfy
$\left|f\left(t\right)\right|<Ae^{B\left|t\right|}$, for some
constants $A$ and $B$, in this region.  Then, if $z =
\left|z\right|e^{i\theta}$, $-\frac{\pi}{2} - b <\theta
<\frac{\pi}{2}-a$, one has
$$
\int_0^\infty e^{-zt}f\left(t\right)\,dt
\sim\frac{f\left(0\right)}{z}
$$
as $\left|z\right|\to\infty$.
\end{thm}

\begin{thm}[Stationary Phase\cite{Asymp}]\label{state-phase}
Let $I\left(z\right)$ be given by
\begin{equation}
I\left(z\right) = \int_a^b\,e^{-z\,p\left(t\right)}q\left(t\right)\,dt
\end{equation}
and satisfy
{\renewcommand{\labelenumi}{{\rm(\roman{enumi})}}
\begin{enumerate}
\item $p^\prime\left(t\right)$ and $q\left(t\right)$ independent of $z$,
single valued and holomorphic in a domain $\mathbb{T}$
\item The contour given by the integral from $a$ to $b$ is independent
of $z$ and the interval $\left(a,b\right)$ lies in the domain
$\mathbb{T}$
\item In the neighbourhood of $a$ $p\left(t\right)$ and $q\left(t\right)$
admit the following (convergent) expansions
$$
p\left(t\right) = p\left(a\right) +\sum_{n=0}^\infty
p_n\left(t-a\right)^{n+\mu}
$$
$$
q\left(t\right) = \sum_{n=0}^\infty q_n\left(t-a\right)^{n+\lambda -1}
$$
with $\mu\in\R^+$ and $\Re\lambda > 0$. If either $\mu$ or $\lambda$
is fractional then the branch is chosen by continuity.
\item if z satisfies $\theta_1\le\Arg\,z\le\theta_2$ with
$\theta_2 - \theta_1 <\pi$ and is of sufficiently large absolute value
then the integral is absolutely and uniformly convergent
\item For $t\in\left(a,b\right)$ one has $\Re\left(e^{i\theta}p\left(t\right)
-e^{i\theta}p\left(a\right)\right) > 0$ and moreover $\exists M >0$ such that
$\Re\left(e^{i\theta}p\left(t\right)
-e^{i\theta}p\left(a\right)\right)\ge M$ as $t\to b$ and $M$ does not depend
on $\theta$
\end{enumerate}
}
then one has
\begin{equation}
\int_a^b e^{-zp\left(t\right)}q\left(t\right)\,dt
\sim e^{-zp\left(a\right)}\sum_{n=0}^\infty
\Gamma\left(\frac{n+\lambda}{\mu}\right)\frac{a_n}{z^{\frac{n+\lambda}
{\mu}}}
\end{equation}
as $\left|z\right|\to\infty$ with $\theta_1\le\Arg\,z\le\theta_2$
where the $a_i$ depend only on the $p_i$ and the $q_i$.
\end{thm}

\begin{thm}[Saddle Point method\cite{Asymp}]\label{saddle}
Let $I\left(z\right)$ be as in theorem \ref{state-phase} with
$p\left(t\right)$ and $q\left(t\right)$ satisfying the same conditions
with, in addition, there being a stationary point
$t_0\in\left(a,b\right)$ such that $p^\prime\left(t\right) = 0$.
Then
\begin{equation}
\int^b_a e^{-zp\left(t\right)}q\left(t\right)\,dt
\sim2 e^{-zp\left(t_0\right)}\sum_{n=0}^\infty
\Gamma\left(n+\frac{1}{2}\right)\frac{a_{2n}}{z^{s+\frac{1}{2}}}
\end{equation}
as $\left|z\right|\to\infty$ with $\theta_1\le\Arg\,z\le\theta_2$.
Here the coefficients $a_i$ are functions only of $p$ and $q$ and
their derivatives at $t_0$.
\end{thm}

It is instructive to note from table \ref{CGC-summ} that the integrand
converges to a constant or zero at both endpoints when all the
representations are in the discrete series but when one or more
representations are in the continuous series the upper or lower
integrand exhibits a mild divergence.  This is largely unsurprising in
the light of the orthogonality relation of proposition \ref{CGC-orth}
as will be shown in the next result.

\begin{prop}\label{orth}
$\left[\begin{array}{ccc} 
\mathcal{J}_1 & \mathcal{J}_2 & \mathcal{J}_{12}\\
p_1 & p_{12} -p_1 & p_{12} \end{array}\right]$
behaves asymptotically as
$$
\left[\begin{array}{ccc} 
\mathcal{J}_1 & \mathcal{J}_2 & \mathcal{J}_{12}\\
p_1 & p_{12} -p_1 & p_{12} \end{array}\right]\sim
\begin{cases}
\le\frac{1}{p_1} & \text{if }\mathcal{J}_{12}\text{ is in the discrete series}\\
\frac{1}{\sqrt{p_1}} & \text{if }\mathcal{J}_{12}
\text{ is in the continuous series}
\end{cases}
$$
as $p_1$ gets large with all other arguments held fixed.
\end{prop}

Consider a Clebsch-Gordon coefficient of the form,
$\left[\begin{array}{ccc}
\mathcal{J}_1 & \mathcal{J}_2 & \mathcal{J}_{12}\\
p_1 & p_{12} -p_1 & p_{12} \end{array}\right] $. There are two types of
integral that a Clebsch-Gordon coefficient of this form might
correspond to.  If it contains a hyperbolic type integral then
one is interested in asymptotics of the integral
\begin{equation}\label{Ihyp}
I_{\text{hyp}}\left(p_1\right) = \int^1_0 e^{ip_1\ln t}
t^{j_2 - 1 -ip_{12}}\left(1-t\right)^{j_{12}-1+ip_{12}}\,dt
\end{equation}
while a trigonometric integral gives rise to a integral of the form
\begin{equation}\label{Itrig}
I_{\text{trig}}\left(p_1\right) = \int^1_0 e^{ip_1\ln \frac{t}{t-1}}
t^{j_2 - 1 -ip_{12}}\left(1-t\right)^{j_1-1}\,dt
\end{equation}
If $\mathcal{J}_i$ in the discrete series $j_i = k_i \ge 1$ while
$j_i = \frac{1}{2}\pm is_i$ if it is in the continuous series.

For equation \ref{Ihyp} one may apply theorem \ref{state-phase} to
the integral by reversing the integration so that $a = 1$ and $b = 0$.
The function $p\left(t\right) = \ln\left(t\right)$ has a leading term
$-\left(1-t\right)$ as $t\to 1$ by a simple Taylor expansion and so
$\mu =1$. The function $q\left(t\right) =
t^{\alpha}\left(1-t\right)^{\beta} { }_2F_1
\left(\begin{array}{cc} a & b\\ & c\end{array} ; t\right)$ near
$t=1$ clearly has leading term $\left(1-t\right)^{j_{12}-1}$ and so
$\lambda = j_{12}$.

Thus
\begin{equation}\label{hyp-asym}
I_{\text{hyp}}\left(z\right)\sim
\Gamma\left(j_{12}\right)\frac{a_0}{z^{j_{12}}}
\end{equation}

For equation \ref{Itrig} one has, after a substitution of $v = 
\ln\left(\frac{t}{1-t}\right)$ and splitting the integral into
two parts that look like Laplace integrals,
$$
I_1\left(z\right) = \int_{0}^\infty\,
dv\,\exp\left\{izv\right\} f\left(v\right)  
+ \int_{0}^\infty\,
dv\,\exp\left\{-izv\right\} f\left(-v\right)  
$$
with 
$$
f\left(v\right) = e^{v\left(\alpha+1\right)}
\left(1+e^v\right)^{2-\alpha-\beta} 
{ }_2F_1
\left(\begin{array}{cc} a & b\\ & c\end{array} ; 
\frac{e^v}{1+e^v}\right)
$$
The hypergeometric function is analytic everywhere for the range of
integration and the only other possible singularity is in the factor
$\left(1+e^v\right)^{2-\alpha-\beta}$, if $\alpha+\beta>2$, at the
points $v = i\pi\left(2n+1\right)$ for $n\in\Z$. Thus
$f\left(v\right)$ is analytic for
$-\frac{\pi}{2}<\Arg\,v<\frac{\pi}{2}$. Applying theorem \ref{Laplace}
gives, since the coefficient of the variable of integration in the
exponential satisfies $\Arg\,z = \pm\frac{\pi}{2}$,
\begin{equation}
I_{\text{trig}}\left(z\right)\sim
\frac{\left(2\right)^{3-\alpha-\beta}} {z}{ }_2F_1
\left(\begin{array}{cc} a & b\\ & c\end{array} ; 
\frac{1}{2}\right)
\end{equation}
as $\left|z\right|\to\infty$.

Thus when $j_{12}$ is in the discrete series both integrals, and hence
the Clebsch-Gordon coefficient, behave at worst like $\frac{1}{p_1}$
as $p_1$ gets large (in practice this bound is achieved everywhere
except for $\mathcal{D}^+\otimes\mathcal{D}^-\to\mathcal{D}^\pm$).  If
$j_{12}$ is in the continuous series $I_{\text{hyp}}$ behaves like
$\frac{1}{\sqrt{p_1}}$ and since the hyperbolic integrals always
appear in the definition of any such Clebsch-Gordon coefficient with
$j_{12}$ in the continuous series this completes the proof of
proposition
\ref{orth}.

The Clebsch-Gordon coefficient is thus only square integrable when
$\mathcal{J}_{12}$ is in the discrete series; which is precisely the
meaning of the orthogonality relation in proposition \ref{CGC-orth}.

We are now in a position to investigate the convergence of the
integral in equation \ref{2p}
\begin{multline}
\int\,dp_{12}\,dp_{23}\,
\left[\begin{array}{ccc} 
\mathcal{J}_1 & \mathcal{J}_2 & \mathcal{J}_{12}\\ 
p - p_{23} & p_{12} + p_{23} - p & p_{12} \end{array}\right]
\left[\begin{array}{ccc} 
\mathcal{J}_{12} & \mathcal{J}_3 & \mathcal{J}\\
p_{12} & p - p_{12} & p\end{array}\right]\\
\left[\begin{array}{ccc} 
\mathcal{J}_2 & \mathcal{J}_3 & \mathcal{J}_{23}\\
p_{12} + p_{23} - p & p - p_{12} & p_{23}\end{array}\right]
\left[\begin{array}{ccc} 
\mathcal{J}_1 & \mathcal{J}_{23} & \mathcal{J}^\prime\\
p^\prime - p_{23} & p_{23} & p^\prime\end{array}\right]\\
= \delta\left(\mathcal{J},\,\mathcal{J}^\prime\right)
\left\{\begin{array}{ccc}
\mathcal{J}_1 & \mathcal{J}_2 & \mathcal{J}_{12}\\
\mathcal{J}_3 & \mathcal{J} & \mathcal{J}_{23}
\end{array}\right\}
\end{multline}

First change to polar co-ordinates with
\begin{align}
p_{12} & = r\cos\theta & p_{23} &= r\sin\theta
\end{align}
so that the determining integral in equation \ref{2p} may be written
as
\begin{multline}\label{2p-polar}
\int_0^{2\pi}\int_0^\infty
\left[\begin{array}{ccc} 
\mathcal{J}_1 & \mathcal{J}_2 & \mathcal{J}_{12}\\ 
p - r\sin\theta &  r\left(\cos\theta+\sin\theta\right)-p 
&  r\cos\theta \end{array}\right]
\left[\begin{array}{ccc} 
\mathcal{J}_{12} & \mathcal{J}_3 & \mathcal{J}\\
r\cos\theta & p -  r\cos\theta & p\end{array}\right]\\
\left[\begin{array}{ccc} 
\mathcal{J}_2 & \mathcal{J}_3 & \mathcal{J}_{23}\\
r\left(\cos\theta+\sin\theta\right) - p & p -  r\cos\theta 
& r\sin\theta\end{array}\right]
\left[\begin{array}{ccc} 
\mathcal{J}_1 & \mathcal{J}_{23} & \mathcal{J}^\prime\\
p^\prime - r\sin\theta & r\sin\theta & p^\prime\end{array}\right]
\,r\,dr\,d\theta\\
= \delta\left(\mathcal{J},\,\mathcal{J}^\prime\right)
\left\{\begin{array}{ccc}
\mathcal{J}_1 & \mathcal{J}_2 & \mathcal{J}_{12}\\
\mathcal{J}_3 & \mathcal{J} & \mathcal{J}_{23}
\end{array}\right\}
\end{multline}

\begin{prop}\label{other-CGC}
For fixed $\theta$ the Clebsch-Gordon coefficients 
\begin{multline*}
\left[\begin{array}{ccc} 
\mathcal{J}_1 & \mathcal{J}_2 & \mathcal{J}_{12}\\ 
p - r\sin\theta &  r\left(\cos\theta+\sin\theta\right)-p 
&  r\cos\theta \end{array}\right]\text{ and } \\
\left[\begin{array}{ccc} 
\mathcal{J}_2 & \mathcal{J}_3 & \mathcal{J}_{23}\\
r\left(\cos\theta+\sin\theta\right) - p & p -  r\cos\theta 
& r\sin\theta\end{array}\right]
\end{multline*}
tend to $0$ as $r\to\infty$ no slower than $\frac{1}{\sqrt{r}}$.
\end{prop}

The Clebsch-Gordon coefficient $\left[\begin{array}{ccc}
\mathcal{J}_1 & \mathcal{J}_2 & \mathcal{J}_{12}\\ 
p - r\sin\theta & r\left(\cos\theta+\sin\theta\right)-p & r\cos\theta
\end{array}\right]$ is constructed from linear combinations of
integrals of the form
\begin{multline}\label{comp1}
I_{\mathrm{trig}}\left(\mathcal{J}_1,\mathcal{J}_2\right) = 
\int_0^1 \exp\left\{ir\left(\sin\theta\ln\left(1-t\right) -  
\left(\cos\theta+\sin\theta\right)\ln t\right)\right\}
t^{j_2-1+ip }\\
\left(1-t\right)^{j_1-1-ip}
\Hyper{a}{b}{c}{t}\,dt
\end{multline}
and
\begin{multline}\label{comp2}
I_{\mathrm{hyp}}\left(\mathcal{J}_1,\mathcal{J}_2\right) =
\int_0^1 \exp\left\{ir\left(\cos\theta\ln\left(1-t\right) -  
\left(\cos\theta+\sin\theta\right)\ln t\right)\right\}
t^{j_2-1+ip }\\
\left(1-t\right)^{j_{12}-1}\Hyper{a}{b}{c}{t}\,dt
\end{multline}
Since the Clebsch-Gordon coefficient is symmetric (up to a phase)
under interchanging the first two columns one has a further two
integrals $I_{\mathrm{trig}}\left(\mathcal{J}_2,\mathcal{J}_1\right)$
and $I_{\mathrm{hyp}}\left(\mathcal{J}_2,\mathcal{J}_1\right)$ whose
asymptotics must also contribute to the Clebsch-Gordon coefficient.

The Clebsch-Gordon coefficient $\left[\begin{array}{ccc}
\mathcal{J}_2 & \mathcal{J}_3 & \mathcal{J}_{23}\\
r\left(\cos\theta+\sin\theta\right) - p & p - r\cos\theta &
r\sin\theta\end{array}\right]$ is constructed from linear combinations
of integrals of the form
\begin{multline}\label{comp3}
I_{\mathrm{trig}}\left(\mathcal{J}_2,\mathcal{J}_3\right) = 
\int_0^1 \exp\left\{ir\left(
\left(\sin\theta+\cos\theta\right)\ln\left(1-t\right) -  
\cos\theta\ln t\right)\right\}
t^{j_1-1+ip}\\
\left(1-t\right)^{j_2-1-ip}
\Hyper{a}{b}{c}{t}\,dt
\end{multline}
and
\begin{multline}\label{comp4}
I_{\mathrm{hyp}}\left(\mathcal{J}_2,\mathcal{J}_3\right) = 
\int_0^1 \exp\left\{ir\left(\cos\theta\ln\left(1-t\right) -  
\left(\cos\theta+\sin\theta\right)\ln t\right)\right\}
t^{j_3-1+ip }\\
\left(1-t\right)^{j_{23}-1}\Hyper{a}{b}{c}{t}\,dt
\end{multline}

For the first, equation \ref{comp1}, the function 
$$p\left(t\right) =
\sin\theta\ln\left(1-t\right) -
\left(\cos\theta+\sin\theta\right)\ln t
$$ 
has a stationary point at $t = \tan\theta + 1$. When this falls within
the range of integration theorem \ref{saddle} may be applied to
estimate the contribution. There are critical points when $\sin\theta$
vanishes, in which case the integral degenerates to the same form as
$I_{\text{hyp}}$ in equation \ref{Ihyp}, and when $\cos\theta =
-\sin\theta$, which leads to an integral that may be estimated in a
similar way to $I_{\text{hyp}}$, that mark the boundary of the region
where a stationary point exists.  The behaviour of this depends on
whether the particular representations that govern the behaviour at the
end points (in this case $\mathcal{J}_2$ and $\mathcal{J}_{12}$) are in
the discrete or continuous series.

While $\theta$ ranges over the rest of its possible values (where
$\tan\theta + 1\not\in\left[0,1\right]$) the function
$\frac{1}{p^\prime\left(t\right)}$ has no singularities in the range
of integration and thus a substitution is possible with $u =
p\left(t\right)$, $u$ ranging from $+\infty$ to $-\infty$ with the
function $f$ in the statement of theorem \ref{Laplace} being taken as
$$
f\left(u\right) =
\frac{q\left(t\right)}{p^\prime\left(t\right)}$$ 
with 
$$
q\left(t\right) = t^{j_2-1+ip }
\left(1-t\right)^{j_1-1-ip}
\Hyper{a}{b}{c}{t}
$$
As a function of $t$ this is clearly analytic in the annulus $0 <
\left|t\right| < 1$ and it remains to see where it is analytic as a
function of $u$ so that one may apply theorem \ref{Laplace}.

The transformation $u = p\left(t\right)$ may be explicitly written as
a transformation of the complex plane with
\begin{equation}\label{u-equation}
u = \sin\theta\ln\left(1-t\right) -
\left(\cos\theta+\sin\theta\right)\ln t
\end{equation}
This transformation must be inverted and so that $t$ may be written as
a function of $u$ and then the singularities of the function
$f\left(u\right)$ determined.  In general it is not possible to write
this in a closed form, however, one may write is as
\begin{equation}\label{t-equation}
t = 1 - e^{\nu}
\end{equation}
where $\nu = \nu\left(u\right)$ is a solution of the equation
\begin{equation}\label{nu-equation}
\exp\left\{\nu\right\} - 1 + 
\exp\left\{\frac{\sin\left(\theta\right)\nu - u}
{\sin\left(\theta\right)+\cos\left(\theta\right)}\right\} = 0 
\end{equation}
Now $f\left(t\right)$ is singular at the points $t=0$ and $t=1$.
These correspond, respectively, to the regions of the complex plane
$\nu = 2n\pi i$ for $n\in\Z$ and $\Re\nu\to-\infty$,  $\Im\nu$ arbitrary,
by inspection of equation \ref{t-equation}.

The first of these, $\nu = 2n\pi i$, translates to
$\mathrm{sign}\left(\sin\theta +\cos\theta\right)\Re u\to\infty$.  In
other words the origin is mapped to a line at infinity as one would
expect from equation \ref{u-equation}.  The second clearly
requires $-\mathrm{sign}\left(\sin\theta\right) u\to\infty$ to satisfy
equation \ref{nu-equation} and the point $t=1$ gets mapped to a line
at infinity as well.

The other possible problem with $f\left(u\right)$ is that equation
\ref{nu-equation} has no solutions.  This can only happen when
$\cos\theta = 0$ and here one can invert equation \ref{u-equation}
directly to find
$$
t = f\left(u\right) = \frac{1}{1+e^{u}}
$$
and there are singularities along the imaginary axis at $u =
\left(2n+1\right)\pi i$.  In this case theorem \ref{Laplace} may be applied 
to gain an asymptotic contribution of $\frac{1}{r}$.

The other equations are similar and we shall omit the details in
favour of specifying the range of $\theta$ where they behave like
$\frac{1}{\sqrt{r}}$.  One finds

\begin{center}
\begin{tabular}{rcl}
$I_{\mathrm{trig}}\left(\mathcal{J}_1\mathcal{J}_2\right)
\sim\frac{1}{\sqrt{r}}$ & for  & $\theta\in\left(-\frac{\pi}{4},0\right)$\\
$I_{\mathrm{trig}}\left(\mathcal{J}_2\mathcal{J}_1\right)
\sim\frac{1}{\sqrt{r}}$ & for  & $\theta\in\left(0,\frac{\pi}{4}\right)$\\
$I_{\mathrm{hyp}}\left(\mathcal{J}_1\mathcal{J}_2\right)
\sim\frac{1}{\sqrt{r}}$ & for  & 
$\theta\in\left(-\frac{\pi}{2},-\frac{\pi}{4}\right)$\\
$I_{\mathrm{hyp}}\left(\mathcal{J}_2\mathcal{J}_1\right)
\sim\frac{1}{\sqrt{r}}$ & for  & $\theta\in\left(0,\frac{\pi}{2}\right)$\\
$I_{\mathrm{trig}}\left(\mathcal{J}_2\mathcal{J}_3\right)
\sim\frac{1}{\sqrt{r}}$ & for  & 
$\theta\in\left(-\frac{\pi}{2},-\frac{\pi}{4}\right)$\\
$I_{\mathrm{trig}}\left(\mathcal{J}_3\mathcal{J}_2\right)
\sim\frac{1}{\sqrt{r}}$ & for  & 
$\theta\in\left(-\frac{\pi}{2},-\frac{\pi}{4}\right)$\\
$I_{\mathrm{hyp}}\left(\mathcal{J}_2\mathcal{J}_3\right)
\sim\frac{1}{\sqrt{r}}$ & for  & 
$\theta\in\left(0,\frac{\pi}{2}\right)$\\
$I_{\mathrm{hyp}}\left(\mathcal{J}_3\mathcal{J}_2\right)
\sim\frac{1}{\sqrt{r}}$ & for  & 
$\theta\in\left(-\frac{\pi}{4},0\right)$
\end{tabular}
\end{center}

\bigskip

\begin{prop}\label{2-CGCs}
As $r\to\infty$ with every other argument held fixed there is a range
of $\theta$ where
\begin{multline}
\left|\left[\begin{array}{ccc} 
\mathcal{J}_1 & \mathcal{J}_2 & \mathcal{J}_{12}\\ 
p - r\sin\theta & r\left(\cos\theta+\sin\theta\right)-p & r\cos\theta
\end{array}\right]\right.\\
\left.\left[\begin{array}{ccc} 
\mathcal{J}_2 & \mathcal{J}_3 & \mathcal{J}_{23}\\
r\left(\cos\theta+\sin\theta\right) - p & p - r\cos\theta &
r\sin\theta\end{array}\right]\right|\sim\frac{1}{r}
\end{multline}
providing the representations are not all in the positive discrete
series or all in the negative discrete series.
\end{prop}
The proposition is true providing one can find a range of $\theta$
such that the possible products of $I_{\mathrm{trig}}$ and
$I_{\mathrm{hyp}}$ have an overlapping range of $\theta$ where they
behave as $\frac{1}{\sqrt{r}}$.  A glance at the above table shows
this is always possible providing the product is not of the form
$I_{\mathrm{trig}}I_{\mathrm{trig}}$.

However the only case of coupling consisting solely of an integral of
type $I_{\mathrm{trig}}$ is where $j_1$, $j_2$, $j_3$, $j_{12}$ and
$j_{23}$ are all in the positive discrete series or all in the
negative discrete series.  This is sufficient to imply $j$ is in the
same discrete series too.  In this case the product tends to zero no
slower than $r^{-\frac{3}{2}}$ because the regions where each
Clebsch-Gordon coefficient tends to zero as $\frac{1}{\sqrt{r}}$ do
not overlap.

\bigskip

Having applied proposition \ref{2-CGCs} to equation \ref{2p-polar} one
is interested in the convergence of an integral of the form
\begin{multline*}
\int_a^{b}\int_0^\infty
\left[\begin{array}{ccc} 
\mathcal{J}_{12} & \mathcal{J}_3 & \mathcal{J}\\
r\cos\theta & p -  r\cos\theta & p\end{array}\right]
\left[\begin{array}{ccc} 
\mathcal{J}_1 & \mathcal{J}_{23} & \mathcal{J}^\prime\\
p^\prime - r\sin\theta & r\sin\theta & p^\prime
\end{array}\right]\,dr\,d\theta
\end{multline*}
(for the exception to proposition \ref{2-CGCs} when everything is
solely in either the positive or the negative discrete series one must
multiply the integrand by an additional factor of
$\frac{1}{\sqrt{r}}$.)

Here the range $\left(a,b\right)$ for $\theta$ is the same as that
found in proposition \ref{2-CGCs} since for the rest of the integral
in equation \ref{2p-polar} the integrand converges faster to 0 as
$r\to\infty$.

The convergence of this integral (and the original integral in
equation \ref{2p}) thus rests on the behaviour of the two
Clebsch-Gordon coefficients that determine the final states
$\mathcal{J}$ and $\mathcal{J}^\prime$.

From the asymptotics of proposition \ref{orth} one sees that the
integral converges absolutely as $\frac{1}{r^2}$ when they are in the
discrete series, unless everything is in the same discrete series in
which case it converges absolutely as $r^{-\frac{5}{2}}$. 

However when in the continuous series they contribute a factor
$\frac{r^{is-is^\prime}}{r}$ which leads to the $\delta$ function on
the right hand side of equation \ref{2p} and the integral only
converges to a distribution rather than a function, as expected.  The
distributional nature in this latter case is rather mild, being a
genuine function (the Racah coefficient itself) producted with a
$\delta$ function. This concludes the proof of theorem
\ref{Racah-converg}.

\section{The Clebsch-Gordon Coeffient in a discrete basis}\label{rac-CGC}

Before calculating the Racah coefficient for the discrete series it is
convenient to calculate the Clebsch-Gordon coefficient in a discrete
basis.  It is a lot simpler to calculate the Racah coefficient in this
case of coupling using these discrete basis Clebsch-Gordon
coefficients than the continuous basis ones met previously and since
the Racah coefficient is independent of basis this freedom is open to
us.

Some symmetries will be derived for these Clebsch-Gordon coefficients
as well.  All sums without a specified range are in integer steps over
the range for which all the factorials in the summand are defined.

From section \ref{rep}, \su{1,1} is characterised by the following
action of its generators on the Hilbert spaces $\mathcal{H}_j$ with
basis $\{\ket{j,m} |\; j,m\in\frac{1}{2}\N \}$
\begin{align}
K_3\ket{j,m} & = m\ket{j,m}\nonumber\\
K_\pm\ket{j,m} & = \sqrt{(m\pm j)(m\mp j\pm 1)}\ket{j,m\pm 1}\label{action}
\end{align}

The Clebsch-Gordon coefficients for two representation series may be
defined as follows
\begin{equation}\label{CGC-def-O2}
\ket{j\left(j_1 j_2\right),m} = \sum_{m_1,m_2} \left[
\begin{array}{ccc} j_1 & j_2 & j\\ m_1 & m_2 & m \end{array} \right]
\ket{j_1,m_1}\otimes\ket{j_2,m_2}
\end{equation}
with $m = m_1 + m_2$, and $j_1$, $j_2$, $j$ taking values appropriate
for the series under discussion.  

The normalisation is taken to agree with that used previously in
equation \ref{CGC-orth}, thus
\begin{equation}\label{CGC-orth-O2}
\sum_{m_1,\,m_2}
\left[\begin{array}{ccc} j_1 & j_2 & j\\ 
m_1 & m_2 & m \end{array} \right]
\left[\begin{array}{ccc} j_1 & j_2 & j^\prime\\
m_1 & m_2 & m^\prime\end{array} \right] =
\delta\left(\mathcal{J},\mathcal{J^\prime}\right)\delta_{m\,m^\prime}
\end{equation}

An explicit formula for the Clebsch-Gordon coefficient may be derived
by adapting Racah's approach to the \su{2} Clebsch-Gordon
problem\cite{RacahII}.

Consider the operators $K_\pm$ acting on equation \ref{CGC-def-O2} as
in equation \ref{action}.  Acting with $K_+$ yields
\begin{multline}\label{K+recurse}
\sqrt{\left(m+j\right)\left(m-j+1\right)}
\left[\begin{array}{ccc} j_1 & j_2 & j\\ m_1 & m_2 & m + 1\end{array}
\right] = \\
\sqrt{\left(m_1-j_1\right)\left(m_1+j_1-1\right)}
\left[\begin{array}{ccc} j_1 & j_2 & j\\ m_1 -1 & m_2 & m \end{array} \right]\\
+ \sqrt{\left(m_2-j_2\right)\left(m_2+j_2-1\right)}
\left[\begin{array}{ccc} j_1 & j_2 & j\\ m_1 & m_2 -1 & m \end{array} \right]
\end{multline}
while acting with $K_-$ gives
\begin{multline}\label{K-recurse}
\sqrt{\left(m-j\right)\left(m+j-1\right)}
\left[\begin{array}{ccc} j_1 & j_2 & j\\ m_1 & m_2 & m - 1\end{array}
\right] = \\ 
\sqrt{\left(m_1+j_1\right)\left(m_1-j_1+1\right)}
\left[\begin{array}{ccc} j_1 & j_2 & j\\ m_1 + 1 & m_2 & m \end{array}
\right]\\
+ \sqrt{\left(m_2+j_2\right)\left(m_2-j_2+1\right)}
\left[\begin{array}{ccc} j_1 & j_2 & j\\ m_1 & m_2 + 1 & m \end{array} \right]
\end{multline}

Equations \ref{K+recurse} and \ref{K-recurse} fix the Clebsch-Gordon
coefficient up to a function of the $j$'s and then equation
\ref{CGC-orth-O2} fixes it up to a phase that can depend only on the $j$'s.

To investigate some symmetries, define
\begin{multline}\label{g}
\left[\begin{array}{ccc} j_1 & j_2 & j\\ m_1 & m_2 & m \end{array}\right] = 
\sqrt{\frac{\left(m+j-1\right)!\,\left(m_1+j_1-1\right)!\,
\left(m_2+j_2-1\right)!}
{\left(m-j\right)!\,\left(m_1-j_1\right)!\,\left(m_2-j_2\right)!}}\\
g\left(\begin{array}{ccc} j_1 & j_2 & j\\ m_1 & m_2 & m \end{array}\right)
\end{multline}
where the factorials are understood in terms of gamma functions when
$j$ is in the continous series.  When $j$ is in the discrete series
one has $\left|m\right| \ge j$ and the sign of $m$ determines whether
it is in the postive or negative discrete series.

Subsitution of equation \ref{g} into the recursion relations in
equations \ref{K+recurse} and \ref{K-recurse} yields, respectively
\begin{multline}\label{K+recurse1}
\left(m+j\right)g\left(\begin{array}{ccc} j_1 & j_2 & j\\ 
m_1 & m_2 & m + 1\end{array}\right) = \\
\left(m_1-j_1\right) g\left(\begin{array}{ccc} j_1 & j_2 & j\\ 
m_1 -1 & m_2 & m \end{array} \right)
+ \left(m_2-j_2\right) g\left(\begin{array}{ccc} j_1 & j_2 & j\\ 
m_1 & m_2 -1 & m \end{array} \right)
\end{multline}
and
\begin{multline}\label{K-recurse1}
\left(m-j\right)
g\left(\begin{array}{ccc} j_1 & j_2 & j\\ 
m_1 & m_2 & m - 1\end{array}\right) = \\ 
\left(m_1+j_1\right)g\left(\begin{array}{ccc} j_1 & j_2 & j\\ 
m_1 + 1 & m_2 & m \end{array}\right)
+ \left(m_2+j_2\right)g\left(\begin{array}{ccc} j_1 & j_2 & j\\ 
m_1 & m_2 + 1 & m \end{array} \right)
\end{multline}

Now passively transform this equation by writing $j$ for $j_1$, $j_1$
for $j_2$, $j_2$ for $j$, $-m$ for $m_1$, $m_1$ for $m_2$ and $-m_2$
for $m$.  Thus equation \ref{K+recurse1} and \ref{K-recurse1} become,
respectively

\begin{multline}\label{K+recurse2}
\left(-m_2+j_2\right)g\left(\begin{array}{ccc} j & j_1 & j_2\\ 
-m & m_1 & -m_2 + 1\end{array}\right) = 
\left(-m-j\right) g\left(\begin{array}{ccc} j & j_1 & j_2\\ 
-m -1 & m_1 & -m_2 \end{array} \right)\\
+ \left(m_1-j_1\right) g\left(\begin{array}{ccc} j & j_1 & j_2\\ 
-m & m_1 -1 & -m_2 \end{array} \right)
\end{multline}
and
\begin{multline}\label{K-recurse2}
\left(-m_2-j\right)
g\left(\begin{array}{ccc} j & j_1 & j_2\\ 
-m & m_1 & -m_2 - 1\end{array}\right) =  
\left(-m+j\right)g\left(\begin{array}{ccc} j & j_1 & j_2\\ 
-m + 1 & m_1 & -m_2 \end{array}\right)\\
+ \left(m_1+j_1\right)g\left(\begin{array}{ccc} j & j_1 & j_2\\ 
-m & m_1 + 1 & -m_2 \end{array} \right)
\end{multline}
It is clear, however, these are simply equations \ref{K+recurse1} and
\ref{K-recurse1} with the function $g$ transformed via
$$
g\left(\begin{array}{ccc} j_1 & j_2 & j\\ 
m_1 & m_2 & m \end{array}\right) \mapsto
g\left(\begin{array}{ccc} j & j_1 & j_2\\ 
-m & m_1 & -m_2 \end{array}\right)
$$
The most general relation between the two functions is thus
\begin{equation}\label{g-sym-ptr}
g\left(\begin{array}{ccc} j_1 & j_2 & j\\ 
m_1 & m_2 & m \end{array}\right) = T\left(j_1,j_2,j\right)
g\left(\begin{array}{ccc} j & j_1 & j_2\\ 
-m & m_1 & -m_2 \end{array}\right)
\end{equation}
with $T$ some function.  However since $g$ is only defined up to a
function of the $j$'s which is then fixed up to phase by orthogonality
we may as well write
\begin{equation}\label{g-sym}
\left|g\left(\begin{array}{ccc} j_1 & j_2 & j\\ 
m_1 & m_2 & m \end{array}\right)\right| =
\left|g\left(\begin{array}{ccc} j & j_1 & j_2\\ 
-m & m_1 & -m_2 \end{array}\right)\right|
\end{equation}

Now consider the Clebsch-Gordon coefficient proper, and in particular
the quantity
\begin{equation}
\left|\frac{\left[\begin{array}{ccc} j_1 & j_2 & j\\ 
m_1 & m_2 & m \end{array}\right]}
{\left[\begin{array}{ccc} j & j_1 & j_2\\ 
-m & m_1 & -m_2 \end{array}\right]}\right|
\end{equation}

From equations \ref{g} and \ref{g-sym} this is simply
\begin{multline}
\left|\sqrt{\frac{\left(m+j-1\right)!\,\left(m_1+j_1-1\right)!\,
\left(m_2+j_2-1\right)!
\left(-m-j\right)!\,\left(m_1-j_1\right)!\,\left(-m_2-j_2\right)}
{\left(m-j\right)!\,\left(m_1-j_1\right)!\,\left(m_2-j_2\right)!
\left(-m+j-1\right)!\,\left(m_1+j_1-1\right)!\,\left(-m_2+j_2-1\right)!
}}\right|
\\= \left|\sqrt{\frac{\left(m+j-1\right)!\,\left(m_2+j_2-1\right)!
\left(-m-j\right)!\,\left(-m_2-j_2\right)}
{\left(m-j\right)!\,\left(m_2-j_2\right)!
\left(-m+j-1\right)!\,\left(-m_2+j_2-1\right)!
}}\right|
\end{multline}

Noting that
\begin{multline}
\frac{(a+n)!}{a!} =
\left(a+n\right)\left(a+n-1\right)\cdots\left(a+1\right)\\
= \left(-1\right)^n
\left(-a-n\right)\left(-a-n+1\right)\cdots\left(-a-1\right)
= \frac{\left(-a-1\right)!}{\left(-a-n-1\right)!}
\end{multline}
one finally derives
\begin{equation}\label{rotate}
\left|\left[\begin{array}{ccc} j_1 & j_2 & j\\ 
m_1 & m_2 & m \end{array}\right]\right| =\left| 
\left[\begin{array}{ccc} j & j_1 & j_2\\ 
-m & m_1 & -m_2 \end{array}\right]\right|
\end{equation}

It is simple enough to also derive the relation
\begin{equation}\label{swap}
\left|\left[\begin{array}{ccc} j_1 & j_2 & j\\ 
m_1 & m_2 & m \end{array}\right]\right| = 
\left|\left[\begin{array}{ccc} j_2 & j_1 & j\\ 
m_2 & m_1 & m \end{array}\right]\right|
\end{equation}
in a similar way from equations \ref{K+recurse1} and \ref{K-recurse1}.

While equations \ref{K+recurse1} and \ref{K-recurse1} are convenient
for exhibiting symmetries, they are less so for undertaking a
calculation of the Clebsch-Gordon coefficient for the discrete series.
There are six cases of coupling in the discrete series:
$$
\begin{array}{rl}
\mathcal{J}^+\otimes\mathcal{J}^+ & \longrightarrow\mathcal{J}^+\\
\mathcal{J}^+\otimes\mathcal{J}^- & \longrightarrow\mathcal{J}^+\\
\mathcal{J}^-\otimes\mathcal{J}^+ & \longrightarrow\mathcal{J}^+\\
\mathcal{J}^+\otimes\mathcal{J}^- & \longrightarrow\mathcal{J}^-\\
\mathcal{J}^-\otimes\mathcal{J}^+ & \longrightarrow\mathcal{J}^-\\
\mathcal{J}^-\otimes\mathcal{J}^- & \longrightarrow\mathcal{J}^-
\end{array}
$$

Thus, let all three representations be in the positive discrete series
and define $f$ so that
\begin{multline}\label{Ffunc}
\left[\begin{array}{ccc} j_1 & j_2 & j\\ 
m_1 & m_2 & m \end{array}\right] = 
\left(-1\right)^{j_1-\left|m_1\right|}\frac{1}
{\sqrt{\left(\left|m\right|+j-1\right)!\,
\left(\left|m\right|-j\right)!\,\left(\left|m_1\right|+j_1-1\right)!}}\\
\times\frac{1}
{\sqrt{\left(\left|m_1\right|-j_1\right)!\,
\left(\left|m_2\right|+j_2-1\right)!\,\left(\left|m_2\right|-j_2\right)!}}
\;f\left(\begin{array}{ccc} j_1 & j_2 & j\\ 
m_1 & m_2 & m \end{array}\right)
\end{multline}
Substitution into \ref{K+recurse} and \ref{K-recurse} yields,
respectively,
\begin{multline}\label{K+recurse3}
f\left(\begin{array}{ccc} j_1 & j_2 & j\\ 
m_1 & m_2 & m + 1\end{array}\right) = 
-\left(m_1-j_1\right)\left(m_1+j_1-1\right) 
f\left(\begin{array}{ccc} j_1 & j_2 & j\\ 
m_1 -1 & m_2 & m \end{array} \right)\\
+ \left(m_2-j_2\right)\left(m_2+j_2-1\right) 
f\left(\begin{array}{ccc} j_1 & j_2 & j\\ 
m_1 & m_2 -1 & m \end{array} \right)
\end{multline}
and
\begin{multline}\label{K-recurse3}
\left(m-j\right)\left(m+j-1\right)
f\left(\begin{array}{ccc} j_1 & j_2 & j\\ 
m_1 & m_2 & m - 1\end{array}\right) = \\ 
-f\left(\begin{array}{ccc} j_1 & j_2 & j\\ 
m_1 + 1 & m_2 & m \end{array}\right)
+ f\left(\begin{array}{ccc} j_1 & j_2 & j\\ 
m_1 & m_2 + 1 & m \end{array} \right)
\end{multline}

Now choose $m = j$ and equation \ref{K-recurse3} gives
$f\left(\begin{array}{ccc} j_1 & j_2 & j\\ 
m_1 & m_2 & j \end{array}\right)$ does not depend on $m_1$ or $m_2$.
Now use equation \ref{K+recurse3} $t$ times on this and one finds
\begin{multline}
f\left(\begin{array}{ccc} j_1 & j_2 & j\\ 
m_1 & m_2 & j + t\end{array}\right) = f\left(\begin{array}{ccc} j_1 & j_2 & j\\ 
m_1 & m_2 & j \end{array}\right)\\
\sum_z\frac{\left(-1\right)^z\,u!\,\left(m_1+j_1-1\right)!\,\left(m_1-j_1\right)!}
{z!\,\left(u-z\right)!\,\left(m_1+j_1-1-z\right)!\,\left(m_1-j_1-z\right)!}\\
\times\frac{\left(m_2+j_2-1\right)!\,\left(m_2-j_2\right)!}
{\left(m_2+j_2-1+z-u\right)!\,\left(m_2-j_2+z-u\right)!}
\end{multline}
and hence that
\begin{multline}
\left[\begin{array}{ccc} j_1 & j_2 & j\\ 
m_1 & m_2 & m\end{array}\right]_{+++} = \left(-1\right)^{m_1-j_1}
f\left(\begin{array}{ccc} j_1 & j_2 & j\\ 
m_1 & m_2 & j \end{array}\right)\\
\sqrt{\frac{\left(m_1+j_1-1\right)!\,\left(m_1-j_1\right)!\,
\left(m_2+j_2-1\right)!\,\left(m_2-j_2\right)!\,\left(m-j\right)!}
{\left(j+m-1\right)!}}\\
\sum_z\frac{\left(-1\right)^z}
{z!\,\left(m-j-z\right)!\,\left(m_1+j_1-1-z\right)!\,\left(m_1-j_1-z\right)!}\\
\times\frac{1}{\left(j+j_2-m_1-1+z\right)!\,\left(j-j_2-m_1+z\right)!}
\end{multline}
where the subscript $+++$ is used to denote the representations are in
the positive discrete series.

Now consider the orthogonality relation of equation \ref{CGC-orth-O2},
taking $m = j$ so that sums over $z$ are immediate, one has
\begin{multline}
\sum_{m_1}\left[\begin{array}{ccc} j_1 & j_2 & j\\ 
m_1 & j - m_1 & j\end{array}\right]_{+++}
\left[\begin{array}{ccc} j_1 & j_2 & j\\ 
m_1 & j - m_1 & j\end{array}\right]_{+++}
=f\left(\begin{array}{ccc} j_1 & j_2 & j\\ 
m_1 & m_2 & j \end{array}\right)^2\\ \times
\sum_{m_1}\frac{1}{\left(2j-1\right)!\,\left(m_1-j_1\right)!\,
\left(m_1+j_1-1\right)!\,\left(j_2+j-m_1-1\right)!\,\left(j-j_2-m_1\right)!}
\end{multline}

The sum can be evaluated by means of
\begin{lem} \label{lem2}
$$
\sum_n \frac{1}{(x-n)!\,(y+n-1)!\,(z-n)!\,n!} = 
\frac{(x+y+z-1)!}{x!\,z!\,(x+y-1)!\,(y+z-1)!}
$$
\end{lem}
which follows simply from the addition theorem for binomial
coefficients by expanding both sides of $(a+b)^n (a+b)^m =
(a+b)^{n+m}$ and equating powers of $a$ and $b$.

One finds
\begin{multline}
\frac{1}{2j-1} = \sum_{m_1}\left[\begin{array}{ccc} j_1 & j_2 & j\\ 
m_1 & j - m_1 & j\end{array}\right]_{+++}
\left[\begin{array}{ccc} j_1 & j_2 & j\\ 
m_1 & j - m_1 & j\end{array}\right]_{+++}\\
= \frac{f\left(\begin{array}{ccc} j_1 & j_2 & j\\ 
m_1 & m_2 & j \end{array}\right)^2}
{\left(2j-1\right)\,\left(j+j_1+j_2-2\right)!\,
\left(j+j_1-j_2-1\right)!\,
\left(j-j_1+j_2-1\right)!\,
\left(j-j_1-j_2\right)!}
\end{multline}
which determines the value of $f\left(\begin{array}{ccc} j_1 & j_2 &
j\\ m_1 & m_2 & j \end{array}\right)$, so that
\begin{multline}\label{O+++}
\left[\begin{array}{ccc} j_1 & j_2 & j\\ 
m_1 & m_2 & m\end{array}\right]_{+++} = \left(-1\right)^{m_1-j_1}\\ \times
\sqrt{\left(j+j_1+j_2-2\right)!\,
\left(j+j_1-j_2-1\right)!\,
\left(j-j_1+j_2-1\right)!\,
\left(j-j_1-j_2\right)!}\\ \times
\sqrt{\frac{\left(m_1+j_1-1\right)!\,\left(m_1-j_1\right)!\,
\left(m_2+j_2-1\right)!\,\left(m_2-j_2\right)!\,\left(m-j\right)!}
{\left(j+m-1\right)!}}\\ \times
\sum_z\frac{\left(-1\right)^z}
{z!\,\left(m-j-z\right)!\,\left(m_1+j_1-1-z\right)!\,\left(m_1-j_1-z\right)!}\\
\times\frac{1}{\left(j+j_2-m_1-1+z\right)!\,\left(j-j_2-m_1+z\right)!}
\end{multline}

Note that using relation \ref{rotate} this formula can be extended, up
to phase, to cover the three cases of coupling in the discrete series
that include either no or two negative discrete series.  To find the
remaining three cases one must start from $\left[\begin{array}{ccc}
j_1 & j_2 & j\\ m_1 & m_2 & m\end{array}\right]_{---}$ and extend that
using equation
\ref{rotate}.

Take $f$ as in equation \ref{Ffunc} then substitution into
\ref{K+recurse} and \ref{K-recurse} yields, respectively,
\begin{multline}\label{K+recurse4}
\left(-m-j\right)\left(-m+j-1\right)
f\left(\begin{array}{ccc} j_1 & j_2 & j\\ 
m_1 & m_2 & m + 1\end{array}\right) = \\ 
-f\left(\begin{array}{ccc} j_1 & j_2 & j\\ 
m_1 - 1 & m_2 & m \end{array}\right)
+ f\left(\begin{array}{ccc} j_1 & j_2 & j\\ 
m_1 & m_2 - 1 & m \end{array} \right)
\end{multline}
and
\begin{multline}\label{K-recurse4}
f\left(\begin{array}{ccc} j_1 & j_2 & j\\ 
m_1 & m_2 & m - 1\end{array}\right) = 
-\left(-m_1-j_1\right)\left(-m_1+j_1-1\right) 
f\left(\begin{array}{ccc} j_1 & j_2 & j\\ 
m_1 + 1 & m_2 & m \end{array} \right)\\
+ \left(-m_2-j_2\right)\left(-m_2+j_2-1\right) 
f\left(\begin{array}{ccc} j_1 & j_2 & j\\ 
m_1 & m_2 + 1 & m \end{array} \right)
\end{multline}

The calculation then proceeds as in the previous case and yields a
formula similar to the $+++$ case, but with $m_i\mapsto -m_i$.  Indeed
\begin{equation}\label{reflect}
\left[\begin{array}{ccc} j_1 & j_2 & j\\ 
m_1 & m_2 &m\end{array}\right]_{+++} =  
\left[\begin{array}{ccc} j_1 & j_2 & j\\
-m_1 & -m_2 & -m\end{array}\right]_{---}
\end{equation}

This final order 2 reflection symmetry is all that is needed, together
with the permutation symmetry of equation \ref{rotate} to generate all
the cases of coupling for all representations in the discrete series.

\section{The Racah Coefficients for the Discrete Series} \label{rac}

In this section explicit calculations will be made for the case where
all the representations are in the discrete series.  All sums without
a specified range are in integer steps over the range for which all
the factorials in the summand are defined.

In contrast to section \ref{CGC} the Clebsch-Gordon coefficients will
be exhibited in the $SO\left(2\right)$ discrete basis since the
calculations are easier in this form and the Racah coefficient is
basis independent as discussed in section \ref{Racah}.

In addition to lemma \ref{lem2} from the previous section the
following lemma will be required
\begin{lem} \label{lem1}
$$
\sum_n (-1)^n\frac{(x+n-1)!}{(z-n)!\,(y+n-1)!\,n!}=(-1)^z
\frac{(x-1)!\,(x-y)!}{z!\,(y+z-1)!\,(x-y-z)!}
$$
\end{lem}

Lemma \ref{lem1} follows from Gauss' formula for summing
the $ _2F_1$ hypergeometric series\cite{HypGeo}
$$
\sum_n \frac{(a+n-1)!\,(b+n-1)!\,(c-1)!}
{(a-1)!\,(b-1)!\,(c+n-1)!\,n!} =
\frac{(c-a-b-1)!\,(c-1)!}{(c-a-1)!\,(c-b-1)!} $$
with $a=x$, $b = -z$, $c= y$.

The discrete basis version of the recoupling identity in equation
\ref{racah} may be expressed as
\begin{multline}\label{racah-discrete}
\sum_{m_{12}} 
\left[ \begin{array}{ccc} j_1 & j_2 & j_{12}\\ 
m_1 & m_2 & m_{12} \end{array} \right]
\left[ \begin{array}{ccc} j_{12} & j_3 & j\\ 
m_{12} & m_3 & m \end{array} \right]\\
= \sum_{j_{23}\;m_{23}} (2j_{23}-1) 
\left\{ \begin{array}{ccc} j_1 & j_2 & j_{12}\\ j_3 & j & j_{23} 
\end{array}\right\}
\left[ \begin{array}{ccc} j_2 & j_3 & j_{23}\\
m_2 & m_3 & m_{23} \end{array} \right]
\left[ \begin{array}{ccc} j_1 & j_{23} & j\\ 
m_1 & m_{23} & m \end{array} \right]
\end{multline}
where $\left[ \begin{array}{ccc} j_1 & j_2 & j_{12}\\  m_1 & m_2 &
m_{12} \end{array} \right]$ are the Clebsch-Gordon coefficients for
the coupling of two unitary irreducible representations of $SU(1,1)$. 

Equation \ref{CGC-orth-O2} may be used to bring equation
\ref{racah-discrete} into the following form
\begin{multline} \label{racdef}
\sum_\beta\left[ \begin{array}{ccc} j_1 & j_2 & j_{12}\\ 
m_1 & \beta & m_1 +\beta \end{array} \right]
\left[ \begin{array}{ccc} j_{12} & j_3 & j\\ 
m_1 +\beta & j_{23}-\beta & j_{23} + m_1 \end{array} \right]
\left[ \begin{array}{ccc} j_2 & j_3 & j_{23}\\ 
\beta & j_{23}-\beta & j_{23} \end{array} \right]\\
= \left\{ \begin{array}{ccc} j_1 & j_2 & j_{12}\\ 
j_3 & j & j_{23} \end{array}\right\}
\left[ \begin{array}{ccc} j_1 & j_{23} & j\\ 
m_1 & j_{23} & m_1 + j_{23} \end{array} \right]
\end{multline}
where the choice $m_{23} = j_{23}$ has been made.

Now consider the representations $j_1$, $j_2$ and $j_3$.  Formally
there are $2^3 = 8$ different combinations of positive and negative
discrete representations, however these come in dual pairs so there
are only four unique possibilities.  It is now a simply matter to
enumerate the possibilities for the remaining discrete series
representations in the Racah coefficent in table \ref{posneg-racah}.
The symmetry given by equation \ref{reflect} applied to all four
Clebsch-Gordon coefficients allows these twelve cases to be extended
to cover the full twenty four possible permutations.

\begin{table}[hbt!]
\begin{center}
\begin{tabular}{ccccccc}
Case & $j_1$ & $j_2$ & $j_3$ & $j_{12}$ & $j_{23}$ & $j$\\
(1) & + & + & + & + & + & +\\
(2) &$-$& + & + & + & + & +\\
(3) &$-$& + & + &$-$& + & +\\
(4) &$-$& + & + &$-$& + &$-$\\
(5) & + & + &$-$& + & + & +\\
(6) & + & + &$-$& + &$-$&$-$\\
(7) & + & + &$-$& + &$-$& +\\
(8) &$-$& + &$-$& + & + & +\\
(9) &$-$& + &$-$& + & + &$-$\\
(10)&$-$& + &$-$&$-$& + &$-$\\
(11)&$-$& + &$-$& + &$-$&$-$\\
(12)&$-$& + &$-$&$-$&$-$&$-$\\
\end{tabular}
\end{center}
\caption{The possible unique distributions of positive and negative discrete 
representations in the Racah coefficient\label{posneg-racah}}
\end{table}

It now remains to calculate the Racah cofficient for the twelve cases
given in table \ref{posneg-racah}.  Let the Racah coefficient for each
of the twelve cases of coupling be denoted by a subscript $N$ for
$N=1\dots 12$.  Explicitly the calculation for case 1 proceeds from a
substitution of equation \ref{O+++} into \ref{racdef}

\begin{multline} \label{rac2}
\left\{ \begin{array}{ccc} j_1 & j_2 & j_{12}\\ j_3 & j & j_{23} \end{array} 
\right\}_1 = \frac{\Delta(j_1j_2j_{12})\Delta(j_2j_3j_{23})
\Delta(j_{12}j_3j)}{\Delta(j_1j_{23}j)}\\ 
\times (j-j_1-j_{23})!\,(j+j_1-j_{23}-1)!\,(j_{23}+\alpha
-j)!\,\mathcal{I}\left(\alpha\right))
\end{multline}
where
\begin{multline*}
\mathcal{I}(\alpha) = 
\sum_{\beta\; t\; u}\frac{(-1)^{t+u}(\alpha +\beta -j_{12})!}
{t!(\alpha +\beta -j_{12}-t)!\,(j_{12}-j_2-\alpha +t)!\,
(j_{12}+j_2-\alpha +t-1)!}\\
\times\frac{1}{u!\,(\alpha+j_{23}-j-u)!\,(j-j_3-\alpha -\beta+u)!\,
(j+j_3-\alpha -\beta+u-1)!}
\\ \times\frac{1}{
(\alpha +\beta -j_{12}-u)\,!(\alpha +\beta +j_{12}-u-1)!\,
(\alpha - j_1 -t)!\,(\alpha + j_1-t-1)!}
\end{multline*}

$\mathcal{I}\left(\alpha\right)$ may be reduced to a single summation 
as follows. Introduce two new summation variables, $m$ and $n$, in 
place of $\beta$ and $u$ such that
\begin{align*}
u & = \alpha +\beta - j_{12} -n\\
\beta & = j_{12} - \alpha+m+n
\end{align*}
Then
\begin{multline}
\mathcal{I}\left(\alpha\right) = 
\sum_{m\; n\; t}\frac{(-1)^{t+m} (m+n)!}{t!\,(t+m+n)!\,(j_{12}-j_2-\alpha+t)!
\,(j_{12}+j_2-\alpha+t-1)!}\\ 
\times \frac{1}{m!\,(j_{23}-j+\alpha -m)!\,
(j+j_3-j_{12}-1-n)!\,n!\,(2j_{12}-1+n)!\,(j-j_3-j_{12}-n)!}\\
\times \frac{1}{(\alpha -j_1 -t)!\,(\alpha +j_1 -t-1)!}
\end{multline}

The sum over $m$, using lemma \ref{lem1}, is found to be
\begin{multline*}
\sum_m \frac{(-1)^m\,(m+n)!}{m!\,(m+n-t)!\,(j_{23}-j+\alpha -m)!}\\
= \frac{(-1)^{j_{23}-j+\alpha} n!\,t!}
{(j_{23}-j+\alpha)!\,(n+j_{23}-j+\alpha-t)!\,(t-j_{23}+j-\alpha)!}
\end{multline*}
and the sum may be written as
\begin{multline} \label{I1}
\mathcal{I}\left(\alpha\right) = 
\sum_{n\; t}\frac{(-1)^{j_{23}-j+\alpha -t}}
{(j_{12}-j_2-\alpha +t)!\,(j_{12}+j_2-\alpha +t-1)!\,
(\alpha -j_1 -t)!}
\\ \times \frac{1}
{(2j_{12}-1+n)!\,(j_{23}-j+\alpha)!\,
(n+j_{23}-j+\alpha -t)!\,(t-j_{23}+j-\alpha)!}\\
\frac{1}{(\alpha +j_1-t-1)!\,(j+j_3-j_{12}-1-n)!\,(j-j_3-j_{12}-n)!}
\end{multline}
Now, transforming with $n = -j_{12}-j_3+j -s$, we may rewrite equation
\ref{I1} as
\begin{multline}
\mathcal{I}(\alpha)=\sum_{t\; s}\frac{(-1)^{j_{23}+\alpha-t-j}}
{(j_{12}-j_2-\alpha+t)!\,(j_{12}+j_2-\alpha +t-1)!\,
(\alpha -j_1-t)!}\\
\times\frac{1}{(j_{12}-j_3+j-1-s)!\,(j_{23}-j+\alpha)!
(j_{23}-j_{12}-j_3-s+\alpha -t)!\,s!}\\
\times\frac{1}{(\alpha +j_1-t-1)!\,(2j_3-1+s)!\,(t-j_{23}+j-\alpha)!}
\end{multline}
The sum over $s$, using lemma \ref{lem2}, is found to be
\begin{multline*}
\sum_s\frac{1}{(2j_3-1+s)!\,(j_{12}-j_3+j-1-s)!\,
(j_{23}-j_{12}-j_3-s+\alpha -t)!s!}\\ = \frac{(j+j_{23}-2-t+\alpha)!}
{(j_{12}-j_3+j-1)!\,(j_{23}-j_{12}-j_3-t+\alpha)!}\\
\times\frac{1}{(j_{12}+j_3+j-2)!\,(j_{23}-j_{12}+j_3-t+\alpha-1)!}
\end{multline*}
and $\mathcal{I}\left(\alpha\right)$ is reduced to a single summation
\begin{multline}
\mathcal{I}(\alpha)=\sum_t\frac{(-1)^{\alpha -t+j_{23}-j}
(j+j_{23}-2-t+\alpha)!}
{(j_{12}-j_2-\alpha +t)!\,(j_{23}+j_2-\alpha +t-1)!\,
(\alpha-j_1-t)!\,(\alpha +j_1-t-1)!}\\ \times
\frac{1}{(t-j_{23}+j-\alpha)!\,(j_{12}-j_3+j-1)!\,
(j_{23}-j_{12}-j_3-t+\alpha)!\,(j_{12}+j_3+j-2)!}\\
\times\frac{1}{(j_{23}-j+\alpha)!\,
(j_{23}-j_{12}+j_3-t+\alpha-1)!}
\end{multline}
If the summation variable is rewritten as $z=\alpha -j_1 -t$ and
substituted into equation \ref{rac2} we find
\begin{multline} 
\left\{ \begin{array}{ccc} j_1 & j_2 & j_{12}\\ 
j_3 & j & j_{23} \end{array}_1 \right\}_1 = 
\frac{(-1)^{j_{23}+j_1-j}\Delta(j_1j_2j_{12})\Delta(j_2j_3j_{23})
\Delta(j_{12}j_3j)}
{\Delta(j_1j_{23}j)(c+d+e-2)!\,(c-d+e-1)!}\\ \times
\sum_z\frac{(-1)^z (j+j_{23}-2+j_1+z)!}
{z!\,(j_{12}-j_2-j_1-z)!\,(j_{12}+j_2-j_1-1-z)!\,(j-j_{23}-j_1-z)!}\\ \times
\frac{(j-j_1-j_{23})!\,(j+j_1-j_{23}-1)!}
{(2j_1-1+z)!(j_{23}-j_{12}-j_3+j_1+z)!(j_{23}-j_{12}+j_3+j_1+z-1)!}
\end{multline}
 
The sum may then be brought into the following, more symmetrical, form
\begin{multline} \label{RacahSU11}
\left\{ \begin{array}{ccc} j_1 & j_2 & j_{12}\\ 
j_3 & j & j_{23} \end{array} 
\right\}_1 = \left(-1\right)^{2j-1}
\frac{\Delta(j_1j_2j_{12})\Delta(j_2j_3j_{23})
\Delta(j_{12}j_3j)(j-j_1-j_{23})!}
{\Delta(j_1j_{23}j)(j_{12}+j_3+j-2)!\,(j_{12}-j_3+j-1)!}\\ \times
\sum_r\frac{(-1)^r (r+1)!\,(j+j_1-j_{23}-1)!}
{(j_{12}-j_2+j+j_{23}-3-r)!\,
(j_{12}+j_2+j+j_{23}-4-r)!!\,(r-j_1-j-j_{23}+3)!}\\ \times
\frac{1}{(r+j_1-j-j_{23}+2)!\,
(r-j_1-j-j_{23}+3)!\,(r+j_3-j_{12}-j+3)\,(2j-3-r)!}
\end{multline}

As for the Clebsch-Gordon coefficient one may use symmetries to
generate the twelve cases of coupling by inducing symmetries on the
Racah coeffcient.  Consider equation \ref{racdef} in its most generic
form
\begin{multline}\label{racdef-gen}
\sum_{m_2\,m_3\,m_{12}}
\left[ \begin{array}{ccc} j_1 & j_2 & j_{12}\\ 
m_1 & m_2 & m_{12} \end{array} \right]
\left[ \begin{array}{ccc} j_{12} & j_3 & j\\ 
m_{12} & m_3 & m \end{array} \right]
\left[ \begin{array}{ccc} j_2 & j_3 & j_{23}\\ 
m_2 & m_3 & m_{23} \end{array} \right]\\
= \left\{ \begin{array}{ccc} j_1 & j_2 & j_{12}\\ 
j_3 & j & j_{23} \end{array}\right\}
\left[ \begin{array}{ccc} j_1 & j_{23} & j\\ 
m_1 & m_{23} & m \end{array} \right]
\end{multline}

Now consider the passive transformation that maps 
$$
\left(\begin{array}{ccc}
j_3 & j_2 & j_{12}
\end{array}\right)\longmapsto\left(\begin{array}{ccc}
j_{12} & j_3 & j_2
\end{array}\right) 
$$
and
$$
\left(\begin{array}{ccc}
j_1 & j & j_{23}
\end{array}\right)\longmapsto\left(\begin{array}{ccc}
j_{23} & j_1 & j
\end{array}\right) 
$$
with corresponding transformations for the $m_i$, applied to equation
\ref{racdef-gen}.  One has
\begin{multline}
\sum_{m_2\,m_3\,m_{12}}
\left[ \begin{array}{ccc} j_{23} & j_3 & j_2\\ 
m_{23} & m_3 & m_2 \end{array} \right]
\left[ \begin{array}{ccc} j_2 & j_{12} & j_1\\ 
m_2 & m_{12} & m_1 \end{array} \right]
\left[ \begin{array}{ccc} j_3 & j_{12} & j\\ 
m_3 & m_{12} & m \end{array} \right]\\
= \left\{ \begin{array}{ccc} j_{23} & j_3 & j_2\\ 
j_{12} & j_1 & j \end{array}\right\}
\left[ \begin{array}{ccc} j_{23} & j & j_1\\ 
m_{23} & m & m_1 \end{array} \right]
\end{multline}
and hence by equations \ref{rotate} and \ref{reflect} one has, up to
an overall phase on the Racah coefficient
\begin{multline}
\sum_{m_2\,m_3\,m_{12}}
\left[ \begin{array}{ccc} j_3 & j_2 & j_{23}\\ 
m_3 & -m_2 & -m_{23} \end{array} \right]
\left[ \begin{array}{ccc} j_1 & j_2 & j_{12}\\ 
m_1 & -m_2 & m_{12} \end{array} \right]
\left[ \begin{array}{ccc} j_3 & j_{12} & j\\ 
m_3 & m_{12} & m \end{array} \right]\\
= \left\{ \begin{array}{ccc} j_{23} & j_3 & j_2\\ 
j_{12} & j_1 & j \end{array}\right\}
\left[ \begin{array}{ccc} j_1 & j_{23} & j\\ 
m_1 & -m_{23} & m \end{array} \right]
\end{multline}
which is simply then defining relation for a Racah coefficient with a
different pattern of positive and negative discrete series
representations.  This symmetry induces the following automorphism of
the twelve different Racah coefficients
\begin{equation}\label{aut1}
\left|\left\{ \begin{array}{ccc} j_1 & j_2 & j_{12}\\ 
j_3 & j & j_{23} \end{array}\right\}_i\right| = 
\left|\left\{ \begin{array}{ccc} j_{23} & j_3 & j_2\\ 
j_{12} & j_1 & j \end{array}\right\}_{\sigma_1\left(i\right)}\right|
\end{equation}
where $\sigma_1$ is the following permutation
\begin{multline*}
\sigma_1 : \left( \begin{array}{cccccccccccc}
1 & 2 & 3 & 4 & 5 & 6 & 7 & 8 & 9 & 10 & 11 & 12
\end{array}\right) \\
\longmapsto\left( \begin{array}{cccccccccccc}
10 & 12 & 1 & 2 & 9 & 5 & 8 & 11 & 6 & 3 & 7 & 4
\end{array}\right)
\end{multline*}

Now consider the recoupling identity in equation \ref{racah-discrete}
\begin{multline}\label{rec}
\sum_{m_{12}} 
\left[ \begin{array}{ccc} j_1 & j_2 & j_{12}\\ 
m_1 & m_2 & m_{12} \end{array} \right]
\left[ \begin{array}{ccc} j_{12} & j_3 & j\\ 
m_{12} & m_3 & m \end{array} \right]\\
= \sum_{j_{23}\;m_{23}} (2j_{23}-1) 
\left\{ \begin{array}{ccc} j_1 & j_2 & j_{12}\\ j_3 & j & j_{23} 
\end{array}\right\}
\left[ \begin{array}{ccc} j_2 & j_3 & j_{23}\\
m_2 & m_3 & m_{23} \end{array} \right]
\left[ \begin{array}{ccc} j_1 & j_{23} & j\\ 
m_1 & m_{23} & m \end{array} \right]
\end{multline}
There are essentially two independent passive transformations one may
make without disrupting the sums, swapping $j_1$ with $j_2$ and $j_3$
with $j$ or swapping $j_1$ with $j_3$ and $j_2$ with $j$.  The first
of these transforms equation \ref{rec} to
\begin{multline}
\sum_{m_{12}} 
\left[ \begin{array}{ccc} j_2 & j_1 & j_{12}\\ 
m_2 & m_1 & m_{12} \end{array} \right]
\left[ \begin{array}{ccc} j_{12} & j & j_3\\ 
m_{12} & m & m_3 \end{array} \right]\\
= \sum_{j_{23}\;m_{23}} (2j_{23}-1) 
\left\{ \begin{array}{ccc} j_2 & j_1 & j_{12}\\ j & j_3 & j_{23} 
\end{array}\right\}
\left[ \begin{array}{ccc} j_1 & j & j_{23}\\
m_1 & m & m_{23} \end{array} \right]
\left[ \begin{array}{ccc} j_2 & j_{23} & j_3\\ 
m_2 & m_{23} & m_3 \end{array} \right]
\end{multline}
Once again using equations \ref{rotate} and \ref{reflect} one may
transform this (up to an overall phase on the Racah coefficient) into
\begin{multline}
\sum_{m_{12}} 
\left[ \begin{array}{ccc} j_1 & j_2 & j_{12}\\ 
m_1 & m_2 & m_{12} \end{array} \right]
\left[ \begin{array}{ccc} j_3 & j_{12} & j\\ 
-m_3 & m_{12} & -m \end{array} \right]\\
= \sum_{j_{23}\;m_{23}} (2j_{23}-1) 
\left\{ \begin{array}{ccc} j_2 & j_1 & j_{12}\\ j & j_3 & j_{23} 
\end{array}\right\}
\left[ \begin{array}{ccc} j_{23} & j_1 & j\\
-m_{23} & m_1 & -m \end{array} \right]
\left[ \begin{array}{ccc} j_3 & j_2 & j_{23}\\ 
-m_3 & m_2 & -m_{23} \end{array} \right]
\end{multline}

As before this induces the following order 2 automorphism on the
twelve Racah coefficients
\begin{equation}\label{aut2}
\left|\left\{ \begin{array}{ccc} j_1 & j_2 & j_{12}\\ 
j_3 & j & j_{23} \end{array}\right\}_i\right| = 
\left|\left\{ \begin{array}{ccc} j_2 & j_1 & j_{12}\\ 
j & j_3 & j_{23} \end{array}\right\}_{\sigma_2\left(i\right)}\right|
\end{equation}
where $\sigma_2$ is the following permutation
\begin{multline*}
\sigma_2 : \left( \begin{array}{cccccccccccc}
1 & 2 & 3 & 4 & 5 & 6 & 7 & 8 & 9 & 10 & 11 & 12
\end{array}\right) \\
\longmapsto\left( \begin{array}{cccccccccccc}
6 & 3 & 2 & 8 & 7 & 1 & 5 & 4 & 10 & 9 & 12 & 11
\end{array}\right)
\end{multline*}

Now consider the passive transformation swapping $j_1$ with $j_3$ and
$j_2$ with $j$.  This transforms equation \ref{rec} to
\begin{multline}
\sum_{m_{12}} 
\left[ \begin{array}{ccc} j_3 & j & j_{12}\\ 
m_3 & m & m_{12} \end{array} \right]
\left[ \begin{array}{ccc} j_{12} & j_1 & j_2\\ 
m_{12} & m_1 & m_2 \end{array} \right]\\
= \sum_{j_{23}\;m_{23}} (2j_{23}-1) 
\left\{ \begin{array}{ccc} j_3 & j & j_{12}\\ j_1 & j_2 & j_{23} 
\end{array}\right\}
\left[ \begin{array}{ccc} j & j_1 & j_{23}\\
m & m_1 & m_{23} \end{array} \right]
\left[ \begin{array}{ccc} j_3 & j_{23} & j_2\\ 
m_3 & m_{23} & m_2 \end{array} \right]
\end{multline}
As before using equations \ref{rotate} and \ref{reflect} one may
transform this (up to an overall phase on the Racah coefficient) into
\begin{multline}
\sum_{m_{12}} 
\left[ \begin{array}{ccc} j_{12} & j_3 & j\\ 
-m_{12} & m_3 & -m \end{array} \right]
\left[ \begin{array}{ccc} j_1 & j_2 & j_{12}\\ 
m_1 & -m_2 & -m_{12} \end{array} \right]\\
= \sum_{j_{23}\;m_{23}} (2j_{23}-1) 
\left\{ \begin{array}{ccc} j_3 & j & j_{12}\\ j_1 & j_2 & j_{23} 
\end{array}\right\}
\left[ \begin{array}{ccc} j_1 & j_{23} & j\\
m_1 & -m_{23} & -m \end{array} \right]
\left[ \begin{array}{ccc} j_2 & j_3 & j_{23}\\ 
-m_2 & m_3 & -m_{23} \end{array} \right]
\end{multline}

This induces the another order 2 automorphism on the twelve Racah
coefficients
\begin{equation}\label{aut3}
\left|\left\{ \begin{array}{ccc} j_1 & j_2 & j_{12}\\ 
j_3 & j & j_{23} \end{array}\right\}_i\right| = 
\left|\left\{ \begin{array}{ccc} j_3 & j & j_{12}\\ 
j_1 & j_3 & j_{23} \end{array}\right\}_{\sigma_3\left(i\right)}\right|
\end{equation}
where $\sigma_3$ is the following permutation
\begin{multline*}
\sigma_3 : \left( \begin{array}{cccccccccccc}
1 & 2 & 3 & 4 & 5 & 6 & 7 & 8 & 9 & 10 & 11 & 12
\end{array}\right) \\
\longmapsto\left( \begin{array}{cccccccccccc}
8 & 2 & 3 & 6 & 5 & 4 & 7 & 1 & 12 & 11 & 10 & 9
\end{array}\right)
\end{multline*}

It is not too difficult to see that one may generate the appropriate
Racah coefficient for all twelve cases of coupling from case 1 by
suitable applications of $\sigma_1$, $\sigma_2$ and $\sigma_3$.

\chapter{The Quantum Geometry of $SU\left(1,1\right)$}\label{Quantum-Geometry}

In this chapter the relationship between the geometry of three
dimensional Lorentzian space and the irreducible unitary
representations of \su{1,1} will be explored.  In sections \ref{state}
and \ref{inv} a state sum model for \su{1,1} is stated and shown to be
a formal topological invariant, in the same way that the \su{2}
Ponzano-Regge model is\cite{PR}.  Section \ref{inv} also places the
symmetries of the \su{1,1} Racah, Clebsch-Gordon coefficient, already
stated and derived in sections \ref{rac} and \ref{rac-CGC}
respectively, in the correct categorical framework.

\section{Quantum Vector Addition and $SU(1,1)$}\label{Geom}

There is an intimate relation between irreducible unitary
representations of certain groups and `quantum vectors' in three
dimensional space.  In the case of $SU\left(2\right)$, since it is the
double cover of $SO\left(3\right)$, the $j$-th representation may be
identified with a Euclidean vector of length $j+\frac{1}{2}$ as
discussed in \cite{PR}.  The Casimir $C =
\sqrt{J_x^2+J_y^2+J_z^2}$ is the `length operator' for these quantum
vectors and acts in the $j$-th representation as
$\sqrt{j\left(j+1\right)}$, leading to the shift of plus a half.

The group $SU\left(1,1\right)$ has both a richer geometry and
representation theory.  In view of it being the double cover of the
three dimensional Lorentz group, $SO\left(2,1\right)$, the different
representations discussed in section \ref{rep} should correspond to
the different types of Lorentzian vector.

The correspondence is discussed in \cite{W88} within the framework of
geometric quantisation\footnote{While this paper has the only explicit
calculation we have been able to find it seems likely this
correspondence was discovered much earlier}.  Here it is shown that
the positive and negative discrete series correspond to, respectively,
future pointing and past pointing time like vectors of length
$j-\frac{1}{2}$ (in this case the Casimir acts as
$\sqrt{j\left(j-1\right)}$ so there is a shift of minus one half in
contrast to the $SU\left(2\right)$ case).  The continuous series
representation given by $j=\frac{1}{2}+is$ corresponds to space like
vectors of length $\left|s\right|$, while the positive and negative
mock discrete series, where $j=\frac{1}{2}$, corresponds to the future
and past light cone.

Since each quantum vector is a separate quantum system with its own
Hilbert space to represent its possible states, one may add such
vectors by tensoring their Hilbert spaces to gain the Hilbert space of
possible states of the resultant vector.

Such a tensor product space is not an irreducible representation in
general and so one needs to project onto the irreducibles present with
the resultant map giving a notion of quantum vector addition, the
result of the addition being the vector associated to the
representation projected onto and the value of the map being the
probability of such a vector addition being observed.

Since these projection maps are precisely the Clebsch-Gordon
coefficients for the representation series one may regard the
Clebsch-Gordon coefficients as `quantum vector addition' in some
sense.

For $SU\left(2\right)$ these results are well known\cite{PR} and the
Clebsch-Gordon coefficient is regarded as the wave function of a
quantum triangle in $\R^3$ equipped with a Euclidean metric.  The
sides are formed with vectors of length $j_1+\frac{1}{2}$ and
$j_2+\frac{1}{2}$ adding to give a result vector of length
$j_{12}+\frac{1}{2}$. The respective $m_i$'s give the appropriate
vector's projection on the $z$ axis\footnote{One usually works in a
basis in which the generator of rotations about the $z$ axis $J_z$
acts as a diagonal matrix}. A geometric representation is shown in
figure \ref{su2-couple}.

\begin{figure}[htb!]
\begin{texdraw}
\drawdim{mm}
\arrowheadtype t:V
\lvec(0 100)\move(-20 0)\lvec(60 0) 
\move(0 0)
\ravec(10 50) \lpatt(1 5)\rlvec(50 30)\lpatt(2 3)
\move(10 50)\lvec(0 50)\move(0 0)\lpatt()
\ravec(50 30) \lpatt(1 5)\rlvec(10 50)\lpatt(2 3)
\move(50 30)\lvec(0 30)\move(0 0)\lpatt()
\ravec(60 80)\lpatt(2 3)\lvec(0 80)
\move(2 90)\htext{z axis}
\move(9 40)\htext{$j_1$}
\move(25 10)\htext{$j_2$}
\move(50 60)\htext{$j$}
\move(-5 49)\htext{$m_1$}
\move(-5 29)\htext{$m_2$}
\move(-5 79)\htext{$m$}
\move(60 40)\htext{$\equiv\;\left[\begin{array}{ccc}
j_1-\frac{1}{2} & j_2-\frac{1}{2} & j-\frac{1}{2}\\
m_1-\frac{1}{2} & m_2-\frac{1}{2} & m-\frac{1}{2}\end{array}\right]$}
\end{texdraw}
\caption{\label{su2-couple}
The $SU\left(2\right)$ Clebsch-Gordon coefficient as
quantum vector addition in $\R^3$}
\end{figure}

For \su{1,1} the tensoring of representations is still regarded as
quantum vector addition and the Clebsch-Gordon coefficient is non-zero
when the three quantum vectors of the specified type can form the
sides of a triangle\footnote{As for the correspondence between the
Racah coefficient and tetrahedra, the Clebsch-Gordon coefficient is
oscillatory so it is only strictly true that the coefficient is zero
when it cannot form the sides of a triangle}. The value of the
associated Clebsch-Gordon coefficient is thus regarded as the `wave
function' of a quantum triangle in three dimensional Lorentzian space
having edge lengths $j_1-\frac{1}{2}$, $j_2-\frac{1}{2}$,
$j_{12}-\frac{1}{2}$.

\begin{figure}[htb!]
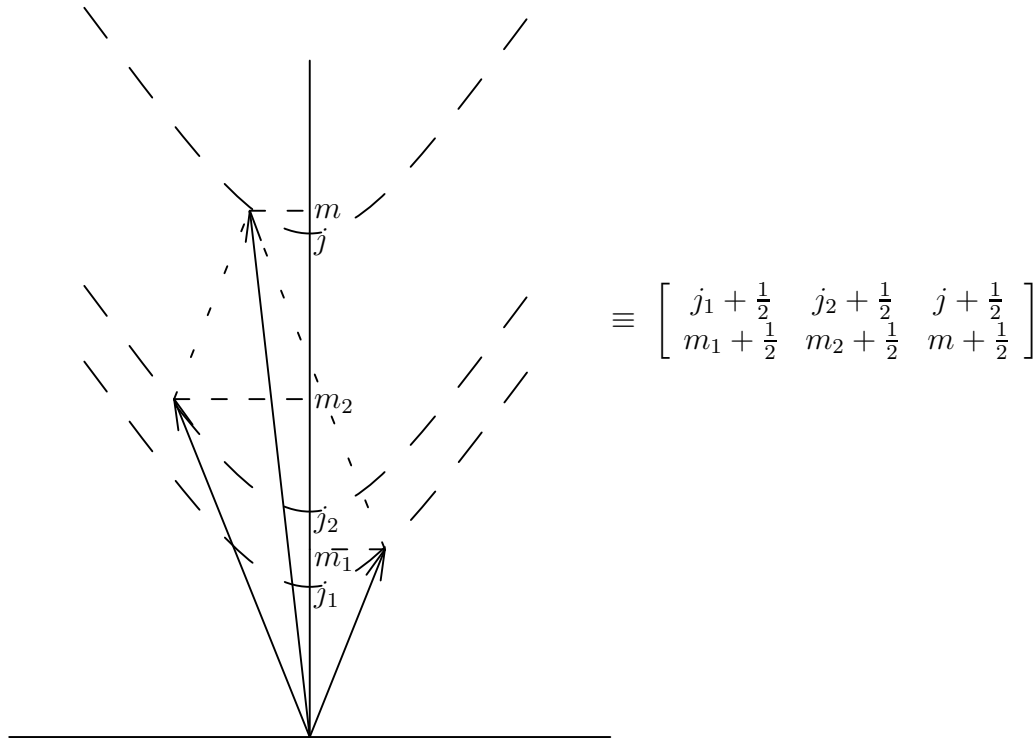

\begin{texdraw}
\drawdim{mm}
\move(0 -10)\lvec(0 80)
\move(-40 -10)\lvec(40 -10)
\lpatt(5 5)
\move(-30 40)\clvec(0 0)(0 0)(30 40)
\move(-30 50)\clvec(0 10)(0 10)(30 50)
\move(-30 87)\clvec(0 47)(0 47)(30 87)
\lpatt() \arrowheadtype t:V
\move(0 -10)\avec(10 15)\lpatt(2 3)\lvec(0 15)\lpatt(1 5)\move(10 15)\rlvec(-18 45)\lpatt()
\move(0 -10)\avec(-18 35)\lpatt(2 3)\lvec(0 35)\lpatt(1 5)\move(-18 35)\rlvec(10 25)\lpatt()
\move(0 -10)\avec(-8 60)\lpatt(2 3)\lvec(0 60)\lpatt()
\move(0.5 7)\htext{$j_1$}
\move(0.5 17)\htext{$j_2$}
\move(0.5 54)\htext{$j$}
\move(0.5 12)\htext{$m_1$}
\move(0.5 33)\htext{$m_2$}
\move(0.5 59)\htext{$m$} 
\move(40 40)\htext{$\equiv\;\left[\begin{array}{ccc}
j_1+\frac{1}{2} & j_2+\frac{1}{2} & j+\frac{1}{2}\\
m_1+\frac{1}{2} & m_2+\frac{1}{2} & m+\frac{1}{2}\end{array}\right]$}
\end{texdraw}
\caption{\label{pp-couple} 
Geometric representation of the coupling of two positive DUR's
to give a third}
\end{figure}

\begin{figure}[htb!]
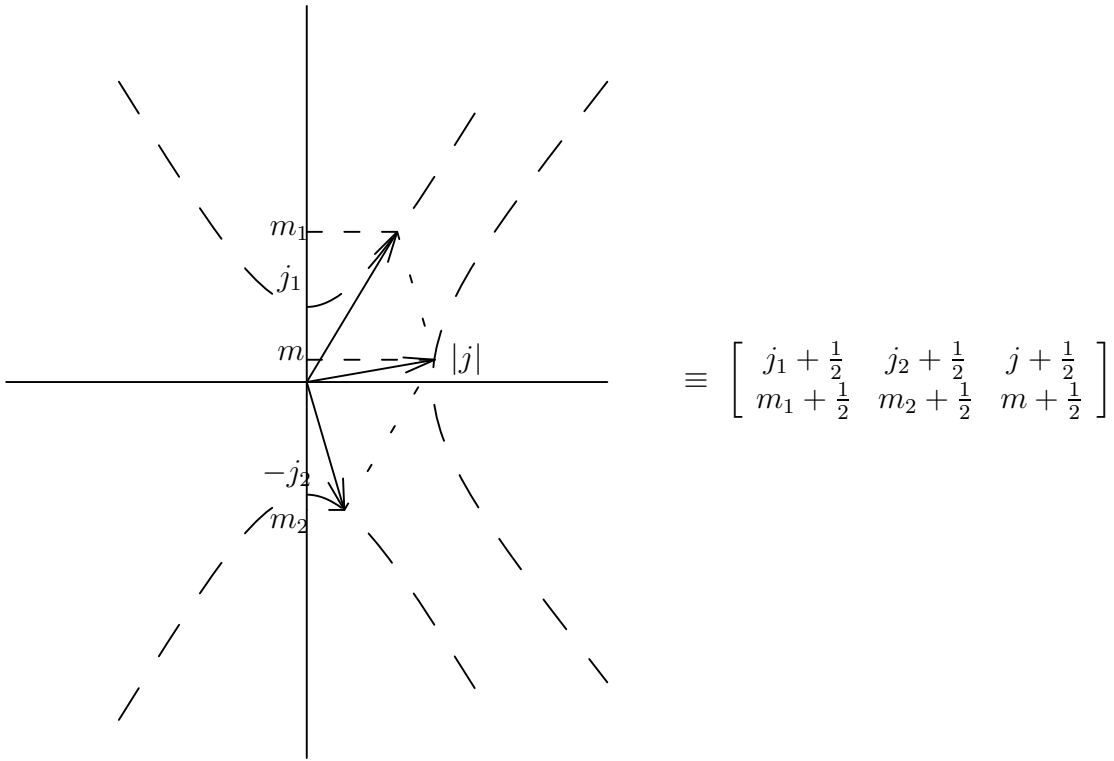

\begin{texdraw}
\drawdim{mm}
\move(0 -50)\lvec(0 50)
\move(-40 0)\lvec(40 0)
\lpatt(5 5)
\move(-25 40)\clvec(0 0)(0 0)(25 40)
\move(-25 -45)\clvec(0 -5)(0 -5)(25 -45)
\move(40 40)\clvec(9 0)(9 0)(40 -40)
\lpatt() \arrowheadtype t:V
\move(0 0)\avec(12 20)\lpatt(1 5)\rlvec(5 -17)\move(12 20)\lpatt(2 3)
\lvec(0 20)\lpatt()
\move(0 0)\avec(5 -17)\lpatt(1 5)\rlvec(12 20)\move(5 -17)\lpatt(2 3)
\lvec(0 -17)\lpatt()
\move(0 0)\avec(17 3)\lpatt(2 3)\lvec(0 3)
\move(-4 12)\htext{$j_1$}
\move(-6 -14)\htext{$-j_2$}
\move(19 1)\htext{$|j|$}
\move(-5 19)\htext{$m_1$}
\move(-5 -20)\htext{$m_2$}
\move(-4 3)\htext{$m$} 
\move(50 -5)\htext{$\equiv\;\left[\begin{array}{ccc}
j_1+\frac{1}{2} & j_2+\frac{1}{2} & j+\frac{1}{2}\\
m_1+\frac{1}{2} & m_2+\frac{1}{2} & m+\frac{1}{2}\end{array}\right]$}
\end{texdraw}
\caption{\label{pn-couple} Geometric representation of the coupling of a 
positive and a negative DUR to give a representation in the principal
series}
\end{figure}

In this case one has essentially two ways to represent the
Clebsch-Gordon coefficients.  The traditional way is as for $\su{2}$
with the generator of the $SO\left(2\right)$ subgroup diagonal. This
leads to the normal discrete basis discussed initially in section
\ref{rep}.  Here the basis elements still project onto the $z$-axis
(or time like axis in the Lorentzian geometry under consideration) and
geometric representations of two different cases of coupling for this
case are given in figures \ref{pp-couple} and \ref{pn-couple}.

The Racah coefficient is thus regarded as the wave function of a quantum
tetrahedron formed by the six quantum vectors specified by its six
representation labels. The faces of the tetrahedron are determined by the
representation labels of the four Clebsch-Gordon coefficients that define
the Racah coefficient.

\section{The State Sum}\label{state}

We are now in a position to define a Ponzano-Regge type state
sum\cite{PR} for the group $SU(1,1)$.

Let $\mathfrak{M}$ be a piecewise linear oriented 3-manifold with
given triangulation.  Then define the partition function as
\begin{equation}\label{statesum}
Z\left(\mathfrak{M}\right) = \dint_{\mathrm{edges
\mathcal{J}}}\,\mathfrak{D}\mathcal{J}
\prod_{\mathrm{tetrahedra }\tau}\left\{\begin{array}{ccc}
\mathcal{J}_1 & \mathcal{J}_2 & \mathcal{J}_{12}\\
\mathcal{J}_3 & \mathcal{J} & \mathcal{J}_{23}
\end{array}\right\}_\tau
\end{equation}

An appropriate Racah coefficient $\left\{\begin{array}{ccc}
\mathcal{J}_1 & \mathcal{J}_2 & \mathcal{J}_{12}\\
\mathcal{J}_3 & \mathcal{J} & \mathcal{J}_{23}
\end{array}\right\}_\tau$ is assigned to each tetrahedron in the
decomposition. The orientation of the faces of the tetrahedron
determines an orientation of the tetrahedron as a whole.  If the
orientation of the tetrahedron agrees with the orientation of the
manifold then the assigned Racah coefficient is taken to be
$\left\{\begin{array}{ccc} \mathcal{J}_1 & \mathcal{J}_2 & \mathcal{J}_{12}\\ \mathcal{J}_3 & \mathcal{J} & \mathcal{J}_{23}
\end{array}\right\}$, otherwise $\left\{\begin{array}{ccc}
\mathcal{J}_1 & \mathcal{J}_2 & \mathcal{J}_{12}\\ \mathcal{J}_3 & \mathcal{J} & \mathcal{J}_{23}
\end{array}\right\}^\star$ is used.

The product is taken over all tetrahedra $\tau$ appearing in the
simplicial decomposition of the manifold.  The sum (integral) is taken
over all irreducible representations in the principal series,
$\mathcal{J}$, weighted with the appropriate Plancherel measure
$\mu\left(\mathcal{J}\right)$, that label edges in the triangulation.

One should note that the trivial and exceptional discrete
representations do not appear in the sum (integral) over
representations since they have Plancherel measure zero. This is
fortunate since these representation labels would correspond to
tetrahedra with a light like edge, and would hence be degenerate.

\begin{thm}\label{StateSum}
$$
Z\left(\mathfrak{M}\right) = \dint_{\mathrm{edges
\mathcal{J}}}\,\mathfrak{D}\mathcal{J}
\prod_{\mathrm{tetrahedra }\tau}\left\{\begin{array}{ccc}
\mathcal{J}_1 & \mathcal{J}_2 & \mathcal{J}_{12}\\
\mathcal{J}_3 & \mathcal{J} & \mathcal{J}_{23}
\end{array}\right\}_\tau
$$
defines a formal invariant of closed,
oriented three manifolds
\end{thm}

Since equation \ref{statesum} is not necessarily finite in all
circumstances the term \emph{formal} invariant is used to distinguish
this Ponzano-Regge type `invariant' from the more rigorous family of
invariants defined in \cite{BW96}.

To prove this theorem one must show that equation \ref{statesum} is
invariant under the Pachner moves\cite{Pa91} that do not increase the
number of vertices in the PL manifold $\mathfrak{M}$ (see for instance
section 4 of \cite{BW96}). Using the identification above between
tetrahedra and Racah coefficients, the Pachner moves that do not
increase the number of vertices are precisely given by proposition
\ref{BE} and it is clear by inspection that 
equation \ref{statesum} is indeed invariant under it.

One must also show it does not depend on the isomorphism class of the
irreducible representations that label the edges and it does not
depend on the labelling of the manifold.  These will require the
machinary of section \ref{inv} for a proof.

Finally it is of interest to investigate the divergence in the
state sum, under the four to one Pachner move, caused by adding
an internal vertex as in figure \ref{4-1-pachner}. 

From equation \ref{4-1} (or alternatively by combining
Biedenharn-Elliot in proposition \ref{BE} and orthogonality in
proposition \ref{O}) one has the following identity for Racah
coefficients which corresponds to the four to one Pachner move
\begin{multline}\label{4-1-Pachner}
\dint
\,\mathfrak{D}\mathcal{J}_{1}
\,\mathfrak{D}\mathcal{J}_{2}
\,\mathfrak{D}\mathcal{J}_{23}\,\mathfrak{D}\mathcal{J}_{234}
\left\{\begin{array}{ccc}
\mathcal{J}_1 & \mathcal{J}_2 & \mathcal{J}_{12}\\ 
\mathcal{J}_3 & \mathcal{J}_{123} & \mathcal{J}_{23}
\end{array}\right\}
\left\{\begin{array}{ccc}
\mathcal{J}_1 & \mathcal{J}_{23} & \mathcal{J}_{123}\\
\mathcal{J}_4 & \mathcal{J} & \mathcal{J}_{234}
\end{array}\right\}\\
\left\{\begin{array}{ccc}
\mathcal{J}_2 & \mathcal{J}_3 & \mathcal{J}_{23}\\ 
\mathcal{J}_4 & \mathcal{J}_{234} & \mathcal{J}_{34}
\end{array}\right\}
\left\{\begin{array}{ccc}
\mathcal{J}_1 & \mathcal{J}_2 & \mathcal{J}_{12}\\
\mathcal{J}_{34} & \mathcal{J} & \mathcal{J}_{234}
\end{array}\right\}^\star
= \dint\,\mathfrak{D}\mathcal{J}_{1}
\,\mathfrak{D}\mathcal{J}_{2}
\left\{\begin{array}{ccc}
\mathcal{J}_{12} & \mathcal{J}_3 & \mathcal{J}_{123}\\ 
\mathcal{J}_4 & \mathcal{J} & \mathcal{J}_{34}
\end{array}\right\}\delta\left(0\right)
\end{multline}

While the right hand side is divergent, and correspondingly
meaningless, one can use this divergence to formally regularise
the state sum in equation \ref{statesum}.

Thus define the dimension of the category via the mathematically
meaningless quantity
\begin{equation}\label{quantdim}
\dim\mathcal{C} = 
\left(\dint\,\mathfrak{D}\mathcal{J}\right)^2\delta\left(0\right)
\end{equation}
Then, formally, one has that
\begin{equation}
Z\left(\mathfrak{M}\right) = \frac{1}{\left(
\dim\mathcal{C}\right)^\nu}
\dint\,\mathfrak{D}\mathcal{J}
\prod_{\tau}\left\{\begin{array}{ccc}
\mathcal{J}_1 & \mathcal{J}_2 & \mathcal{J}_{12}\\
\mathcal{J}_3 & \mathcal{J} & \mathcal{J}_{23}
\end{array}\right\}_\tau
\end{equation}
is invariant under both the three to two and the four to one Pachner
moves, where $\nu$ is the number of vertices in the simplicial
decomposition.  While these are very formal manipulations the form of
equation \ref{quantdim} indicates the quantity that must be made
finite in order to have a finite state sum.

\section{Labelling Invariance and Spin Networks} \label{inv}

The main purpose of this section is to prove the following theorems
needed for section \ref{state}.

\begin{figure}[htb!]
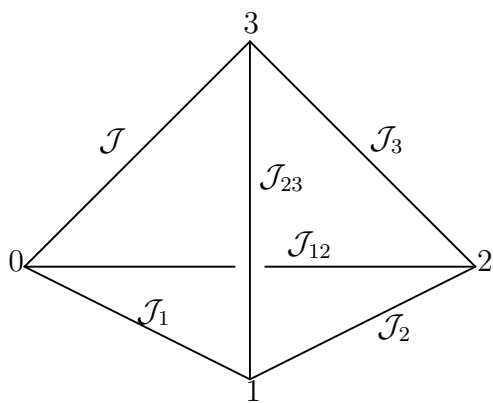

\begin{center}
\begin{texdraw}
\drawdim{mm}
\lvec(-30 15) \move(-15 7) \htext{$\mathcal{J}_1$} 
\move(0 0) \lvec(30 15) \move(17 5) 
\htext{$\mathcal{J}_2$} \move(30 15)
\lvec(2 15)\move(-2 15)\lvec(-30 15) 
\move(5 16) \htext{$\mathcal{J}_{12}$} \move(-30 15)
\lvec(0 45) \move(-20 30) \htext{$\mathcal{J}$} \move(0 45) 
\lvec(0 0) \move(1.3 25) \htext{$\mathcal{J}_{23}$} \move(0 0)
\move(0 45) \lvec(30 15) \move(16 30)\htext{$\mathcal{J}_3$}
\move(-0.5 -3) \htext{1}
\move(-32 14.2) \htext{0}
\move(30.3 14.2) \htext{2}
\move(-0.8 46) \htext{3}
\end{texdraw}
\end{center}
\caption{A labelled tetrahedron\label{tetrahedron}}
\end{figure}

\begin{thm}\label{label-inv}
Let the $\su{1,1}$ Racah coefficient $\left\{\begin{array}{ccc}
\mathcal{J}_1 & \mathcal{J}_2 & \mathcal{J}_{12}\\
\mathcal{J}_3 & \mathcal{J} & \mathcal{J}_{23}
\end{array}\right\}$ be defined by the geometric tetrahedron
with vertices labelled $0$, $1$, $2$ and $3$ and edges labelled
$e_{01}$, $e_{12}$, $e_{23}$, $e_{02}$, $e_{13}$ and $e_{03}$. The
following gives the correspondence between the two labelling conventions
(see figure \ref{tetrahedron})
\begin{eqnarray}
e_{01} = \mathcal{J}_1 & e_{12} = \mathcal{J}_2 & e_{23} =
\mathcal{J}_3\nonumber\\ e_{02} = \mathcal{J}_{12} & e_{13} =
\mathcal{J}_{23} & e_{03} = \mathcal{J}\nonumber
\end{eqnarray}
with $e_{10} = \hat{\mathcal{J}}_1$, etc where $\hat{ }$ denotes the dual
representation to $\mathcal{J}$.

Then the value of the Racah coefficient is invariant up to phase under
permutations of $0$, $1$, $2$ and $3$ which label the $e_{ij}$'s.
\end{thm}
For spherical categories this result is given by theorem 3.9 in
\cite{BW96}.

\begin{thm}\label{isoclass}
Let $a$, $b$, $c$, $d$, $e$ and $f$ be irreducible Principal series
representations of \su{1,1} and let $\phi_a : a\rightarrow a^\prime$,
$\dots\;\phi_f : f\rightarrow f^\prime$ be a series of isomorphisms
with irreducible Principal series representations of \su{1,1}.  Then
$$
\left\{\begin{array}{ccc}
a & b & c\\ d & e & f\end{array}\right\} =
\left\{\begin{array}{ccc}
a^\prime & b^\prime & c^\prime\\ 
d^\prime & e^\prime & f^\prime\end{array}\right\}
$$
\end{thm}
For spherical categories this result is given by proposition 3.3 of
\cite{BW96} and the proof is the same for here and so omitted.

In order to prove theorem \ref{label-inv} one must adapt some of the
structure maps of spherical categories\cite{BW96} to the non-spherical
category of unitary representations of \su{1,1}

For theorem \ref{label-inv}, firstly note that the invariance of the
Racah coefficient under relabellings (combinatorial isomorphisms) is
equivalent to being able to permute the representations in the space
of homomorphisms $\hom{a\otimes b}{c}$ as in \cite{BW96}.  Indeed any
permutation of the four labels labelling the vertices decomposes as
four permutations of the three labels defining each face, defining a
relabelling of that face, and a permutation of the four faces.

For example, consider the tetrahedron as labelled in figure
\ref{tetrahedron} and the permutation of vertices, $\Phi$, that takes
$\left(0,1,2,3\right)$ to $\left(1,3,2,0\right)$.

$
\left\{\begin{array}{ccc}
e_{01} & e_{12} & e_{02}\\
e_{23} & e_{03} & e_{13}
\end{array}\right\}
$ may be regarded as a map from 
\begin{multline}\label{perm1}
\tau_{0123} = 
\hom{e_{01}\otimes e_{12}}{e_{02}}\otimes
\hom{e_{02}\otimes e_{23}}{e_{03}}\\
\otimes\hom{e_{03}}{e_{01}\otimes e_{13}}\otimes
\hom{e_{13}}{e_{12}\otimes e_{23}}
\end{multline}
to $\C$ as in equation \ref{2p}. Applying the combinatorial
isomorphism to equation \ref{perm1} one gets the following space
\begin{multline}
\tau_{0123}^\prime = 
\hom{e_{13}\otimes\hat{e}_{23}}{e_{12}}\otimes
\hom{e_{12}\otimes\hat{e}_{02}}{\hat{e}_{01}}\\
\otimes\hom{\hat{e}_{01}}{e_{13}\otimes \hat{e}_{03}}\otimes
\hom{\hat{e}_{03}}{\hat{e}_{23}\otimes\hat{e}_{02}}
\end{multline}
for which the Racah coefficient 
$\left\{\begin{array}{ccc}
e_{13} & \hat{e}_{23} & e_{12}\\
\hat{e}_{02} & \hat{e}_{01} & \hat{e}_{03}
\end{array}\right\}$
defines a map to $\C$. But $\tau_{0123}$ is isomorphic to
$\tau_{0123}^\prime$ if the following two propositions are true
\begin{prop}\label{dual}
$$\hom{e_{01}\otimes e_{12}}{e_{02}}\equiv
\hom{\hat{e}_{02}}{\hat{e}_{12}\otimes\hat{e}_{01}}
$$\end{prop}
and
\begin{prop}\label{reorder}
$$\hom{e_{01}\otimes e_{12}}{e_{02}}\equiv
\hom{e_{\sigma\left(0\right)\sigma\left(1\right)}
\otimes e_{\sigma\left(1\right)\sigma\left(2\right)}}
{e_{\sigma\left(0\right)\sigma\left(2\right)}}$$
\end{prop}
for $\sigma$ some even permutation of $\left(0,1,2\right)$.  Note that
proposition \ref{reorder} has the same content as the relation between
Clebsch-Gordon coefficients given by equation
\ref{rotate}, while proposition \ref{dual} is equivalent to equation
\ref{reflect} (although in both cases there are now no ambiguities of
phase).  

To prove proposition \ref{reorder} one needs the analogue of the
$\epsilon$ maps for pivotal categories\cite{BW96} in order to permute
the representations in the Hom space.  For the case of $\su{2}$ one
has obvious maps, for any representation $a$ and $e$ the trivial
representation,
\begin{align*}
\epsilon\left(a\right) : e & \longrightarrow a\otimes\hat{a}\\
\lambda & \longmapsto\lambda\sum_n\ket{a,n}\otimes\bra{a,n}
\end{align*}
where \ket{a,n} is a basis for the carrier space of the representation
$a$, and
\begin{align*}
\epsilon^\star\left(a\right) : a\otimes\hat{a}&\longrightarrow e\\
v\otimes w &\longmapsto w\left(v\right) = v\left(w\right)
\end{align*}

\begin{figure}[hbt!]
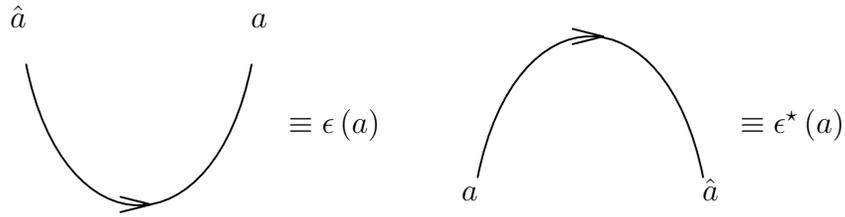

\vspace{2.5cm}
\begin{center}
\begin{texdraw}
\drawdim{mm}
\arrowheadtype t:V
\move(0 0)
\clvec(5 -25)(25 -25)(30 0)
\move(15.6 -18.6)\ravec(1 0)
\move(35 -10)\htext{$\equiv\epsilon\left(a\right)$}
\move(-2 5)\htext{$\hat{a}$}
\move(30 5)\htext{$a$}
\move(60 -15)
\clvec(65 10)(85 10)(90 -15)
\move(75.7 3.7)\ravec(1 0)
\move(95 -10)\htext{$\equiv\epsilon^\star\left(a\right)$}
\move(58 -18)\htext{$a$}
\move(90 -18)\htext{$\hat{a}$}
\end{texdraw}
\end{center}
\caption{A graphical representation of the $\epsilon$ maps
\label{epsilon}}
\end{figure}

The $\epsilon$ maps are traditionally represented diagramatrically as
`cup' and `cap' spin network operators as in figure \ref{epsilon} and
satisfy the following three axioms
\begin{defi}\label{defi-epsilon}
\begin{enumerate}
\item For every $f\in\hom{a}{b}$ the following commutes
$$
\begin{CD}
e @> \epsilon\left(a\right)>> a\otimes\hat{a}\\
@V\epsilon\left(b\right)VV  @VV f\otimes 1 V\\
b\otimes\hat{b} @>>1\otimes\hat{f}> b\otimes\hat{a}
\end{CD}
$$
The content of this is being able to slide a morphism $f$ around the
cup or cap by replacing it by its dual map $\hat{f}$.  This then
provides a definition of a dual morphism $\hat{f}$ from a morphism
$f$.
\item The map $\left(\epsilon\left(\hat{a}\right)\otimes 1\right)
\left(1\otimes\hat{\epsilon}\left(a\right)\right)$ in $\End{\hat{a}}$
is the identity.  Here one is asserting that a `wiggly' spin network
of two $\epsilon$ maps may be `straightened' out. 
\item $\epsilon\left(a\right)\left(1\otimes\epsilon\left(b\right)\otimes 1
\right) = \epsilon\left(a\otimes b\right)$ as maps in $\hom{e}{a\otimes b
\otimes\widehat{a\otimes b}}$.  In spin network terms one is able to
replace two nested cups or caps by two laying side by side.
\end{enumerate}
\end{defi}

This is uncontroversial for a group, such as \su{2}, where the trivial
representation occurs in the decomposition of $a\otimes\hat{a}$ since
the maps are simply the obvious projection on or injection from the
trivial representation from or to the decomposition of
$a\otimes\hat{a}$ in a direct sum.

However when this is not so, as it will be when dealing with the
unitary representations of \su{1,1}, one runs into immediate
difficulties.  Firstly, $\epsilon\left(a\right)$ is not a vector in
the Hilbert space $a\otimes\hat{a}$ since it has a divergent norm.
Also note that $\epsilon^\star\left(a\right)$ is not defined on every
vector in $a$; in particular
$$
v = \sum_n\frac{1}{n}\ket{a,n}\otimes\ket{\hat{a},n}
$$
is a vector in $a\otimes\hat{a}$, but it is clear
$\epsilon^\star\left(a\right)$ is divergent.  Thus the representations
of \su{1,1} do not form a pivotal category, although they are
sufficiently close to one that a state sum may still be defined.

\begin{figure}
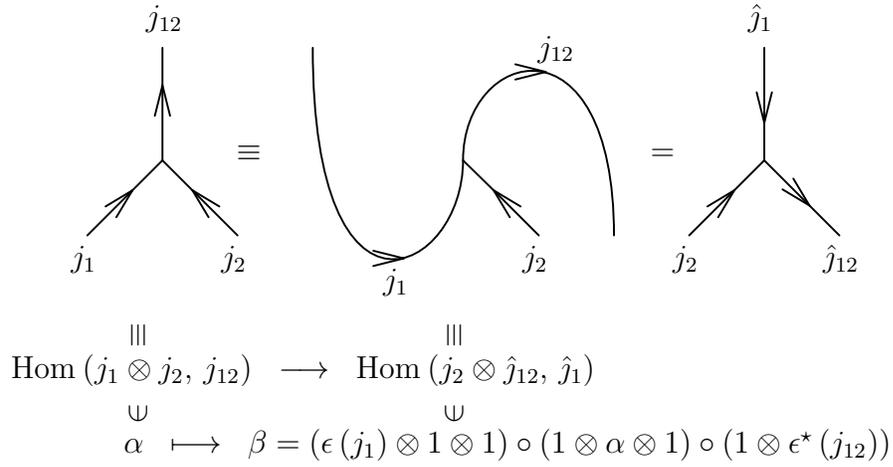

\begin{center}
\begin{texdraw}
\drawdim{mm}
\arrowheadtype t:V
\move(0 -20)\avec(0 -10)\lvec(0 -5)
\move(-2 -3)\htext{$j_{12}$}
\move(-10 -30)\avec(-4 -24)\lvec(0 -20)
\move(-12 -35)\htext{$j_1$}
\move(10 -30)\avec(4 -24)\lvec(0 -20)
\move(8 -35)\htext{$j_2$}
\move(10 -20)\htext{$\equiv$}
\move(40 -20)\clvec(40 -5)(60 0)(60 -30)
\move(50.9 -8.2)\avec(51 -8.2)
\move(50 -7)\htext{$j_{12}$}
\move(40 -20)\clvec(40 -35)(20 -45)(20 -5)
\move(32 -33)\avec(32.1 -33)
\move(29.5 -38)\htext{$j_1$}
\move(50 -30)\avec(44 -24)\lvec(40 -20)
\move(48 -35)\htext{$j_2$}
\move(65 -20)\htext{$=$}
\move(80 -5)\avec(80 -15)\lvec(80 -20)
\move(78 -3)\htext{$\hat{\jmath}_1$}
\move(70 -30)\avec(76 -24)\lvec(80 -20)
\move(68 -35)\htext{$j_2$}
\move(80 -20)\avec(86 -26)\lvec(90 -30)
\move(88 -35)\htext{$\hat{\jmath}_{12}$}
\move(-2 -45)\vtext{$\equiv$}
\move(40 -45)\vtext{$\equiv$}
\move(-5 -60)\htext{$\alpha\;\;\longmapsto\;\;
\beta = \left(\epsilon\left(j_1\right)\otimes 1\otimes 1\right)
\circ\left(1\otimes\alpha\otimes 1\right)
\circ\left(1\otimes\epsilon^\star\left(j_{12}\right)\right)$}
\move(-2 -55)\vtext{$\in$}
\move(40 -55)\vtext{$\in$}
\move(-20 -50)\htext{$\hom{j_1\otimes j_2}{j_{12}}\;\;
\longrightarrow\;\;
\hom{j_2\otimes\hat{\jmath}_{12}}{\hat{\jmath}_{1}}$}
\end{texdraw}
\caption[Permutations of Hom spaces are isomorphic]{An epsilon
map implementing an isomorphism of Hom spaces for \su{2}\label{hom-perm}}
\end{center}
\end{figure}

If one can define the epsilon maps then these can be used to implement
permutations of Hom spaces as discussed in \cite{BW96}, graphically
one such isomorphism is depicted in figure \ref{hom-perm}.  Thus one
doesn't really need the $\epsilon$ maps to be well defined
independently of each other; all one really needs is for them to be
well defined in certain combinations.

Let $\alpha\in\hom{j_1\otimes j_2}{j_{12}}$ and
$\beta\in\hom{j_2\otimes\hat{\jmath}_{12}}{\hat{\jmath}_{1}}$. 
$$
\alpha: j_1\otimes j_2\longrightarrow  j_{12}
$$
is explicitly realised as 
$$
\ket{j_{12},m_{12}}  = \sum_{m_1,\,m_2}\left[\begin{array}{ccc}
j_1 & j_2 & j_{12}\\m_1 & m_2 & m_{12}\end{array}\right]
\ket{j_1,m_1}\otimes\ket{j_2,m_2}
$$

Define $\beta$ formally as in figure \ref{hom-perm} so that
\begin{multline}
\ket{\hat{\jmath}_1,-m_1} 
\xrightarrow{1\otimes\epsilon^\star\left(j_{12}\right)} 
\sum_{m_{12}}\ket{\hat{\jmath}_1,-m_1}
\otimes\ket{j_{12},m_{12}}\otimes\ket{\hat{\jmath}_{12},-m_{12}}\\
\xrightarrow{1\otimes\alpha\otimes 1} 
\sum_{\substack{m_{12}\\m_1^\prime,\,m_2}}
\left[\begin{array}{ccc}
j_1 & j_2 & j_{12}\\m_1^\prime & m_2 & m_{12}\end{array}\right]
\ket{\hat{\jmath}_1,-m_1}\otimes\ket{j_1,m_1^\prime}\otimes\ket{j_2,m_2}
\otimes\ket{\hat{\jmath}_{12},-m_{12}}\\
\xrightarrow{\epsilon\left(j_1\right)\otimes 1\otimes 1} 
\sum_{m_{12},\,m_2}
\left[\begin{array}{ccc}
j_1 & j_2 & j_{12}\\m_1 & m_2 & m_{12}\end{array}\right]
\ket{j_2,m_2}\otimes\ket{\hat{\jmath}_{12},-m_{12}}
\end{multline}

Now although the two intermediate steps do not give decompositions of
the respective spaces in terms of vectors in $\hat{\jmath}_1$, the
final result is a vector in the Hilbert space $\hat{\jmath}_1$.
Indeed the orthogonality relations for Clebsch-Gordon coefficients
guarantee it has finite norm.  It is clear this also defines the
Clebsch-Gordon coefficient that implements $\beta$ and so one has
recovered equation
\ref{rotate} (although with no ambiguity in phase this time)
$$
\left[\begin{array}{ccc}
j_1 & j_2 & j_{12}\\m_1 & m_2 & m_{12}\end{array}\right] = 
\left[\begin{array}{ccc}
j_2 & \hat{\jmath}_{12} & \hat{\jmath}_1\\
m_2 & -m_{12} & -m_1\end{array}\right]
$$
and proved proposition \ref{reorder}.  Proposition \ref{dual} is
similar, except here one uses three epsilon maps to reverse the
direction of every leg on the trivalent vertex.  Thus define
$\gamma\in\hom{\hat{\jmath}_{12}}
{\hat{\jmath}_1\otimes\hat{\jmath}_2}$ by
\begin{multline*}
\ket{\hat{\jmath}_1,-m_1}\otimes\ket{\hat{\jmath}_2,-m_2} 
\xrightarrow{1\otimes\epsilon^\star\left(j_{12}\right)\otimes 1} 
\sum_{m_{12}}\ket{\hat{\jmath}_1,-m_1}
\otimes\ket{j_{12},m_{12}}\otimes\ket{\hat{\jmath}_{12},-m_{12}}
\otimes\ket{\hat{\jmath}_2,-m_2}\\
\xrightarrow{1\otimes1\otimes\alpha\otimes 1} 
\sum_{\substack{m_{12}\\m_1^\prime,\,m_2^\prime}}
\left[\begin{array}{ccc}
j_1 & j_2 & j_{12}\\m_1^\prime & m_2^\prime & m_{12}\end{array}\right]
\ket{\hat{\jmath}_1,-m_1}\otimes\ket{j_1,m_1^\prime}
\otimes\ket{j_2,m_2^\prime}\\
\otimes\ket{\hat{\jmath}_{12},-m_{12}}\otimes\ket{\hat{\jmath}_2,-m_2}\\
\end{multline*}
\begin{multline*}
\longrightarrow\sum_{\substack{m_{12}\\m_1^\prime,\,m_2^\prime}}
\left[\begin{array}{ccc}
j_1 & j_2 & j_{12}\\m_1^\prime & m_2^\prime & m_{12}\end{array}\right]
\ket{\hat{\jmath}_1,-m_1}\otimes\ket{j_1,m_1^\prime}
\otimes\ket{j_2,m_2^\prime}\\
\otimes\ket{\hat{\jmath}_2,-m_2}\otimes\ket{\hat{\jmath}_{12},-m_{12}}\\
\xrightarrow{\epsilon\left(\hat{\jmath}_1\right)
\otimes\epsilon\left(j_2\right)\otimes 1} 
\sum_{m_{12}}
\left[\begin{array}{ccc}
j_1 & j_2 & j_{12}\\m_1 & m_2 & m_{12}\end{array}\right]
\ket{\hat{\jmath}_{12},-m_{12}}
\end{multline*}

Again while the intermediate steps give `vectors' with divergent
norms, the final result is a convergent vector in
$\hat{\jmath}\otimes\hat{\jmath}_2$ and the proof of proposition
\ref{dual} by showing 
$$
\left[\begin{array}{ccc}
j_1 & j_2 & j_{12}\\m_1 & m_2 & m_{12}\end{array}\right] = 
\left[\begin{array}{ccc}
\hat{\jmath}_1 & \hat{\jmath}_2 & \hat{\jmath}_{12}\\
-m_1 & -m_2 & -m_{12}\end{array}\right]^\star
$$
as required.

This completes the proof of theorem \ref{label-inv}.

A trace, $\tr$, is a map from each endomorphism, $f$, of an object $a$
to the endomorphisms of the trivial object (representation) $e$ given
in two different ways
$$
f\mapsto\tr_L\, f \text{ or }\tr_R\, f
$$
explicitly by
$$
\tr_R\,f = \epsilon\left(a\right)\circ\left(f\otimes 1\right)
\circ\epsilon^\star\left(a\right)
$$
and
$$
\tr_L\,f = \epsilon\left(\hat{a}\right)\circ\left(1\otimes f\right)
\circ\epsilon^\star\left(\hat{a}\right)
$$
For the spherical categories in \cite{BW96} these so-called left and
right traces are required to be equal so it makes sense to just talk
of a trace.  The trace of the identity morphism of $a$ is then defined
as the (quantum) dimension $\mathrm{dim}\left(a\right)$ of $a$

In the example of \su{2} one is looking to assign a number to each
endomorphism of a representation $V$ of \su{2}; this canonical trace
is just the obvious trace of matrices and $\mathrm{dim}\left(a\right)$
is the dimension of the representation under consideration.

It is clear that this would not be satisfactory for \su{1,1} since the
dimension of every unitary representation is necessarily infinite, so
a more subtle prescription is needed.  This has already been used
implicitly in the definition of the \su{1,1} Racah coefficient in
equation \ref{RAC} in section \ref{Racah}.  

\begin{figure}
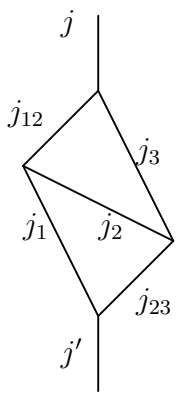

\begin{center}
\begin{texdraw}
\drawdim{mm}
\move(0 -10)\lvec(0 0)\lvec(-10 20)\lvec(0 30)\lvec(0 40)
\move(0 0)\lvec(10 10)\lvec(0 30)\move(10 10)\lvec(-10 20)
\move(-5 -7)\htext{$j^\prime$}
\move(-10 10)\htext{$j_1$}
\move(0 10)\htext{$j_2$}
\move(5 20)\htext{$j_3$}
\move(5 0)\htext{$j_{23}$}
\move(-12 25)\htext{$j_{12}$}
\move(-5 37)\htext{$j$}
\end{texdraw}
\end{center}
\caption{A tetrahedral spin network\label{tetnet}}
\end{figure}

Consider the definition of the Racah coefficient in the \su{2} case;
in terms of spin networks one talks of the closure of the so-called
tetrahedral net in figure \ref{tetnet} by taking the trace in the
$j$-th representation.  The value given is then defined to be the
Racah coefficient $\left\{\begin{array}{ccc} j_1 & j_2 & j_{12}\\ j_3
& j & j_{23} \end{array}\right\}$

Thus the value of the operator given by the spin network in figure
\ref{tetnet} must be
\begin{equation}\label{RR1}
\frac{\delta_{j,j^\prime}}{\mathrm{dim}\left(j\right)}
\left\{\begin{array}{ccc} j_1 & j_2 & j_{12}\\ j_3
& j & j_{23} \end{array}\right\} \mathbb{I}_j
\end{equation}
where $\mathbb{I}_j$ is the identity in $\End{j}$.  One should compare
this equation with equation \ref{RAC} and note that the Plancherel
measure in equation \ref{RAC} is playing the role of the dimension
weighting factor in equation \ref{RR1}.

Thus we have defined a pseudo-trace on \su{1,1} that maps the identity
endomorphism for each irreducible\footnote{Irreducible representations
are the simple representations of \cite{BW96}.  The semi-simple
condition in definition 2.10 of \cite{BW96} is then just an assertion
that the Plancherel decomposition exists.}  $a$ to the morphism of the
trivial representation given by the Plancherel measure
$\mu\left(a\right)$.

\begin{defi}
For a morphism $f\in\hom{a}{a^\prime}$, $f = \lambda\mathbb{I}_a$ for
$\lambda\in\C$ define the pseudo-trace $\Tilde{\tr}$ to be
$$
\tilde{\tr} f = 
\lambda\,\mu\left(a\right)
\delta\left(\mathcal{A},\mathcal{A}^\prime\right)
\text{ for $a$ irreducible}\\
$$
\end{defi}
This is certainly a trace in respect of being able to permute linear
maps.  Let $f,\,g,\,h\in\End{a}$ then it is clear one must have
$$
\Tilde{\tr}\left( fgh\right) = \Tilde{\tr}\left( hfg\right)
$$
simply by virtue of every $f\in\End{a}$ being a multiple of the
identity.

We may use this to define a pairing along the lines of definition 2.7
in \cite{BW96}
\begin{prop}
The pseudo trace defines a non degenerate pairing
$\Theta\left(f,g\right)$ for $f\in\hom{a}{b}$, $g\in\hom{b}{a}$, on
morphisms of Principal series representations of
\su{1,1} given by
$$
\Theta\left(f,g\right) = \Tilde{\tr}\left(fg\right) = 
\Tilde{\tr}\left(gf\right)
$$
for $a$ and $b$ irreducible.
\end{prop}
That it is non degenerate is trivial since the Plancherel measure is
non zero for every irreducible unitary representation in the Principal
series of \su{1,1}.

One should note that our pseudo trace is undefined on non irreducible
representations (in particular tensor product representations) so one
isn't able to cut a closed spin network arbitrarily as is possible for
\su{2} spin networks.   

The pseudo-trace allows the Racah coefficient to be written in the
following fashion, as in \cite{BW96},
\begin{multline}
\left\{\begin{array}{ccc}
a & b & c\\ d & e & f
\end{array}\right\} :
\hom{e}{d\otimes c}\otimes\hom{d\otimes b}{f}\\
\otimes\hom{f}{e\otimes a}\otimes\hom{c\otimes a}{b}
\longrightarrow\C\\
\alpha\otimes\beta\otimes\gamma\otimes\delta\longmapsto\Tilde{\tr}
\left(\gamma\left(\alpha\otimes 1\right)\left(1\otimes\delta\right)
\beta\right)
\end{multline}

and the proof of theorem \ref{isoclass} follows exactly as in
\cite{BW96} since one only requires the ability to permute maps inside
the trace, providing one avoids worrying about the intermediate cases
where the trace is over a non irreducible representation.  This is
similar to our use of the $\epsilon$ maps to define the permutation
invariance of the Clebsch-Gordon coefficients where intermediate steps
made only formal mathematical sense, even though the final result was
meaningful.

This then completes the proof of theorem \ref{StateSum} in section
\ref{state}.

\chapter{Asymptotics of the Discrete Series}\label{ASYMPT}

In this chapter the connections between the \su{2} Ponzano-Regge
state sum model and the \su{1,1} state sum model of the previous
chapter are developed in order to derive an asymptotic formula for the
special case of the \su{1,1} Racah coefficient where all
representations are in the positive discrete series.

In section \ref{ext} the notion of an \emph{extended} Racah
coefficient is introduced within which both the \su{2} and \su{1,1}
Racah coefficients are contained as specific regions of the more
general coefficient.  Section \ref{geom} is devoted to a detailed
exploration of the geometry associated to the \su{1,1} region of the
generalised Racah coefficient and investigation of the symmetries of
the generalised coefficient that map between the \su{1,1} and \su{2}
regions.  Finally the asymptotic formula is stated and proved in
section \ref{asymp}.

\section{Extensions of 6j Symbols} \label{ext}

In \cite{Pon74_3} and \cite{Pon74_6} the symmetries of \su{2} 3j
(Clebsch-Gordon) and 6j (Racah) coefficients were extended beyond the
usual symmetries, which respect the triangle inequality, to a new
domain, which satisfies an anti-triangle inequality. The extension of
the Racah coefficient is related to the Racah coefficient for the
discrete unitary representation series of \su{1,1}.

To be more precise, the extension of the 3j symbol discussed in
\cite{Pon74_3} corresponds, within a phase, to the explicitly
calculated 3j symbol for the coupling of two elements of the discrete
series of SU(1,1) given in \cite{HoBi66}.  For the Racah coefficient,
the regions associated with the extension to anti-triangle
inequalities, discussed in \cite{Pon74_6}, have been conjectured to be
related to the Racah coefficient for the discrete unitary
representation series of SU(1,1).  

In this section we shall explicitly compute a transformation of the
Racah coefficient to the region conjectured to be associated to these
discrete unitary representations using the symmetries in
\cite{Pon74_6}. We start with some definitions.

\begin{defi}
We shall use the symbol $\left| \begin{array}{ccc} a & b & c \\ d & e
& f \end{array} \right|_{SU(2)}$ to denote an ordered set of real
numbers for which the ordered sets of real numbers $|abc|_{SU(2)}$,
$|cde|_{SU(2)}$, $|afe|_{SU(2)}$ and $|bdf|_{SU(2)}$ each satisfy
mutual triangle inequalities (that is $\pm a\pm b\pm c\ge 0$ where two
plus signs are chosen).  We shall use the symbol $\left|
\begin{array}{ccc} a & b & c \\ d & e & f \end{array}
\right|_{SU(1,1)}$ in a similar way, but here $|abc|_{SU(1,1)}$,
etc. satisfy $c\ge a+b+1$, $a\le b+c$ and $b\le a+c$ instead of mutual
triangle inequalities. Both will satisfy the sum of the three elements
being at least -1.\footnote{ For the symbols $|abc|_{SU(2)}$, etc this
last condition is redundant since one can show that the mutual
triangle inequalities imply the non negativity of $a$, $b$ and $c$}
\end{defi}

\begin{defi}
The Racah coefficient defines a map\footnote{strictly the Racah
coefficient is only defined for certain combinations of non negative
positive half integers but here it is considered to be zero outside 
this domain of definition so one may consider it defined on the whole
of $\R^6$.}
$$
\R^6\to\R
$$
given by
$$
\left| \begin{array}{ccc}
a & b & c \\ d & e & f \end{array} \right|_{SU(2)}
\mapsto
\left\{ \begin{array}{ccc}
a & b & c \\ d & e & f \end{array} \right\}_{SU(2)}
$$
while what we shall call the \emph{extension} defines another map $\R^6\to\R$
given by
$$
\left| \begin{array}{ccc}
a & b & c \\ d & e & f \end{array} \right|_{SU(1,1)}\mapsto
\left\{ \begin{array}{ccc}
a & b & c \\ d & e & f \end{array} \right\}_{ext}
$$
The details of these two maps will be given later. 
\end{defi}

\begin{defi} \label{S}
Define a map $S:\R^6 \to\R^6$ 
\begin{align}
a & = \frac{1}{2}\left(a^\prime+b^\prime-d^\prime+e^\prime\right) \label{b1} \\
b & = \frac{1}{2}\left(-a^\prime-b^\prime-d^\prime+e^\prime\right)-1\label{b2} \\
c & = c^\prime\label{b3} \\
d & = \frac{1}{2}\left(-a^\prime+b^\prime+d^\prime+e^\prime\right) \label{b4} \\
e & = \frac{1}{2}\left(a^\prime-b^\prime+d^\prime+e^\prime\right) \label{b5} \\
f & = f^\prime\label{b6}
\end{align}
It should be noted that if one shifts all
the values of the variables by $+\frac{1}{2}$ then this transformation is an
\emph{orthogonal} linear map. The inverses to
equations \ref{b1} - \ref{b6} are
\begin{align}
a^\prime & = \frac{1}{2}\left(a-b-d+e-1\right) \label{a} \\
b^\prime & = \frac{1}{2}\left(a-b+d-e-1\right) \label{b} \\
c^\prime & = c  \label{c} \\
d^\prime & = \frac{1}{2}\left(-a-b+d+e-1\right) \label{d} \\
e^\prime & = \frac{1}{2}\left(a+b+d+e+1\right) \label{e} \\
f^\prime & = f  \label{f}
\end{align}
\end{defi}

\begin{prop}
For $S$ defined in definition \ref{S} we have
\begin{equation}
S:\left| \begin{array}{ccc}
a^\prime & b^\prime & c^\prime \\ d^\prime & e^\prime & f^\prime
\end{array} \right|_{SU(1,1)} \to
\left| \begin{array}{ccc}
a & b & c \\ d & e & f \end{array} \right|_{SU(2)}
\end{equation}
\end{prop}

To prove this, consider the map acting on the ordered sets
$|abc|_{SU(2)}$ associated to $\left| \begin{array}{ccc}
a & b & c \\ d & e & f \end{array} 
\right|_{SU(2)}$. We find
\begin{align}
a+b-c & = e^\prime-d^\prime-c^\prime-1\label{eq1}\\
a-b+c & = a^\prime+b^\prime+c^\prime+1\\
-a+b+c & = -a^\prime-b^\prime+c^\prime-1\\
a+b+c+1 &= e^\prime-d^\prime+c^\prime
\end{align} \begin{align}
c+d-e & = c^\prime+b^\prime-a^\prime\\
c-d+e & = a^\prime-b^\prime+c^\prime\\
-c+d+e & = d^\prime+e^\prime-c^\prime\\
c+d+e+1 & = e^\prime+d^\prime+c^\prime+1\label{eq8}
\end{align} 

\begin{figure}[htb]
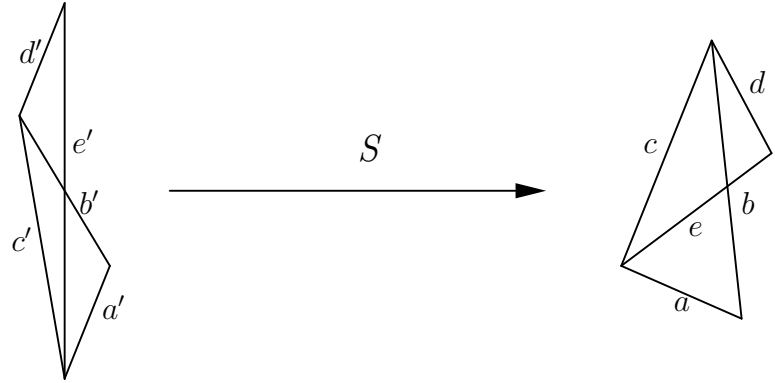

\begin{center}
\begin{texdraw}
\drawdim{mm}
\lvec(0 50) \rmove(1 -20) \htext{$e^\prime$} \rmove(-1 20) 
\rlvec(-6 -15) \rmove(0 7) \htext{$d^\prime$} \rmove(0 -7)
\rlvec(6 -35) \rmove(-7 17) \htext{$c^\prime$} \rmove(7 -17) 
\rlvec(6 15) \rmove(-1 -7) \htext{$a^\prime$} \rmove(1 7)
\rlvec(-12 20) \rmove(8 -13) \htext{$b^\prime$} \rmove(-8 13)

\rmove(20 -10)
\arrowheadtype t:F 
\ravec(50 0) \rmove(-50 0) 
\rmove(25 4) \htext{\textit{\large S}} \rmove(-25 -4)
\rmove(-20 10)

\rmove(80 -20)
\rlvec(20 15) \rmove(-11 -11) \htext{$e$} \rmove(11 11)
\rlvec(-8 15) \rmove(5 -7) \htext{$d$} \rmove(-5 7)
\rlvec(-12 -30) \rmove(3 15) \htext{$c$} \rmove(-3 -15)
\rlvec(16 -7) \rmove(-9 1) \htext{$a$} \rmove(9 -1)
\rlvec(-4 37) \rmove(4 -23) \htext{$b$} 

\end{texdraw}
\end{center}
\caption{A graphic representation of equations \ref{eq1} - \ref{eq8}\label{halftrans}}
\end{figure}

One should note that equations \ref{eq1} - \ref{eq8} specify a
transformation of five of the six variables amongst
themselves. Geometrically we may associate triangles, for some choice
of metric, to each symbol $|abc|$ and can, thus, show the above
equations graphically in figure \ref{halftrans} where the left hand
side is embedded into a space with a Minkowski signature metric and
the edges are regarded as time like vectors. We shall discuss the
geometry in more detail in section \ref{geom}.

Eight similar equations may be derived connecting $a,b,d,e,f$ and
$a^\prime,b^\prime,d^\prime,e^\prime,f^\prime$ to which may be
associated a very similar geometry to figure \ref{halftrans}. Here $f
= f^\prime$ is the shared edge.

The left hand side of equations \ref{eq1} - \ref{eq8}, and the
analogous equations connecting $a,b,d,e,f$ to
$a^\prime,b^\prime,d^\prime,e^\prime,f^\prime$, being positive is
equivalent to the symbol $\left| \begin{array}{ccc} a & b & c \\ d & e
& f \end{array} \right|_{SU(2)}$ being defined, while positivity of
the right hand side is equivalent to the symbol $\left|
\begin{array}{ccc} a^\prime & b^\prime & c^\prime \\ d^\prime &
e^\prime & f^\prime \end{array} \right|_{SU(1,1)}$ being defined.  So
the map is well defined and by definition the following anti-triangle
inequalities are enforced
\begin{align}
c^\prime \ge & a^\prime+b^\prime+1 \label{tri1} \\
e^\prime \ge & d^\prime+c^\prime+1 \label{tri2} \\
e^\prime \ge & a^\prime+f^\prime+1 \label{tri3} \\
f^\prime \ge & b^\prime+d^\prime+1 \label{tri4} 
\end{align}

We may also define the extension $\left\{ \begin{array}{ccc} a^\prime
& b^\prime & c^\prime \\ d^\prime & e^\prime & f^\prime \end{array}
\right\}_{ext}$ of the SU(2) Racah coefficient to the anti-triangle
inequality domain via the map $S$.

\begin{defi}
\begin{equation}
\left\{ \begin{array}{ccc}
a^\prime & b^\prime & c^\prime \\
d^\prime & e^\prime & f^\prime \end{array} \right\}_{ext}
:= \left\{ \begin{array}{ccc}
a & b & c \\ d & e & f \end{array} \right\}_{SU(2)}
\end{equation}
where 
\begin{multline} \label{6jdef}
\left\{ \begin{array}{ccc}
a & b & c \\ d & e & f \end{array} \right\}_{SU(2)} =
\left(-1\right)^{a+b+d+e}\tilde{\Delta}(abc)\tilde{\Delta}(cde)
\tilde{\Delta}(bdf)\tilde{\Delta}(aef)\\ \times
\sum_n \frac{(-1)^n (n+1)!}{(n-a-b-c)!(n-c-d-e)!(n-b-d-f)!(n-a-e-f)!}\\ \times
\frac{1}{(a+b+d+e-n)!(a+c+d+f-n)!(b+c+e+f-n)!}
\end{multline}
and $\Delta(abc) = \sqrt{\frac{(a+b-c)!(a-b+c)!(-a+b+c)!}{(a+b+c+1)!}}$ 

When any of the factorials are undefined $\left\{ \begin{array}{ccc} a
& b & c
\\ d & e & f \end{array} \right\}_{SU(2)}$ is defined to be zero. 
This requirement ensures the sum over $n$ is finite, restricts the
indices to non negative half integers and ensures that $a+b+c$, etc
are always integer. \end{defi}

All symmetries of the `extended' Racah coefficient may be reduced to
permutations and sign changes in certain variables\cite{Pon74_6}.
Thus for the Racah coefficient $\left\{ \begin{array}{ccc} a & b & c
\\ d & e & f \end{array} \right\}$, we define the variables

\begin{align*}
s_1 & = a+d+1 & s_0 & = d-a \\
s_3 & = b+e+1 & s_2 & = e-b \\
s_5 & = c+f+1 & s_4 & = f-c 
\end{align*}

Then all permutations of the $s_i$, or sign changes of an even number
of the $s_i$, give the total number of extended symmetries of the
associated 6j symbol. The Regge symmetries\footnote{By which we mean
the 144 symmetries that preserve the triangle
inequalities}\cite{Reg59} correspond to permutations of $\left (
s_0,s_2,s_4\right)$ or $\left(s_1,s_3,s_5\right)$, and sign changes of
any two of $\left(s_0,s_2,s_4\right)$.

Let $s^\prime_{\rho\left(i\right)} = s_i$, then the symmetry that
corresponds to the map $S$ above is simply the following permutation,
$\rho$,
\begin{equation}\label{rho}
\rho =   
\left( \begin{array}{cccccc}
0 & 1 & 2 & 3 & 4 & 5 \\
0 & 2 & 3 & 1 & 4 & 5
\end{array} \right)
\end{equation}
and from equations \ref{tri1} - \ref{tri4} it is easy to see the
transformation $S$ takes us into the region characterised by anti
triangle inequalities, related to the Racah coefficient for the
discrete unitary representations of $SU(1,1)$.

Finally one may state the relationship between $\left\{
\begin{array}{ccc} a & b & c \\ d & e & f \end{array} \right\}_{ext}$
and the Racah coefficient for the discrete series as derived
in section \ref{rac} and given by equation \ref{RacahSU11}, $\left\{
\begin{array}{ccc} a & b & c \\ d & e & f \end{array} \right\}_1$.
\begin{thm} \label{extend-thm}
$$
\left\{\begin{array}{ccc} a & b & c \\ d & e & f \end{array} \right\}_{ext}
= \left\{\begin{array}{ccc} a+1 & b+1 & c+1 \\ d+1 & e+1 & f+1
\end{array}\right\}_1
$$
\end{thm}

The proof is to transform the Racah coefficient for the
$SU\left(1,1\right)$ positive discrete series in equation
\ref{RacahSU11}, after shifting its indices by one.  Note that the
shift in the indices is just another manifestation that the Casimir
for $\su{1,1}$ acts as $\sqrt{j(j-1)}$, while that for \su{2} acts as
$\sqrt{j(j+1)}$.

To derive the other twelve cases of coupling one simply notes that the
symmetries $\sigma_1$, $\sigma_2$ and $\sigma_3$, given respectively
by equations \ref{aut1}, \ref{aut2} and \ref{aut3}, may easily be
related to permutations in the $s_i$ (in fact they are just examples
of Regge symmetries).  The phase ambiguities may now be resolved
insisting these transformations hold exactly and not just up to phase.

Thus one finds $\sigma_1$ is just the symmetry in terms of the $s_i$
given by $s_i = s_{\rho_1\left(i\right)}$ where 
$$
\rho_1 =   
\left( \begin{array}{cccccc}
0 & 1 & 2 & 3 & 4 & 5 \\
-4 & 5 & -0 & 1 & 2 & 3
\end{array} \right)
$$
and we are using the shorthand $s_{-i} = -s_i$.  $\sigma_2$ is the
symmetry given in terms of the $s_i$ by $s_i =
s_{\rho_2\left(i\right)}$ where
$$
\rho_2 =   
\left( \begin{array}{cccccc}
0 & 1 & 2 & 3 & 4 & 5 \\
2 & 3 & 0 & 1 & 4 & 5
\end{array} \right)
$$
Finally $\sigma_3$ is the symmetry in terms of the $s_i$ given by $s_i
= s_{\rho_3\left(i\right)}$ where
$$
\rho_3 =   
\left( \begin{array}{cccccc}
0 & 1 & 2 & 3 & 4 & 5 \\
-0 & 1 & -2 & 3 & 4 & 5
\end{array} \right)
$$

This settles the claim of D'Adda, D'Auria and Ponzano, in
\cite{Pon74_6}, that the extension of the SU(2) Racah coefficient was
related to the SU(1,1) Racah coefficient for the discrete series and
demonstrates the exact relationship.

\section{Geometry} \label{geom}

We wish to explore the geometry of the extended Racah coefficients for
the SU(1,1) region. It is known (see
\cite{PR},\cite{BaFo93}) that the symbol
$\left| \begin{array}{ccc} a & b & c \\ d & e & f \end{array}
\right|_{SU(2)}$ may be identified with a Euclidean, or space like
Lorentzian, tetrahedron with edge lengths equal to $j_{12} =
a+\frac{1}{2}$, etc. Here a space like Lorentzian tetrahedra is one for
which all faces and all edges are space like.  From the results of
chapter \ref{Quantum-Geometry} and theorem \ref{extend-thm} in section
\ref{ext} one expects the $\su{1,1}$
region to be identified with a tetrahedron whose edges are all
time like and future pointing when the \su{2} region gives a standard
Euclidean tetrahedron.

We shall denote such tetrahedra by
$T\left(j_{12},j_{13},j_{14},j_{34},j_{24},j_{23}\right)$, and omit
the edge lengths when these are not relevant. We shall also use
subscripts, SU(2) and SU(1,1), to indicate the region the tetrahedron
is associated to when confusion can arise. Note that we shall impose
the requirement that the edge lengths in the symbol $T$ be positive
for the SU(1,1) case\footnote{ While $j_{12},j_{13}$, etc. are always
positive for $T_{SU(2)}$ the same cannot be said for $T_{SU(1,1)}$.
An easy counter example is gained by mapping a regular tetrahedron to
the SU(1,1) domain with equations \ref{a} - \ref{f}. So this
assumption is necessary.}.

\begin{figure}[htb!]
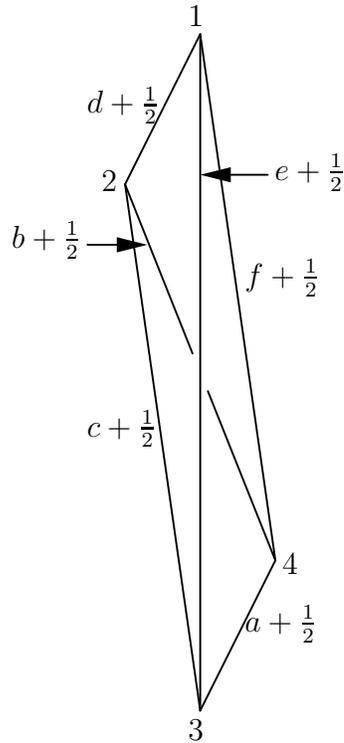

\begin{center}
\begin{texdraw}
\drawdim{mm}
\lvec(10 20) 
\move(6 9)\htext{$a+\frac{1}{2}$} 
\move(0 0) \lvec(0 90)
\move(10 69) \htext{$e+\frac{1}{2}$}
\move(9 71.3)
\arrowheadtype t:F \avec(0 71.3)
\move(0 0) \lvec(-10 70)
\move(-15 35) \htext{$c+\frac{1}{2}$}
\move(10 20) \lvec(0 90)
\move(6 55)  \htext{$f+\frac{1}{2}$}
\move(10 20) \lvec(1 42.5)
\move(-1 47.5) \lvec(-10 70)
\move(-25 60) \htext{$b+\frac{1}{2}$}
\move(-15 62)
\arrowheadtype t:F \avec(-7 62)
\move(-10 70) \lvec(0 90)
\move(-15 78)  \htext{$d+\frac{1}{2}$}
\move(-1.5 -4)\htext{3}
\move(-1.5 91)\htext{1}
\move(-13 69)\htext{2}
\move(11 18)\htext{4}
\end{texdraw}
\end{center}
\caption{A Lorentzian tetrahedron with all edges and all faces time like.
Time increases vertically up the page.\label{tltri}}
\end{figure}

To fix notation we shall denote the length of the edge $(h,k)$, formed
by deleting the $h$-$th$ and $k$-$th$ vertex (see figure \ref{tltri}),
as $j_{hk}$.  The area $A_h$ denotes the area of the face,
$\mathcal{T}_{h}$, obtained by deleting the $h$-th vertex from the
tetrahedron. It is clear we may associate a geometric triangle,
$\mathcal{T}$, to each symbol $|abc|$.

We shall denote by $\theta_{hk}$ the (exterior) dihedral angle on the
edge $(h,k)$ between the two outward normals of the faces
$\mathcal{T}_h$ and $\mathcal{T}_k$. In Euclidean space these are
always bone fide real angles; for Lorentzian space the situation is
more subtle since the `angles' can turn out to be complex. This
situation has been analysed in some detail in
\cite{BaFo93} and we shall say more about this in section \ref{Cayneg}.

Associated to each $T$ is a number, $V^2$, given by the
Cayley determinant which defines the volume squared of the tetrahedron.
\begin{equation}
V^2 = \frac{1}{2^3\left(3!\right)^2}
\left|\begin{array}{ccccc}
0 & j_{34}^2 & j_{24}^2 & j_{23}^2 & 1\\
j_{34}^2 & 0 & j_{14}^2 & j_{13}^2 & 1\\
j_{24}^2 & j_{14}^2 & 0 & j_{12}^2 & 1\\
j_{23}^2 & j_{13}^2 & j_{12}^2 & 0 & 1\\
1 & 1 & 1 & 1 & 0\\
\end{array}\right|
\end{equation}

$T_{SU(2)}$ is Euclidean if, and only if, the Cayley determinant is
positive and Minkowskian when it is negative. For edge lengths that
are positive half integers the Cayley determinant cannot vanish.

For $T_{SU(1,1)}$, as already stated, it may be identified with a
tetrahedron whose faces are time like and edges are either all
space like or all time like. These time like triangles have one `long'
side and two `short' sides. As such they obey anti-triangle
inequalities along the lines of $$ c \ge a + b
$$ where c is the `long' side. The normals to such triangles are
space like, and the triangles possess two interior `angles', which are
complex and may thus be identified with Lorentzian boosts as in
\cite{BaFo93} (opposite the `a' and `b' sides), with the third
interior angle being undefined\footnote{If the edges are time like this
interior angle would involve boosting from the future light cone to
the past light cone, which can't be done. If the edges are space like
it involves boosting through either the past, or future, light cone.},
and one exterior `angle' (for the vertex opposite the `c' side) which
may, again, be identified with a Lorentzian boost.  The area squared
defined by $A^2 =\frac{1}{16}\left(a+b+c\right)
\left(a+b-c\right)\left(a-b+c\right)\left(-a+b+c\right)$
is negative. The area, as in the triangle inequality case, may be
defined by taking the square root of the area squared, so that
$A=i\sqrt{\left|A^2\right|}$.

Equations \ref{tri1} - \ref{tri4} specify how to fit four such
time like triangles together.  The resulting object has one `super
long' edge ($j_{24}$), two `long' edges ($j_{14}$ and $j_{23}$) and
the remaining three are `short' edges. An embedding of such an object
into Minkowski space is shown in figure \ref{tltri}.

Figure \ref{tltri} is the general form for such a tetrahedron. If the
edges are time like there must be a strict time ordering (up to time
reversal) of the vertices.  Once we have chosen such an ordering (say
1,2,4,3\footnote{Our choice of numbering comes from attempting to
preserve conventions with
\cite{PR}} from future to past) the `super long' edge connects vertex 1 to
vertex 3, the two long edges connect vertex 1 to vertex 4 and vertex 2
to vertex 3, and the remaining vertices are connected by short edges.

One should note that if the symbol $\left| \begin{array}{ccc} a & b &
c \\ d & e & f \end{array} \right|_{SU(2)}$ has a `degenerate'
triangle (ie $a+b=c$ for some triangle $|abc|$) then the corresponding
tetrahedron, $T_{SU(2)}$ has an `almost degenerate' triangle, that is
$j_{12} + j_{13}= j_{14}+\frac{1}{2}$.  The +1's in equations
\ref{tri1} - \ref{tri4} ensure the same is true for the SU(1,1) case.

We now state a proposition relating $T_{SU(1,1)}$ and $T_{SU(2)}$.

\begin{prop} \label{p1}
Let $T\left(j_{12},j_{13},j_{14},j_{34},j_{24},j_{23}\right)_{SU(2)}$ and 
$T\left(j_{12}^\prime,j_{13}^\prime,j_{14}^\prime,
j_{34}^\prime,j_{24}^\prime,j_{23}^\prime\right)_{SU(1,1)}$
be related by equations \ref{a} - \ref{f}. 
 
Then the transformation preserves the Cayley determinant
and the product of the associated face areas.
\end{prop}
The proof is straightforward, if laborious, algebra.

\bigskip

There are now two geometric cases to consider depending on whether the
Cayley determinant is positive or negative.

\subsection{The case where $V^2 > 0$} \label{Caypos}

If the Cayley determinant is positive we choose an embedding of
$T_{SU(1,1)}$ in Lorentzian space with metric signature $(+,-,-)$ so
that the time like edges have a positive length squared.  Moreover,
since the normals to the faces span a space like plane, all the
dihedral angles are defined, in contrast to the space like case
discussed in \cite{BaFo93}.

We now wish to consider how the dihedral angles of the tetrahedra
transform under equations \ref{b1} - \ref{b6} in this case.  In
contrast to the Regge symmetries the sum of dihedral angles times edge
lengths does not remain constant.

\begin{thm} \label{dihedral_pos}
Under equations \ref{b1} - \ref{b6} the dihedral angles transform as:
\begin{align}
\theta_{12} & = \pi-\frac{1}{2}\left(\theta_{12}^\prime+\theta_{13}^\prime-
\theta_{34}^\prime+\theta_{24}^\prime\right) \label{a1}\\
\theta_{13} & = -\frac{1}{2}\left(-\theta_{12}^\prime-\theta_{13}^\prime-
\theta_{34}^\prime+\theta_{24}^\prime\right)\\
\theta_{14} & = \pi - \theta_{14}^\prime \\
\theta_{34} & = \pi - \frac{1}{2}\left(-\theta_{12}^\prime+\theta_{13}^\prime+
\theta_{34}^\prime+\theta_{24}^\prime\right) \\
\theta_{24} & = 2\pi - \frac{1}{2}\left(\theta_{12}^\prime-\theta_{13}^\prime+
\theta_{34}^\prime+\theta_{24}^\prime\right)\\
\theta_{23} & = \pi -\theta_{23}^\prime \label{a2}
\end{align}
for $V^2>0$.
\end{thm}

The proof involves the following Euclidean trigonometric relations
between dihedral angles and edge lengths:

\begin{align} 
-C_{rs} = 16 A_r A_s\cos\theta_{rs} & \;\;\;  & r\not= s \label{cos}\\
\frac{3}{2}Vj_{rs} = A_r A_s\sin\theta_{rs} & & r\not= s \label{sin}
\end{align}

where $j_{rs}$ is the shared side for the triangles whose areas are
given by $A_r$ and $A_s$, $\theta_{rs}$ is the (exterior) dihedral
angle between the outward normals to the faces $\mathcal{T}_r$ and
$\mathcal{T}_s$, and $C_{rs}$ is the $(r,s)$ algebraic minor of the
Cayley determinant formed by deleting the row and the column common to
the $(r,s)$ matrix entry. Note that equation \ref{sin} does not
distinguish exterior and interior dihedral angles, whereas equation
\ref{cos} does.

To derive equation \ref{sin} for the Lorentzian case one must choose a 
square root of the identity
\begin{equation}
V^2 = \frac{4 A_h^2 A_k^2}{9j_{hk}^2}\sin^2\theta_{hk} 
\;\;\;\;\;\;\;\;\;\;\;\;\;\;\;\;\;\;  h\not= k
\end{equation}
so that the dihedral angle has the correct range, that is 
$0\le\theta_{hk}\le\pi$.
Thus, since $\left(A_h^\prime\right)^2 < 0$, we must choose
\begin{equation}
V = \frac{2 |A^\prime_h| |A^\prime_k|}{3j^\prime_{hk}}\sin\theta^\prime_{hk}
\;\;\;\;\;\;\;\;\;\;\;\;\;\;\;\;\;\;  h\not= k\label{sin_lor}
\end{equation}

Now, since we want to use the fact that, from 
proposition \ref{p1}, 
\begin{equation}
 A_1 A_2 A_3 A_4 = A^\prime_1 A^\prime_2 A^\prime_3 A^\prime_4 = 
|A^\prime_1| |A^\prime_2| |A^\prime_3| |A^\prime_4|
\end{equation}
in the following proof,
we must rewrite equation \ref{cos} in a 
similar way. Thus, for $T_{SU(1,1)}$
\begin{align}
-C_{rs}^\prime = & 16 A^\prime_r A^\prime_s\cos\theta^\prime_{rs}\nonumber\\
= &  -16 |A^\prime_r| |A^\prime_s|\cos\theta^\prime_{rs}\nonumber\\
= & 16 |A^\prime_r| |A^\prime_s|\cos\left(\pi - \theta^\prime_{rs}\right)
   \label{cos_lor}
\end{align}

Note that equation \ref{cos_lor} now gives \emph{interior} dihedral
angles.  In the following we shall use the Euclidean formulae,
equations \ref{cos} and \ref{sin}, for $T_{SU(2)}$ on the left hand
side of the following equations and the Lorentzian formulae, equations
\ref{sin_lor} and \ref{cos_lor}, for $T_{SU(1,1)}$ on the right side
of the following equations, thus we get interior rather than exterior
angles for the SU(1,1) case. To prevent confusion we shall denote an
interior dihedral angle as $\bar{\theta}_{hk}$ and so we have $\pi
-\bar{\theta}_{hk} = \theta_{hk}$

We may show
\begin{align}
\sin\left(\theta_{12}+\theta_{34}\right) = &
\sin\left(\bar{\theta}_{13}^\prime+\bar{\theta}_{24}^\prime\right)\label{ex}\\
\sin\left(\theta_{12}-\theta_{34}\right) = &
\sin\left(\bar{\theta}_{12}^\prime-\bar{\theta}_{34}^\prime\right)\\
\sin\left(\theta_{13}+\theta_{24}\right) = &
\sin\left(\bar{\theta}_{24}^\prime-\bar{\theta}_{13}^\prime\right)\\
\sin\left(\theta_{13}-\theta_{24}\right) = &
\sin\left(-\bar{\theta}_{12}^\prime-\bar{\theta}_{34}^\prime\right)\\
\sin\theta_{14} = & \sin\bar{\theta}_{14}^\prime\\
\sin\theta_{23} = & \sin\bar{\theta}_{23}^\prime\label{ex1}
\end{align}

The proof is simple, if laborious, algebra; for instance, by using equations 
\ref{cos}, \ref{sin}, \ref{sin_lor}, \ref{cos_lor}
and proposition \ref{p1}, equation \ref{ex} may be reduced to showing
\begin{equation}
j_{12}C_{34}+j_{34}C_{12} = j_{13}^\prime C_{24}^\prime + j_{24}^\prime C_{13}^\prime
\end{equation}
which follows directly from algebra.

The same equations, with sines replaced by cosines, may be derived in a
similar way; so we conclude, since all the $\theta_{ij},
\theta_{ij}^\prime\in\left[0,\pi\right]$,

\begin{align}
\theta_{12}+\theta_{34} & = 
\bar{\theta}_{13}^\prime+\bar{\theta}_{24}^\prime \label{di1}\\
\theta_{12}-\theta_{34} & = 
\bar{\theta}_{12}^\prime-\bar{\theta}_{34}^\prime \label{di2}\\
\theta_{13}+\theta_{24} & = 
\bar{\theta}_{24}^\prime-\bar{\theta}_{13}^\prime + 2n_1\pi\label{di3}\\
\theta_{13}-\theta_{24} & = 
-\bar{\theta}_{12}^\prime-\bar{\theta}_{34}^\prime + 2n_2\pi\label{di4}\\
\theta_{14} & = \bar{\theta}_{14}^\prime\\
\theta_{23} & = \bar{\theta}_{23}^\prime
\end{align}
where the $n_i = 1$ or $0$.

And hence that
\begin{align}
\theta_{12} & = \pi - \frac{1}{2}\left(\theta_{12}^\prime+\theta_{13}^\prime-
\theta_{34}^\prime+\theta_{24}^\prime\right)\label{ang1}\\
\theta_{13} & = -\pi-\frac{1}{2}\left(-\theta_{12}^\prime-\theta_{13}^\prime-
\theta_{34}^\prime+\theta_{24}^\prime\right)+\left(n_1+n_2\right)\pi \\
\theta_{14} & = \pi-\theta_{14}^\prime\label{ang2}\\
\theta_{34} & = \pi-\frac{1}{2}\left(-\theta_{12}^\prime+\theta_{13}^\prime+
\theta_{34}^\prime+\theta_{24}^\prime\right)\\
\theta_{24} & = \pi-\frac{1}{2}\left(\theta_{12}^\prime-\theta_{13}^\prime+
\theta_{34}^\prime+\theta_{24}^\prime\right)+\left(n_1-n_2\right)\pi\label{ang3}\\
\theta_{23} & = \pi-\theta_{23}^\prime\label{ang6}
\end{align}
where we are now relating the \emph{exterior} dihedral angles.

Now, the sum of the interior dihedral angles around any vertex for a
Euclidean tetrahedron are greater than $\pi$, while those for the top
and bottom vertices of the SU(1,1) tetrahedron are less than
$\pi$. Indeed for every vertex of a Euclidean tetrahedron one may
associate a spherical triangle whose interior angles correspond to the
tetrahedron's interior dihedral angles; each of the three triangles
meeting at a given vertex defines a plane and the intersection of
these planes with a sphere defines the triangle. For a $T_{SU(1,1)}$
the top and bottom vertices define hyperbolic triangles via an
intersection with hyperbolic space in much the same way.

Thus, from equations \ref{ang1}, \ref{ang2} and \ref{ang3},
\begin{equation} \label{te1}
2\pi > \left(\theta_{12}+\theta_{24}+\theta_{14}\right)
 = 3\pi - \left(\theta_{12}^\prime+\theta_{24}^\prime+\theta_{14}^\prime\right)
+\left(n_1-n_2\right)\pi
\end{equation}
where $$\theta_{12}^\prime+\theta_{24}^\prime+\theta_{14}^\prime > 2\pi$$

Now consider a long thin $T_{SU(1,1)}$ that is on the verge of
degenerating into a line. We have $j_{14}^\prime+j_{34}^\prime\approx
j_{24}^\prime\approx j_{12}^\prime+j_{23}^\prime$
with $\theta_{12}^\prime+\theta_{24}^\prime+\theta_{14}^\prime\approx
2\pi$ which implies for $T_{SU(2)}$ $j_{12}+j_{13}\approx j_{14}$ and
$j_{13}+j_{34}\approx j_{23}$ so that
$\theta_{12}+\theta_{24}+\theta_{14}\approx 2\pi$

Thus, in this case, we have $n_1 = 1$ and $n_2 = 0$. Now vary the edge
lengths $j_{hk}^\prime$ continuously. Since the dihedral angles depend
continuously on the edge lengths, the angles will vary continuously
between 0 and $\pi$.  Thus, by continuity, the result holds generally;
which concludes the proof of theorem \ref{dihedral_pos}.

\subsection{The case where $V^2 < 0$} \label{Cayneg}

If the Cayley determinant is negative then we do not have the above
embedding into Minkowski space. It is clear the metric has signature
$(+,+,-)$ or $(-,-,-)$, but the latter, being equivalent to an
embedding into Euclidean space, cannot happen. Thus geometrically we
embed in a spacetime with metric $(+,+,-)$ and regard the edges of the
tetrahedron as space like, while the faces must still be time like since
they satisfy anti-triangle inequalities.

If we define the dihedral angles in the same way to the previous
discussion then, in both cases, they are complex. These complex angles
will be called exterior or interior depending on whether the defining
equation gave exterior or interior dihedral angles in section
\ref{Caypos}.

The possible Lorentzian boosts that take the place of the dihedral
angles in this case come in two flavours, either an interior boost is
defined with no possible exterior boost, or vice versa.  Since the
normals to the faces and the edges are space like, the normals span a
plane in Minkowski space and there will be no exterior boost defined
when two normals are separated by the light cone. A similar criterion
determines the existence of interior boosts.

There are only two patterns that may occur. Either one has three
interior boosts, around one face, with the remainder exterior. Here
opposite edges have different flavours of boost. Or, one has two
exterior boosts and four interior boosts, with opposite edges having
the same flavour. This should be compared to the space like Lorentzian
case for $T_{SU(2)}$\cite{BaFo93} where an identical situation arises
for analogous reasons. In the following the first case will be
referred to as a \emph{type 1} tetrahedron and the second as a
\emph{type 2} tetrahedron for both the $T_{SU(2)}$ and $T_{SU(1,1)}$
cases.

We use the following conventions in making sense of these complex
dihedral angles\cite{BaFo93} that arise when one tries to use the
Euclidean formula to define the dihedral angles.

For $T_{SU(2)}$ we choose an embedding into Lorentzian spacetime with
metric $(-,+,+)$ (so that the sign of the Cayley determinant is
preserved by the transformation). Thus an interior dihedral boost is
given by
$$ \Theta_{hk} = \cosh^{-1}\left(n_h\cdot n_k\right) $$
while an exterior dihedral boost is given by 
$$ \Theta_{hk} = -\cosh^{-1}\left(-n_h\cdot n_k\right) $$

where $n_i$ is the outward normal to the $i$-th triangle.  In the
first case the complex angle $\theta$, given by the usual Euclidean
formula, has the form $\theta_{hk} = \pi + i\Im\theta_{hk}$, while for
the second it is pure imaginary.

For $T_{SU(1,1)}$ we embed into a spacetime as above. Here an exterior
dihedral boost is given by
$$ \Theta_{hk}^\prime = -\cosh^{-1}\left(n_h^\prime\cdot
n_k^\prime\right) 
$$ 
while the interior dihedral boost is given by
$$ \Theta_{hk}^\prime = \cosh^{-1}\left(-n_h^\prime\cdot n_k^\prime\right) 
$$
since the normals are space like and $n^2 = 1$ for a space like unit
vector $n$.  Similarly we have $\theta_{hk}^\prime$ as pure imaginary
for exterior angles, while $\theta_{hk}^\prime = \pi +
i\Im\theta_{hk}^\prime$ for interior angles.

In view of this we make the obvious identification $\Theta_{hk} =
\Im\theta_{hk}$, where $\Theta_{hk}$ is a Lorentzian boost. Such a boost is an
interior dihedral boost when it arises as the imaginary part of a
complex angle given by the usual Euclidean formula for interior
angles. Otherwise it will be called an exterior dihedral boost.

We now state and prove a theorem about the transformation of these
Lorentzian boosts.
\begin{thm} \label{dihedral_neg}
Under equations \ref{b1} - \ref{b6} the boosts transform as:
\begin{align}
\Theta_{12} & = -\frac{1}{2}\left(\Theta_{12}^\prime+\Theta_{13}^\prime-
\Theta_{34}^\prime+\Theta_{24}^\prime\right)\\
\Theta_{13} & = -\frac{1}{2}\left(-\Theta_{12}^\prime-\Theta_{13}^\prime-
\Theta_{34}^\prime+\Theta_{24}^\prime\right)\\
\Theta_{14} & = -\Theta_{14}^\prime \\
\Theta_{34} & = -\frac{1}{2}\left(-\Theta_{12}^\prime+\Theta_{13}^\prime+
\Theta_{34}^\prime+\Theta_{24}^\prime\right) \\
\Theta_{24} & = -\frac{1}{2}\left(\Theta_{12}^\prime-\Theta_{13}^\prime+
\Theta_{34}^\prime+\Theta_{24}^\prime\right)\\
\Theta_{23} & = -\Theta_{23}^\prime
\end{align}
for $V^2<0$.
\end{thm}

Our starting point will be the following equations relating complex
exterior angles on the left to complex interior angles on the right,
as in the previous case with the complex angles still given by the
normal Euclidean formula

\begin{align}
\sin\left(\theta_{12}+\theta_{34}\right) = &
\sin\left(\bar{\theta}_{13}^\prime+\bar{\theta}_{24}^\prime\right)\label{z1}\\
\sin\left(\theta_{12}-\theta_{34}\right) = &
\sin\left(\bar{\theta}_{12}^\prime-\bar{\theta}_{34}^\prime\right)\\
\sin\left(\theta_{13}+\theta_{24}\right) = &
\sin\left(\bar{\theta}_{24}^\prime-\bar{\theta}_{13}^\prime\right)\\
\sin\left(\theta_{13}-\theta_{24}\right) = &
\sin\left(-\bar{\theta}_{12}^\prime-\bar{\theta}_{34}^\prime\right)\\
\sin\left(\theta_{14}\right) = &
\sin\left(\bar{\theta}_{14}^\prime\right)\\
\sin\left(\theta_{23}\right) = &
\sin\left(\bar{\theta}_{23}^\prime\right)\label{z2}
\end{align}

As before, the same equations with sine replaced by cosine are also
valid.  These follow from algebra using the expressions for the sine
and cosine of dihedral angles as in section \ref{Caypos}. We may then
expand these using the standard trigonometric formula for angle sums
and discard the real part of equations \ref{z1} - \ref{z2} (which is
clearly identically zero for both sides).

Hence we are left with the following:
\begin{multline}
\cos\left(\Re\theta_{12}+\Re\theta_{34}\right)
\sinh\left(\Im\theta_{12}+\Im\theta_{34}\right) = \\
\cos\left(\Re\bar{\theta}_{13}^\prime+\Re\bar{\theta}_{24}^\prime\right)
\sinh\left(\Im\bar{\theta}_{13}^\prime+
\Im\bar{\theta}_{24}^\prime\right)\label{dih1}
\end{multline}
\begin{multline}
\cos\left(\Re\theta_{12}-\Re\theta_{34}\right)
\sinh\left(\Im\theta_{12}-\Im\theta_{34}\right) = \\
\cos\left(\Re\bar{\theta}_{12}^\prime-\Re\bar{\theta}_{34}^\prime\right)
\sinh\left(\Im\bar{\theta}_{12}^\prime-
\Im\bar{\theta}_{34}^\prime\right)\label{dih2}
\end{multline}
\begin{multline}
\cos\left(\Re\theta_{13}+\Re\theta_{24}\right)
\sinh\left(\Im\theta_{13}+\Im\theta_{24}\right) = \\
\cos\left(\Re\bar{\theta}_{24}^\prime-\Re\bar{\theta}_{13}^\prime\right)
\sinh\left(\Im\bar{\theta}_{24}^\prime-\Im\bar{\theta}_{13}^\prime\right)
\end{multline}
\begin{multline}
\cos\left(\Re\theta_{13}-\Re\theta_{24}\right)
\sinh\left(\Im\theta_{13}-\Im\theta_{24}\right) = \\
\cos\left(-\Re\bar{\theta}_{12}^\prime-\Re\bar{\theta}_{34}^\prime\right)
\sinh\left(-\Im\bar{\theta}_{12}^\prime-\Im\bar{\theta}_{34}^\prime\right)
\end{multline}
\begin{align}
\cos\left(\Re\theta_{14}\right)
\sinh\left(\Im\theta_{14}\right) = &
\cos\left(\Re\bar{\theta}_{14}^\prime\right)
\sinh\left(\Im\bar{\theta}_{14}^\prime\right)\\
\cos\left(\Re\theta_{23}\right)
\sinh\left(\Im\theta_{23}\right) = &
\cos\left(\Re\bar{\theta}_{23}^\prime\right)
\sinh\left(\Im\bar{\theta}_{23}^\prime\right)
\end{align}

We also gain the same equations with sinh replaced by cosh by taking
the real part of the cosine versions of equations \ref{z1} -
\ref{z2}. It is clear, in the second case, that the result of the
cosine must have the same sign for each side of the equations. From
which we can deduce that the tetrahedron type is preserved by the
transformation and derive (once we have replaced the interior complex
angles on the right hand side by exterior complex angles)

\begin{align}
\Im\theta_{12} & = -\frac{1}{2}\left(\Im\theta_{12}^\prime+\Im\theta_{13}^\prime-
\Im\theta_{34}^\prime+\Im\theta_{24}^\prime\right)\\
\Im\theta_{13} & = -\frac{1}{2}\left(-\Im\theta_{12}^\prime-\Im\theta_{13}^\prime-
\Im\theta_{34}^\prime+\Im\theta_{24}^\prime\right)\\
\Im\theta_{14} & = -\Im\theta_{14}^\prime\\
\Im\theta_{34} & = -\frac{1}{2}\left(-\Im\theta_{12}^\prime+\Im\theta_{13}^\prime+
\Im\theta_{34}^\prime+\Im\theta_{24}^\prime\right)\\
\Im\theta_{24} & = -\frac{1}{2}\left(\Im\theta_{12}^\prime-\Im\theta_{13}^\prime+
\Im\theta_{34}^\prime+\Im\theta_{24}^\prime\right)\\
\Im\theta_{23} & = -\Im\theta_{23}^\prime 
\end{align}
which concludes the proof of theorem \ref{dihedral_neg}.

For the transformation of the real part of the complex dihedral angle (as
defined by the Euclidean formula) we have the following result

\begin{thm} \label{real}
Under equations \ref{b1} - \ref{b6} the real parts of the dihedral `angles' 
transform as:
\begin{align}
\Re\theta_{12} & = \pi - \frac{1}{2}\left(\Re\theta_{12}^\prime+\Re\theta_{13}^\prime-
\Re\theta_{34}^\prime+\Re\theta_{24}^\prime\right)\\
\Re\theta_{13} & = -\frac{1}{2}\left(-\Re\theta_{12}^\prime-\Re\theta_{13}^\prime-
\Re\theta_{34}^\prime+\Re\theta_{24}^\prime\right)\\
\Re\theta_{14} & = \pi - \Re\theta_{14}^\prime \\
\Re\theta_{34} & = \pi - \frac{1}{2}\left(-\Re\theta_{12}^\prime+\Re\theta_{13}^\prime+
\Re\theta_{34}^\prime+\Re\theta_{24}^\prime\right) \\
\Re\theta_{24} & = 2\pi-\frac{1}{2}\left(\Re\theta_{12}^\prime-\Re\theta_{13}^\prime+
\Re\theta_{34}^\prime+\Re\theta_{24}^\prime\right)\\
\Re\theta_{23} & = \pi - \Re\theta_{23}^\prime
\end{align}
for $V^2<0$.
\end{thm}

Indeed it is almost obvious that the real parts must transform in the
same way as the dihedral angles for the tetrahedra with positive
Cayley determinant. The real parts of the angles correspond to a least
degenerate geometric configuration of the edges for an embedding into
the space in which we may legitimately embed the associated positive
Cayley determinant tetrahedra.

Thus for $T_{SU(2)}$, type 1 tetrahedra are characterised in Euclidean
space by three of the faces lying flat on one face and failing to meet
at a vertex. It is clear that rotating the faces upwards in Euclidean
space simply makes the configuration more degenerate. Thus the
Euclidean `dihedral angles' \emph{are} given by the real part. The
type 2 tetrahedra in this case consist of a pair of triangles lying
flat on another pair of triangles in a least degenerate configuration
as well. Again we find the Euclidean `dihedral angles' given by the
real part.

For $T_{SU(1,1)}$ we have an analogous situation. For instance a type
1 tetrahedron embedded into $(+,-,-)$ Lorentzian space consists of
three overlapping faces lying flat on one face. It is clear that
boosting the faces outwards makes them more degenerate since they
overlap more.  Thus we may apply theorem \ref{dihedral_pos} to the
real parts of the dihedral angles by regarding it as simply a
transformation of two degenerate positive Cayley determinant
tetrahedra to gain theorem \ref{real} as a corollary.

\section{Asymptotics} \label{asymp}

It is of interest to see if one can find a similar asymptotic formula
to the Ponzano-Regge formula for the $SU(2)$ Racah coefficient. Their
formula for $V^2 >0$, from \cite{PR}, is
\begin{equation} 
\left(-1\right)^{a+b+d+e}
\left\{ \begin{array}{ccc}
a & b & c \\ d & e & f \end{array} \right\}\sim 
\frac{1}{\sqrt{12\pi V}}\cos\left(\sum^4_{h,k=4}
j_{hk}\theta_{hk}+\frac{\pi}{4}\right) \label{PR_pos1}
\end{equation}
where $\theta_{hk}$ is defined as previously, $V$ is the volume and
each $j_{ik} = r + \frac{1}{2}$ for $r$ the appropriate index of the
\su{2} Racah coefficient as defined in equation \ref{6jdef}.

\begin{figure}[htb]
\epsfbox{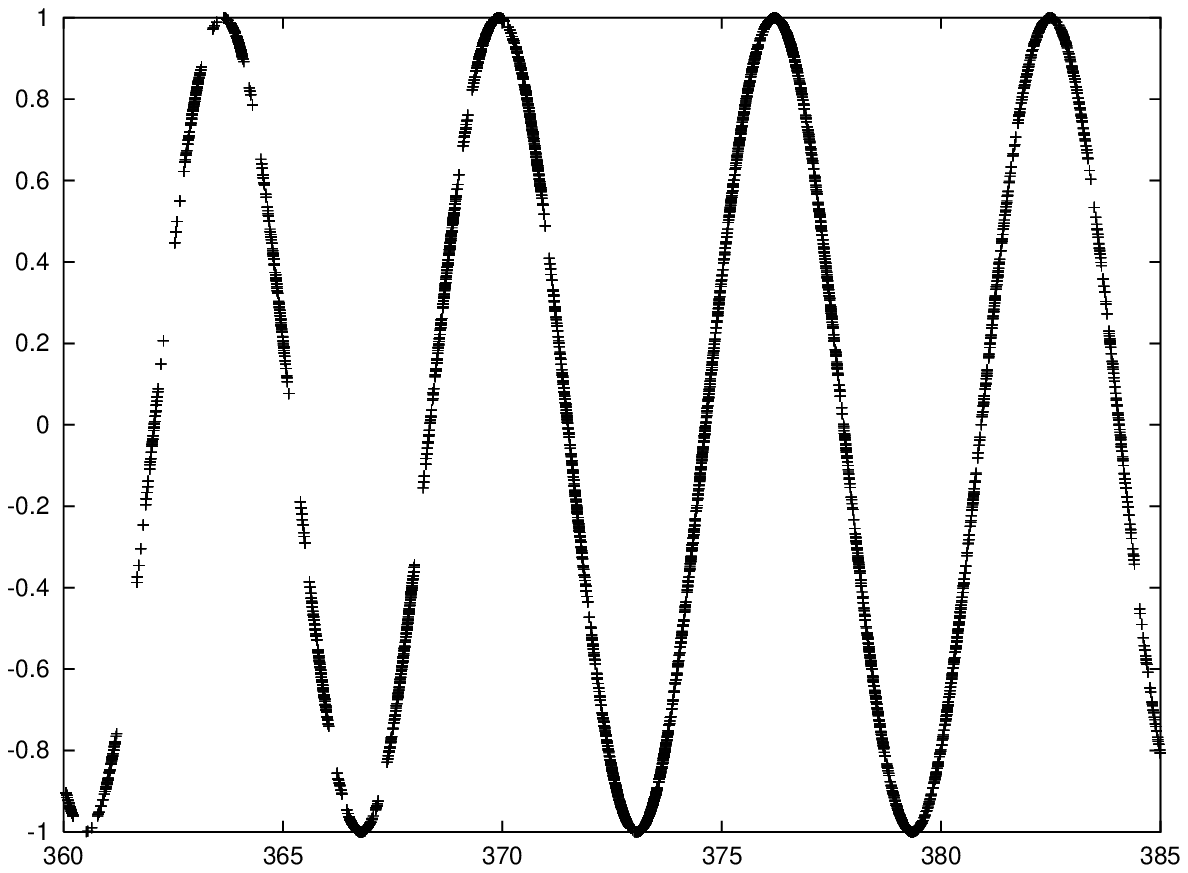}
\caption[Asymptotics of the \su{2} Racah coefficient]
{Asymptotics of the \su{2} Racah coefficient: a plot of
$\sum^4_{h,k=0} j_{hk}\theta_{hk}$ (x-axis) versus $\sqrt{12\pi
V}\left\{\begin{array}{ccc} a & b & c\\ d & e & f
\end{array}\right\}$ (y-axis)\label{su2}}
\end{figure}

There has never been a direct proof of the validity of this formula
but a formula asymptotic to equation
\ref{PR_pos1} has been proven in \cite{Flude,Rob98} and numerical
results give a good indication of its validity. Indeed we have plotted
some values in figure \ref{su2}, which gives a clear cosine shape.

For the $SU(1,1)$ Racah coefficient we have been considering, one may,
subject to the validity of equation \ref{PR_pos1}, derive the
following
\begin{prop}
\begin{equation}
\left\{ \begin{array}{ccc}
a & b & c \\ d & e & f \end{array} \right\}_1 
\sim\frac{1}{\sqrt{12\pi V}}
\left(-1\right)^{j^\prime_{12} +j_{14}^\prime +j_{34}^\prime
+j_{23}^\prime}\cos\left(\sum^4_{h,k=4}
j_{hk}^\prime\theta_{hk}^\prime -\frac{\pi}{4}\right)
\end{equation}
for $V^2>0$
\end{prop}
where here each $j_{ik} = r - \frac{1}{2}$ for $r$ the appropriate
index of the \su{1,1} Racah coefficient defined in equation
\ref{RacahSU11}.

In view of theorem \ref{dihedral_pos}, one should consider how the
quantity $\sum^4_{h,k=0} j_{hk}\theta_{hk}$ transforms under
equations \ref{a} - \ref{f}. Using equations \ref{a1} - \ref{a2} and
the orthogonality of the transformation from $T_{SU(2)}$ to
$T_{SU(1,1)}$ given by equations \ref{a} - \ref{f}, it is easy to
show that, for $V^2>0$,

\begin{equation}
\sum^4_{h,k=0}j_{hk}\theta_{hk} = 
-\sum^4_{h,k=0}j_{hk}^\prime \theta_{hk}^\prime
+\left(j^\prime_{12} +j_{14}^\prime +j_{34}^\prime+ 2j^\prime_{24}
+j_{23}^\prime\right)\pi
\end{equation}

\begin{figure}[htb]
\epsfbox{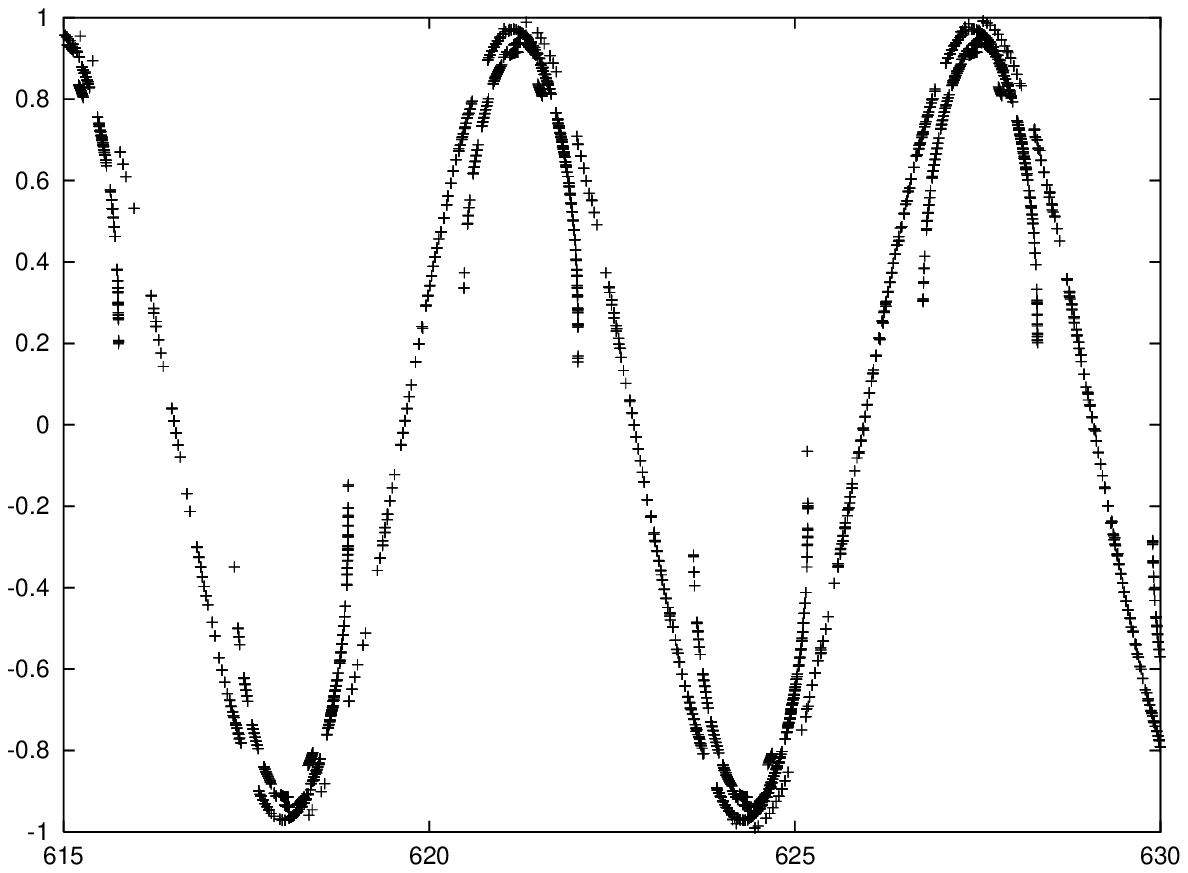}
\caption[Asymptotics of the \su{1,1} positive discrete series
Racah coefficient for $V^2>0$]{The asymptotics of the \su{1,1}
positive discrete series Racah coefficient for $V^2 >0$: a plot of
$-\sum^4_{h,k=0}j_{hk}^\prime
\theta_{hk}^\prime +\left(j^\prime_{12} +j_{14}^\prime +j_{34}^\prime+
2j^\prime_{24} +j_{23}^\prime\right)\pi$ (x-axis) versus $\sqrt{12\pi
V}\left\{\begin{array}{ccc} a & b & c\\ d & e & f
\end{array}\right\}$ (y-axis)\label{sl2}}
\end{figure}

This completes the proof.

We show the validity of this result in figure \ref{sl2}. One might be
concerned by the regions that fall off more steeply than a cosine in
the figure, however numerical results indicate that the tetrahedra in
these regions have at least one face that is reasonably close to being
degenerate, and as such we might expect the above asymptotic formula
to be a worse approximation here.

For the case $V^2 <0$ Ponzano and Regge's exponentially
decaying asymptotic formula for the $SU(2)$ Racah coefficient is:

\begin{equation} \label{PR_neg}
\left(-1\right)^{a+b+c+d}
\left\{ \begin{array}{ccc}
a & b & c \\ d & e & f \end{array} \right\}\sim 
\frac{1}{2\sqrt{12\pi\left| V\right|}}\cos\phi\exp\left(-\left|\sum^4_{h,k=0}
j_{hk}\mathrm{Im}\theta_{hk}\right|\right)
\end{equation}
where
\begin{equation}
\cos\phi = \left(-1\right)^{\sum\left(j_{hk}-\frac{1}{2}\right)m_{hk}}
\end{equation}
and $m_{hk}$ is 1 if $\theta_{hk}$ is an interior angle, and 0
otherwise.  Here each $j_{ik} = r + \frac{1}{2}$ for $r$ the
appropriate index of the \su{2} Racah coefficient.

There has been no proof of the validity of this formula, although
numerical results provide substantial agreement. Assuming its validity
we may derive the following for the $\su{1,1}$ Racah coefficient
\begin{prop}
\begin{equation}
\left\{ \begin{array}{ccc}
a & b & c \\ d & e & f \end{array} \right\}_1  
\sim\left(-1\right)^{2j_{24}}
\frac{1}{2\sqrt{12\pi\left| V\right|}}
\cos\phi^\prime\exp\left(-\left|\sum^4_{h,k=0}
j_{hk}^\prime\Theta_{hk}^\prime\right|\right)
\end{equation}
for $V^2<0$ and $\phi^\prime$ as in equation \ref{phiprime}, and each
$j_{ik} = r - \frac{1}{2}$ for $r$ the appropriate index of the
\su{1,1} Racah coefficient.
\end{prop}

Applying theorem \ref{dihedral_neg} and using the orthogonality up to
sign of the transformation as before we see

\begin{equation}
\left|\sum^4_{h,k=0}j_{hk}\Theta_{hk}\right|
 = \left|\sum^4_{h,k=0}j_{hk}^\prime \Theta_{hk}^\prime\right|
\end{equation}

\begin{figure}[htb]
\epsfbox{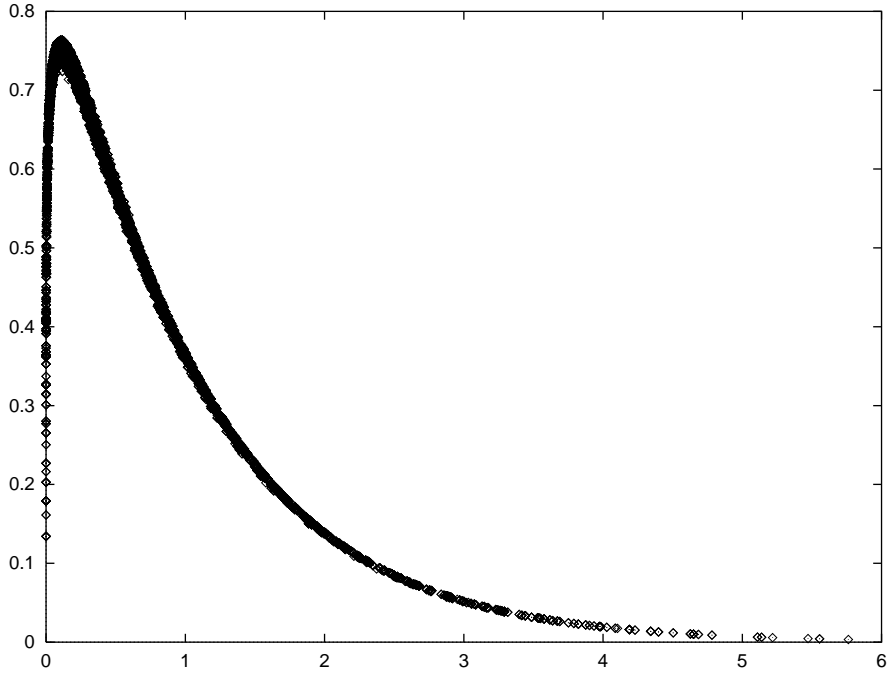}
\caption[The asymptotics of the \su{1,1} positive discrete series 
Racah coefficient with $V^2<0$]{The asymptotics of the \su{1,1} Racah
coefficient for $V^2<0$: a plot of $\left|\sum^4_{h,k=0}
j_{hk}^\prime\theta_{hk}^\prime\right|$ (x-axis) versus
$\frac{2\sqrt{12\pi\left|V\right|}}{\cos\phi^\prime}
\left\{\begin{array}{ccc}a & b & c\\ d & e & f
\end{array}\right\}$ (y-axis)
\label{sl2.exp}}
\end{figure}

For the quantity
\begin{equation}
\phi = \sum\left(j_{hk}-\frac{1}{2}\right)\Re\theta_{hk}
\end{equation}
we may use theorem \ref{real} for the transformation of the real part
and apply equations \ref{a} - \ref{f} to the edge lengths. The
resulting transformations are orthogonal up to a shift depending on
the edge lengths and we may derive the following
\begin{equation} \label{phiprime}
\phi^\prime = -\sum_{hk}\left(j_{hk}^\prime-\sigma_{hk}\right)\Re\theta_{hk}^\prime
+(j^\prime_{12} +j_{14}^\prime +j_{34}^\prime+ 2j^\prime_{24}
+j_{23}^\prime - 3)\pi
\end{equation}
where 
\begin{equation} 
\sigma_{hk} = \left\{ \begin{array}
{r@{\quad\mathrm{for}\quad}l} 
0 & (h,k) = (1,2),(1,3),(3,4)\\
-\frac{1}{2} & (h,k) = (1,4),(2,3)\\
-1 & (h,k) = (2,4)
\end{array}\right.
\end{equation}

We have plotted some values for this in figure \ref{sl2.exp} to show the
validity of this result.

\chapter{Conclusions}\label{CON}

While the state sum model defined by equation \ref{statesum} has been
shown to possess the correct geometry for a theory of 2+1 Lorentzian
quantum gravity there are a number of questions and problems that
remain unaddressed.  The purpose of this chapter is to briefly discuss
these issues.

The most glaring issue is shared with the Ponzano-Regge model and is
simply the fact that equation \ref{statesum} will fail to converge in
a large number of, if not most, circumstances.  In particular, since
the 4-1 Pachner move given by equation \ref{4-1-Pachner} introduces a
clear divergence, the state sum will not converge when the
triangulation of a given manifold has an internal vertex of this sort.

One should recall from section \ref{pr-sect} that for the analogous
issue in the Ponzano-Regge model one may deform the underlying Lie
algebra, \su{2}, as a Hopf algebra with the deformation parameter
specialised to a root of unity and use this to truncate the category
of representations so that only a finite number of representations
occur but the essential tensor structure of the category, and hence
the geometry of the model, is preserved.

One might hope that a similar thing would help in the \su{1,1} case.
Such a deformation would need to render the dimension of the category
of representations, as given by equation \ref{quantdim}, finite to
achieve its objective.  In order to do this it must certainly make the
continuous part of the principal series representations,
$\mathcal{C}_\epsilon^s$, discrete in order to remove the troublesome
$\delta\left(0\right)$ and most likely perform some truncation of the
Principal series to a finite number of representations.  We do not
know of the existence of such a deformation of \su{1,1}.

A further issue is related to the invariance of equation
\ref{statesum} under a relabelling of the vertices of the simplicial
manifold in question.  In section \ref{inv} we constructed canonical
isomorphisms between the Racah coeffients associated to a given
tetrahedron under two different labelling schemes, however in general
this isomorphism introduces some phase that depends on specifics of
each Racah coefficient.  Thus under a relabelling equation \ref{inv}
isn't actually invariant.  There is, however, a clear recipe for
deriving a relabelled version which will associate the same number to
the relabelled simplicial manifold as equation \ref{inv} associated to
the original labelling of the simplicial manifold.

For the Ponzano-Regge model such a subtlety is resolved by not using
the Racah coefficient, which has this problem, but the so-called 6j
symbol which has its phase convention chosen in such a way that these
phase discrepencies are no longer present when the simplicial manifold
is relabelled and the analogous canoncial isomorphism to that
constructed in section \ref{inv} is the identity.  We do not know if
it is possible to construct such an object for \su{1,1} by using, say,
a different choice of phase conventions in equation \ref{RAC}.

Another issue worthy of further investigation is the derivation of
formulae and asymptotics of the Racah coefficient in other cases of
coupling beyond the limited cases discussed in section \ref{rac}.
While it is clear simply from their definition that they have the
necessary geometry to represent the required types of Lorentzian
tetrahedra, nothing at all is known about their asymptotic behaviour
when their indices become large.  This is crucial in making the
necessary connections with physics that a state sum model for quantum
gravity should have.

\chapter*{Acknowledgments}
\addcontentsline{toc}{chapter}{Acknowledgments}
I wish to thank John Barrett for many useful, patient discussions over
the years, much encouragement and for pointing out many mistakes and
omissions in earlier versions of this work.  Thanks are also due to
Roger Francis Picken for giving me the opportunity to present some of
the ideas in this thesis at the Instituto Superior T\'{e}cnico,
Lisbon, and to Louis Crane for useful discussions on some aspects of
this work.

The work was funded by an EPSRC research studentship.  

\newpage
\addcontentsline{toc}{chapter}{Bibliography}
\bibliographystyle{stefan}
\bibliography{paper}

\end{document}